\documentclass[aps,prd,showpacs,superscriptaddress,twocolumn]{revtex4}
\usepackage{epsfig}
\usepackage[english]{babel}
\usepackage{pgf}
\usepackage{amsfonts,amsmath,amssymb,enumerate,array,amssymb}
\usepackage[latin1]{inputenc}
\usepackage{bm}
\usepackage{fixmath}
\usepackage{amsbsy}
\usepackage{graphicx}
\usepackage{capt-of}
\usepackage{dcolumn}

\usepackage{layouts}
\usepackage{natbib}
\usepackage{abrev-jour-long}

\begin{document}

\title{PPN rotation curves in static distributions with spherical symmetry}

\author{Henrique Matheus Gauy}\email[]{henmgauy@df.ufscar.br}
%\affiliation{Departamento de F\'{i}sica, Universidade
	%Federal de S\~{a}o Carlos, 13565-905, SP, Brazil}

\author{Javier Ramos-Caro}\email[]{javier@ufscar.br}
\affiliation{Departamento de F\'isica, Universidade
Federal de S\~{a}o Carlos, S\~ao Carlos, 13565-905 SP, Brazil}


\begin{abstract}

From a Parametrized Post-Newtonian (PPN) \cite{will1,will2,will3,will4,will5,will6} perspective, we address
 the question of whether or not the new degrees of freedom, represented by the PPN potentials, can
 lead to significant modifications in the dynamics of galaxies in the direction of rendering dark matter obsolete.
 Here, we focus on the study of rotation curves associated with spherically symmetric configurations.

The values for the post-Newtonian parameters, which help us to classify the different metric theories of gravity,
are tightly constrained mainly by
solar system experiments \cite{will2,will3}. Such restrictions renders the modifications of gravitational effects,
with respect to General Relativity (GR), to be insignificant, making attempts to find alternative metrical theories rather
 fruitless. However, in recent years, metric theories characterized by screening mechanisms
 \cite{avilez,eugeny,jose,JOYCE20151} have become popular, due to the fact that
 they lead to the possibility of modifications in larger scales than the solar system while retaining the success of GR on it,
 allowing for violations of the constraints of the Post-Newtonian parameters.

 In such a context, we consider here two kinds of solutions for field equations: (i) Vacuum solutions (i.e.
when no matter fields are present) and (ii) fields in the presence of a politropic distribution of matter.
For the case (i) we find that the post-Newtonian corrections do not lead to modifications significant enough to be considered an
 alternative to the dark matter hypothesis.
 In the case (ii) we find that for a wide range of values for the
 PPN parameters $\gamma$, $\beta$, $\xi$, $\alpha_{3}$, $\zeta_{1}$ and
 $\zeta_{2}$, the need for dark matter is unavoidable, in order to find flat rotation curves.
 It is only for theories in which $\zeta_{3}>0$ that some resemblance of flat rotation curves is found. The latter suggests, at least for the models considered, that these are the only theories capable of replacing dark matter as a possible explanation for the dynamics of galaxies.

\end{abstract}

%\begin{keywords}
%galaxies: spiral -- galaxies: kinematics and dynamics.
%\end{keywords}
\date{\today}

\maketitle

\section{Introduction}

The success of the Theory of General Relativity (TGR) lies both in its conceptual beauty
and in the fact of adjusting satisfactorily to a considerable amount of observations.
In fact, it is widely known that TGR has among the best comparison with experiment of modern science.
In the solar system, mediated by the Parametrized Post-Newtonian formalism (PPN),
it agrees significantly with observations \cite{will1,will2,will3,will4,will5,will6},
and, in more recent years, it has been shown to be in agreement with gravitational radiation experiments,
as measured by Ligo \cite{ligo}.
However, despite the great success of TGR, there are still big questions to be solved as, for example: Is General Relativity valid in galactic scales?

Vera rubin et al. introduced the idea, via galactic rotation curves, that visible matter is  insufficient to explain the
 internal dynamics of a number of galaxies \cite{Rubin1,Rubin2}. This is known as the ``missing mass problem'', the central
  evidence that some form of dark matter must exist in galaxies.
 Some other indirect evidences such as the stability of
galaxies, galaxy cluster dynamics, cosmological structure formation and cosmic microwave background anisotropies all point
out to the existence of such dark component \cite{lensing,tessa,Psaltis2008,BULL201656},

However, no direct evidence about the existence of dark matter is known to this
day \cite{dark1, dark2, dark3, dark4}. This fact, on the other hand, motivates the searching for different
explanations to the missing mass problem, as the formulation of
 alternative theories of gravity capable of making obsolete the
dark matter hypothesis. In the literature there are
several proposals for modifying gravity, most with the intent of explaining the phenomenology of cosmological regimes,
mainly the accelerated expansion (see for example
 \cite{will3,clifton,JOYCE20151}). There exist theories with some success as alternatives to dark matter at galactic scales.
Perhaps the most notorious is Mond (Modified Newton Dynamics)\cite{milgrom1,milgrom2,milgrom3,milgrom5,milgrom6,mond},
 which proposes that the gravitational force behaves differently below a fundamental acceleration.
 But as successful as it is at
  explaining the rotation curves of various galaxies and the Tully-Fisher relation
   \cite{sandersMcgaughARAA2002,stacy1,stacy2,famaeyMcgaugh2012LRR}, recent developments
   have made this interesting possibility highly unlikely \cite{DRodrigues}.

Changing the foundations of a fundamental theory such as gravity is no easy feat.
The vast space of possible competing theories makes treating this problem, case by case, an unsurmountable task.
A more economic way is to have on hand a formalism containing, as particular cases, a large amount of theories.
Based on this idea, several general schemes have been constructed,
 and the Parameterized Post-Newtonian (PPN) formalism , by Will et al \cite{will1,will2,will3,will4,will5,will6},
is probably the most used.
It generalizes the Post-Newtonian expansion for metrical theories of gravity by
the introduction of 10 parameters
(i.e. the so called post-Newtonian parameters) which help us to
 differentiate between the competing theories. In the last decades, highly precise experiments
 in the solar system have lead to ever increasingly smaller bounds for the PPN parameters
 (some being at most $10^{-20}$ \cite{will3}), leading to discard most alternative theories.

However, recent developments suggest that this may not be the final word for a number of alternative metrical theories.
The existence of \emph{screening mechanisms} (SM) allows these theories to pass experiments in solar scales without
maiming its deviations from General Relativity at larger scales \cite{avilez,eugeny,jose,JOYCE20151}.
These theories, with some SM, cannot be perturbatelly expanded from infinity all the way
to the Schwarzschild radius \cite{avilez}, a basic tenet of the classical perturbation theory (and, therefore,
of the PPN approach).
This fact leads to different predictions at solar system scales, than the ones by the PPN formalism.
The reason for this is that the failures of the perturbative expansion only happen
 within a screening radius $r_{V}$ \cite{avilez},
outside of which the classical perturbation theory becomes applicable. Therefore, in accordance with \cite{avilez}, at
least as long we are far from $r_{V}$, the linear theory should be the PPN approach.

Not all theories with some SM can be used to justify modifications at the astrophysical scale. Take
for instance the cubic galileon \cite{avilez}, which predicts the screening for all scales smaller
than galaxy clusters, therefore excluding astrophysical modifications.
On the other hand, the problem on the existence of SMs that allows modifications
at the astrophysical scale, is a question addressed by \cite{cubicdark,salzano}, showing that some
SM mimic the effects of dark matter in galaxy clusters and in our own galaxy.

In this paper we assume the existence of some SM capable of hiding the modifications to TGR
in some scale (say, of the order of the solar system),
but keeping the modifications to TGR at the scale of galaxies or galaxy clusters
(i.e. the existence of a theory decoupling solar system from astrophysical scales).
As long as the analysis is restricted to outside and far from $r_{V}$, the perturbative expansion should be the PPN
scheme\cite{avilez}.
In the face of such prerogative it is pertinent to raise some questions:
what are the corrections to the rotation curves that
arise as consequences of the PPN potentials? Are these corrections enough to explain the
missing mass problem in galaxies?

For the sake of simplicity, one can start to address this problem by obtaining the PPN potentials generated by
static configurations with spherical symmetry, in order to obtain an expression for the corresponding
circular velocity, as done in section \ref{sec3sub2}. There we will find that the existence of circular motion
(and, therefore, the existence of rotation curves) depends strongly of the vanishing of two PPN parameters. In particular,
it is required that $\alpha_1 = \alpha_2 =0$. In the same section we also study the behavior of circular velocity
outside a sphere with finite radius, finding that no choice of the
PPN parameters can lead to the flattening of the rotation curves. Away from the spherical distribution, rotation curves
exhibit the usual Keplerian fall off.

In section \ref{sec3sub3} we analyze the behavior of circular velocity inside static spherical distributions,
focusing on the case of the so-called polytropes,
one of the simplest models with some relevance in galactic dynamics.
To construct  polytropic models in the PPN approximation we use the ``$f$ to $\rho$'' approach of
galactic dynamics\cite{binneytremaineGD,ramos1,ramos2},
in which the matter distribution (given by mass density, $\rho$, in Newtonian gravity)
is obtained from a known distribution function $f$.
In the framework of metrical theories, and thereof the PPN scheme, we assume that the distribution function (DF) satisfies
a generalized version of the collisionless Boltzmann equation (CBE) \cite{Kremer}, also called the Vlasov equation.
This assumption is valid whenever the system in study is sufficiently smooth and encounters can be disregarded,
as in the case of galaxies \cite{binneytremaineGD}.
Then, as a first step we will derive a version of the CBE that accounts for the first PPN
corrections, in a similar fashion as in \cite{ramos1}, focusing on stationary solutions.
To construct the Polytropes, by extension, we will provide an ergodic DF (i.e. depending on energy) with
the same form as in the Newtonian case, determining all the corresponding matter fields.

In section \ref{sec5} we show that, for polytropic models,
only one of the PPN parameters ($\zeta_{3}>0$) can effectively lead to
flattened rotation curves within a certain radius ($\tilde{r}<10$). Outside such region, the usual Keplerian fall off
is still observed, even for the most exotic theories. Everything seems to indicate that, in static spherical distributions,
the modifications encompassed by the PPN scheme are not enough to explain the flatness of rotation curves without the introduction
of additional matter.

The rest of the paper is organized as follows. In Sec. \ref{sec2} we present a brief overview of the PPN formalism, pointing out it's most notable features. In Subsec. \ref{sec3} we introduce the material content we will consider in the modeling. In Sec. \ref{sphericallmodel} we finally specialize to the statical spherically symmetric case presenting the field equations. In Sec. \ref{sec5} we construct the polytropic models and their respective rotations curves for various theories.

Throughout the paper, we will regard Latin indices to run from 1 to 3, i.e. $i=\{1,2,3\}$, and Greek indices to run from 0 to 3, i.e. $\mu=\{0,1,2,3\}$. We will also use, whenever suited, the notation $\partial f/\partial x = f_{,x}$. Terms of different orders of $c$ will carry an indice as follows,
$$
\overset{n}{A}\equiv{}^{n}A\sim c^{-n},
$$
this means, ${}^{n}A$ is of the order of $c^{-n}$.

%--------------------------------------------
%
%(COMPLEMENTAR A INTRODUÇÃO ACIMA)
%
%However, if we want to study the dynamics of huge
%astrophysical ensembles such as galaxies and galaxy clusters,
%physical collisions between the stars are very rare, and
%the effect of gravitational collisions can be neglected for
%times far longer than the age of the universe. Those
%systems are characterized by a relaxation time, trelax, that
%is arbitrarily large in comparison with their crossing time,
%tcross, and this means that they can be approximated as a
%continuum rather than concentrated into nearly pointlike
%stars. The same holds (with some restrictions) in the case of
%collisional systems such as globular clusters, neutron stars
%and white dwarfs, where the relativistic effects of gravitation
%become important. Although trelax here is significantly
%smaller than the system's age, the CBE is still valid over
%periods of time shorter than trelax or when it is recognized
%that the system evolves slowly towards the equilibrium (on
%a timescale of the order of trelax). For example, Taruya and
%Sakagami showed in [13,14] that the evolution of spherically
%symmetric systems in the collisional regime can be
%modeled as a sequence of polytropic states (i.e. described
%by a DF proportional to , which is a static solution of the
%CBE), with increasing polytropic index.
%
%
%----------------------------------------------

\section{The PPN Formalism}\label{sec2}
There is a well known approximation scheme for general relativity when we consider weak fields and slow moving matter,
known as the post-Newtonian approximation.
In the context of alternative metric theories of gravity there is a general formalism developed by Will and Nordtvedt \cite{will1,will2,will3,will4,will5,will6},
known as the Parametrized Post-Newtonian (PPN) approximation, that find a similar approximation scheme,
but introducing parameters that are different for each theory.

In this formalism the metric is determined by perturbations
of a flat background in terms of a parameter $\epsilon\ll 1$, such that
$$
\epsilon\sim v^{2}/c^{2}\sim GM/rc^{2}\sim p/\rho c^2 ,
$$
where $v$, $M$, $r$, $p$ and $\rho$ are the characteristic velocity, mass, length (or separation), pressure and density
in the system, and  G is the
%Modifiquei%
(Newtonian)
%Modifiquei%
 gravitational constant. At first approximation, the metric can be written as
\cite{will2}
\begin{equation}\label{metric00}
\mathrm{g}_{00}=-1+\frac{2U}{c^2}+\frac{\mathcal{W}}{c^4}+O\left(c^{-6}\right),
\end{equation}
\begin{equation}\label{metric0j}
\mathrm{g}_{0j}=\frac{\mathcal{Q}_{j}}{c^3}+O\left(c^{-5}\right),
\end{equation}
\begin{equation}\label{metricij}
\mathrm{g}_{jk}=\left(1+\frac{2\gamma U}{c^2}\right)\delta_{jk}+O\left(c^{-4}\right),
\end{equation}
where $U$ reduces to the Newtonian potential in the limit $c\rightarrow\infty$ and
functions $\mathcal{W}$, $\mathcal{Q}_{j}$ are defined in terms of  post-Newtonian potentials $\Psi$, $X$,
$U_j$ $(j=1,2,3)$, $\phi_k$ ($k=1,...,6$), $\phi_w$, $\Phi^{\text{{PF}}}$, $\Phi^{\text{{PF}}}_{j}$, as
\begin{eqnarray}
\mathcal{W}&\equiv&2\left(\Psi-\beta U^2\right)+\Phi^{\text{{PF}}},\label{W}\\
&&\nonumber\\
\mathcal{Q}_{j}&\equiv&-\left[2\left(1+\gamma\right)+\frac{\alpha_{1}}{2}\right]U_{j}\nonumber\\
&&\nonumber\\
&&-\frac{1}{2}\left[1+\alpha_{2}-\zeta_{1}+2\xi\right]X_{,tj}+\Phi^{\text{{PF}}}_{j},\label{Q}\\
&&\nonumber\\
\Psi&=&\frac{1}{2}\left(2\gamma+1+\alpha_{3}+\zeta_{1}-2\xi\right)\phi_{1}\nonumber\\
&&\nonumber\\
&&-\left(2\beta-1-\zeta_{2}-\xi\right)\phi_2\nonumber\\
&&\nonumber\\
&& +\left(1+\zeta_{3}\right)\phi_{3}+\left(3\gamma+3\zeta_{4}-2\xi\right)\phi_{4}\nonumber\\
&&\nonumber\\
&&-\frac{1}{2}\left(\zeta_{1}-2\xi\right)\phi_{6}-\xi\phi_{w},\label{psi}
\end{eqnarray}
and $\gamma$, $\beta$, $\xi$, $\alpha_{1}$, $\alpha_{2}$, $\alpha_{3}$, $\zeta_{1}$,  $\zeta_{2}$, $\zeta_{3}$, $\zeta_{4}$ are
the post-Newtonian parameters.

Usually, in the PPN scheme the material content is assumed to be a perfect fluid described by
an energy momentum tensor of the form \cite{will2}
\begin{equation}\label{stressppn1}
c^{-2}T^{00}=\rho^{*}\left\{1+\frac{1}{c^2}\left[\frac{1}{2}u^2-\left(3\gamma-2\right)U+\Pi\right]\right\},
\end{equation}
\begin{equation}\label{stressppn2}
c^{-1}T^{0j}=\rho^{*} u^{j}\left\{1+\frac{1}{c^2}\left[\frac{1}{2}u^2-\left(3\gamma-2\right)U+\Pi+\frac{p}{\rho^{*}}\right]\right\},
\end{equation}
\begin{multline}\label{stressppn3}
T^{ij}=\rho^{*} u^{i}u^{j}\left\{1+\frac{1}{c^2}\left[\frac{1}{2}u^2-\left(3\gamma-2\right)U+\Pi+\frac{p}{\rho^{*}}\right]\right\}
\\
+p\left(1-\frac{2\gamma U}{c^{2}}\right)\delta^{ij},
\end{multline}
where $p$ is the pression field, $u^{k}$ is the velocity field, $\Pi=\varepsilon/\rho^{*}$ ($\varepsilon$ is the internal energy) and
 $\rho^{*}$ is the conserved density, which is related to the proper mass density $\rho$
through the relation
$$
\rho=\left[1-\frac{1}{c^{2}}\left(\frac{1}{2}u^{2}+3\gamma U\right)\right]\rho^{*}.
$$
Such an assumption implies that potentials $U$, $\phi_{1}$, $\phi_{2}$, $\phi_{3}$, $\phi_{4}$,
$\phi_{6}$, $X$, $U^{j}$ and $\phi_{w}$ are determined by the field equations,
\begin{equation}\label{invariant1}
\nabla^{2}\left\{U,\phi_{1},\phi_{2},\phi_{3},U^{j}\right\}=-4\pi G \rho^{*}\left\{1,{u}^2,U,\Pi,{u}^{j}\right\}
\end{equation}
\begin{equation}\label{invariant2}
\nabla^{2}\phi_{4}=-4\pi Gp, \qquad\nabla^{2}X=2U
\end{equation}
\begin{equation}\label{appendix1}
\nabla^{4}\left(\phi_{6}-3\phi_{1}\right)=-2G\left(\rho^{*}{u^{i}}{u^{j}}\right)_{,ij}
\end{equation}
\begin{equation}\label{appendix2}
\nabla^2\left(\phi_{w}+2U^{2}-3\phi_{2}\right)=-2 X_{,ij}U_{,ij},
\end{equation}
whereas the preferred-frame potentials, $\Phi^{\text{{PF}}}$ and $\Phi^{\text{{PF}}}_{j}$, can be written in terms
of the velocity of the PPN coordinate frame relative to an (hypotetical) universal preferred frame, denoted by $w^j$:
\begin{eqnarray}
\Phi^{\text{{PF}}}&=&\left(\alpha_{3}-\alpha_{1}\right)w^2U+\alpha_{2}w^{j}w^{k}X_{,kj}\nonumber\\
&&\nonumber\\
&&+\left(2\alpha_{3}-\alpha_{1}\right)w^{j}U_{j},\label{PF1}\\
&&\nonumber\\
\Phi^{\text{{PF}}}_{j}&=&-\frac{1}{2}\alpha_{1}w_{j}U+\alpha_{2}w^{k}X_{,kj}.\label{PF2}
\end{eqnarray}

All of the above fields are defined in a $3$-dimensional Euclidean space and,
 in particular, equations (\ref{appendix1})-(\ref{PF2}) are stated in
%Mudei%
(Quasi-)
%Mudei
Cartesian coordinates \cite{will2,misner1973gravitation}.
Later on, when we address situations involving spherical symmetry, it will be useful to represent the previous equations in
non-Cartesian coordinates.
As long as the coordinate transformation does not involve time or velocities (changes of reference frame),
we can regard it as a diffeomorphism in Euclidean space which does not require to consider
the PPN metric \eqref{metric00}-\eqref{metricij}. For this reason it is not difficult to find the covariant form of relations
 (\ref{appendix1})-(\ref{PF2}), by switching every common derivative by a covariant derivative and by
 introducing the Euclidean metric $\gamma_{ab}$ at every inner product:
\begin{equation}\label{invariantw}
\nabla^2\left(\phi_{w}+2U^{2}-3\phi_{2}\right)=-2\gamma^{ad}\gamma^{be}X_{;ab}\:U_{;de}\:\:\:,
\end{equation}

\begin{eqnarray}
\Phi^{\text{{PF}}}&=&\left(\alpha_{3}-\alpha_{1}\right)w^2U+\alpha_{2}w^{j}w^{k}X_{;kj}\nonumber\\
&&\nonumber\\
&&+\left(2\alpha_{3}-\alpha_{1}\right)w^{j}U_{j}\:\:,\label{invariantpf}\\
&&\nonumber\\
\Phi^{\text{{PF}}}_{j}&=&-\frac{1}{2}\alpha_{1}w_{j}U+\alpha_{2}w^{k}X_{;kj}\:\:\:,\label{invariantpfj}
\end{eqnarray}
\begin{multline}\label{invariant6}
\nabla^{4}\left(\phi_{6}-3\phi_{1}\right)=-\frac{2G}{\Upsilon}\left(\Upsilon\rho^{*}u^{i}u^{j}\right)_{,ij}
\\
-\frac{2G}{\Upsilon}\left(\Upsilon\left\{{}^{\;j\;}_{ik}\right\}\rho^{*}u^{i}u^{k}\right)_{,j},
\end{multline}
where $\Upsilon=\sqrt{\det \left(\gamma_{ij}\right)}$ and
\begin{eqnarray}
    X_{;ab} &\equiv& X_{,ab}-\left\{{}^{\;c\;}_{ab}\right\}X_{c}\:\:\:, \label{covariant1}\\
    && \nonumber\\
    \left\{{}^{\;c\;}_{ab}\right\}&\equiv&\frac{1}{2}\gamma^{c\mu}\left(\gamma_{a\mu,b}+
    \gamma_{b\mu,a}-\gamma_{ab,\mu}\right) \label{covariant2}
\end{eqnarray}

On the other hand, the equations of motion for a free-falling test particle, can be written in  terms of the
fields appearing in the above expressions:
  \begin{multline}\label{eqmotion}
\frac{\mathrm{d}v^{i}}{\mathrm{d}t}+\left\{{}^{\;i\;}_{jk}\right\}v^{j}v^{k}=
\gamma^{ik}\Bigg\{U_{,k}+\frac{1}{c^2}\Bigg[\frac{1}{2}\mathcal{W}_{,k}-\mathcal{Q}_{k,t}\Bigg.\Bigg.\\
-2\gamma UU_{,k}-\left(\mathcal{Q}_{k,j}-\mathcal{Q}_{j,k}\right)v^{j}-\left(2\gamma+1\right)v_{k}U_{,t}
\Bigg.\Bigg.\\\Bigg.\Bigg.
-2\left(\gamma+1\right) U_{,j}v^{j}v_{k}+\gamma U_{,k}v^{2}\frac{}{}\Bigg]\Bigg\}.
\end{multline}
By choosing the material fields $\rho^{*}$, $\rho^{*}\Pi$, $P$, $\rho^{*}u^{2}$, $\rho^{*}u^{i}$ and $\rho^{*}u^{i}u^{j}$
it should be possible, in principle, to solve \eqref{invariant1}, \eqref{invariant2} and
\eqref{invariantw}-\eqref{invariant6} and, in consequence, determine the test particle motion
through equations \eqref{eqmotion}.

\subsection{Statistical description of collisionless systems}\label{sec3}

 The study of huge astrophysical ensembles such as globular clusters, galaxies and galaxy clusters, is
 simplified by adopting a statistical description and introducing some  assumptions about the configurations:
 (i) the particles of the system have the same mass; (ii) collisions or encounters between particles are insignificant; (iii)
 the gravitational fields are all regarded to be smooth and continuous throughout space.
 Thus the system can be described entirely by a probability density or distribution function (DF),
that represents the number density of particles of a point $\left(x^{i},V^{i}\right)$ in the phase-space (here
$V^i$ represents the four-velocity).
Assumptions (i) and (ii) lead us to consider
that the DF, denoted here by $f\left(x^{i},V^{i}\right)$, satisfy the general-relativistic Vlasov equation,
\begin{equation}\label{vlasov}
V^{\mu}\frac{\partial f}{\partial
x^{\mu}}-\Gamma^{i}_{\mu\nu}V^{\mu}V^{\nu}\frac{\partial f}{\partial V^{i}}=0.
\end{equation}

In order to find equation \eqref{vlasov} to PPN order we follow a similar procedure as in \cite{ramos1}, which
take into account the map $\left(x^{\mu},V^{i}\right)\rightarrow\left(x^{\mu},v^{i}\left(x^{\mu},V^{i}\right)\right)$ and
perform an expansion of the left hand side of (\ref{vlasov}) to post-Newtonian order.
After some calculations, we find the Vlasov equation in the  PPN approach (see Appendix \ref{appendix3}):
\begin{multline}\label{vlasovppn}
\left[1+\frac{U}{c^2}+\frac{v^2}{2c^2}\right]\left(\frac{\partial f}{\partial t}+v^{i}
\frac{\partial f}{\partial x^{i}}\right)+U_{,i}\frac{\partial f}{\partial v^{i}}
\\
-\frac{\partial f}{\partial v^{i}}\frac{v^{i}}{c^2}U_{,t}\left(1+2\gamma\right)-
\frac{\partial f}{\partial v^{i}}\frac{v^{i}v^{j}}{c^2}U_{,j}\left(2+2\gamma\right)
\\
-\frac{\partial f}{\partial v^{i}}\left\{-\frac{1}{2}\frac{\mathcal{W}_{,i}}{c^2}+
\frac{\mathcal{Q}_{i,t}}{c^2}+\frac{2\gamma-1}{c^2}UU_{,i}\right\}
\\
-\left[\mathcal{Q}_{j,l}-\mathcal{Q}_{l,j}\right]\frac{v^{l}}{c^2}
\frac{\partial f}{\partial v^{j}}+\left(\gamma+\frac{1}{2}\right) U_{,j}\frac{v^{2}}{c^2}\frac{\partial f}{\partial v^{j}}=0.
\end{multline}
From \eqref{vlasovppn} one can prove that
$$
\frac{\mathrm{d}f}{\mathrm{d}t}=0,
$$
which means that any stationary solution of equation \eqref{vlasovppn} must be a function only of the integrals
of motion of the system and vice versa: any function of the integrals of motion is a solution of \eqref{vlasovppn}.

From the DF we can build its moments, being the most relevant ones, for this work, the components of the
stress-energy tensor \cite{yvone1}:
$$
T^{\mu\nu}=c\int \frac{V^{\mu}V^{\nu}}{-V_{0}} f \sqrt{-g}\mathrm{d}^{3}V.
$$
Since we want the above expression in PPN approximation, we again consider the map
$\left(x^{\mu},V^{i}\right)\rightarrow\left(x^{\mu},v^{i}\left(x^{\mu},V^{i}\right)\right)$ and the expansion
$$
f=\overset{0}{f}+\overset{2}{f}+...,
$$
which lead us to the following relations:
\begin{equation}\label{stressppn4}
c^{-2}T^{00}=\int \overset{0}{f}\mathrm{d}^{3}v+\int \left(\frac{ {\cal K}}{c^{2}}\overset{0}{f}+\overset{2}{f}\right)\mathrm{d}^{3}v,
\end{equation}
\begin{equation}\label{stressppn5}
c^{-1}T^{0k}=\int v^{k}\overset{0}{f}\mathrm{d}^{3}v+\int v^{k}\left(
\frac{ {\cal K}}{c^{2}}\overset{0}{f}+\overset{2}{f}\right)\mathrm{d}^{3}v,
\end{equation}
\begin{equation}\label{stressppn6}
T^{kj}=\int v^{k}v^{j}\overset{0}{f}\mathrm{d}^{3}v+\int v^{k}v^{j}\left(
\frac{ {\cal K}}{c^{2}}\overset{0}{f}+\overset{2}{f}\right)\mathrm{d}^{3}v,
\end{equation}
where
$$
{\cal K}=U\left(3\gamma+5\right)+3v^{2}.
$$
Comparing \eqref{stressppn4}-\eqref{stressppn6} with \eqref{stressppn1}-\eqref{stressppn3} we can, at least in principle,
find all the matter fields ($\rho^{*}$, $\rho^{*}\Pi$, $P$, $\rho^{*}u^{2}$, $\rho^{*}u^{i}$ and $\rho^{*}u^{i}u^{j}$)
of equations \eqref{invariant1}-\eqref{invariant6}, once we know $f$.

\section{PPN potentials for Static Configurations with Spherical Symmetry }\label{sphericallmodel}

When we are dealing with static spherically symmetric systems, several assumptions are needed.
First we require matter to be static, i.e. with a velocity field $u^{i}=0$ or, in other words, with a DF such that $T^{0k}=0$.
This means that equations \eqref{stressppn1}-\eqref{stressppn3} are simplified:
$$
c^{-2}T^{00}=\rho^{*}\left\{1+\frac{1}{c^2}\left[\Pi-\left(3\gamma-2\right)U\right]\right\},
$$
$$
c^{-1}T^{0j}=0,
$$
$$
{{}^{2}{T^{ij}}}=P\delta^{ij}.
$$
By comparing the previous equations with \eqref{stressppn4}-\eqref{stressppn6},
we find the expression of the matter fields as functionals of the DF,
\begin{equation}\label{stressppn10}
\rho^{*}=4\pi\int \overset{0}{f}v^{2}\mathrm{d}^3v,
\end{equation}
\begin{equation}\label{stressppn11}
P=\frac{4\pi}{3}\int \overset{0}{f}v^{4}\mathrm{d}^3v,
\end{equation}
\begin{equation}\label{stressppn12}
\rho^{*}\Pi=\left(3\gamma-2\right)\rho^{*}U+4\pi\int \left(\overset{0}{f} k+\overset{2}{f}c^{2}\right)v^{2}\mathrm{d}^3v.
\end{equation}
%where $v_{e}$ is the escape velocity, i.e. the one such that the particle is no longer bounded by the system.

We also require that the fields $U$, $\phi_{1}$, $\phi_{2}$, $\phi_{3}$, $\phi_{4}$, $\phi_{6}$, $X$, $U^{j}$ and $\phi_{w}$
are all static, spherically symmetric and well behaved throughout the space. As a consequence of
eqs. (\ref{invariant1}) and the assumption  that
the space-time is asymptotically flat, the fields $U^{j}$ and $\phi_{1}$ can be written as
$$
U^{j}=\frac{A^{j}}{r},\qquad\phi_{1}=\frac{B}{r},
$$
for  $A^{j}$ and $B$ constants (we have introduced  spherical coordinates
$(r,\theta,\varphi)$). Since we require that the fields are well behaved for all $r$, we have to choose $A^{j}=0$ and $B=0$,
in order to avoid the singularity at $r=0$,
which means that $U^{j}=\phi_{1}=0$. Now, from equation (\ref{appendix1}), we have $\phi_{6}=3\phi_{1}$, leading to $\phi_{6}=0$.

Under such assumptions the field equations take a much simpler form:
\begin{equation}\label{fieldU}
\nabla^2U = - 4\pi G \rho^{*},
\end{equation}
\begin{equation}\label{fieldX}
\nabla^2X = 2U,
\end{equation}
\begin{multline}\label{fieldPsi}
\nabla^{2}\Psi=\left(2\beta-1-\zeta_{2}-6\xi\right)4\pi G \rho^{*}U
\\
\quad-\left(1+\zeta_{3}\right)4\pi G \rho^{*}\Pi-\left(3\gamma+3\zeta_{4}-2\xi\right)4\pi G P
\\
+4\xi\left[
4\pi G \rho^{*}\frac{X_{,r}}{r}+U_{,r}\left(U_{,r}+\frac{3X_{,r}}{r^{2}}-\frac{2U}{r}\right)\right],
\end{multline}
\begin{multline}\label{preferredframe1}
\Phi^{\text{{PF}}}=w^{2}\left[\left(\alpha_{3}-\alpha_{1}\right)U+\alpha_{2}\frac{X_{,r}}{r}\right]
\\
+\alpha_{2}\left(w^{r}\right)^2\left(X_{,rr}-\frac{X_{,r}}{r}\right),
\end{multline}
\begin{equation}\label{preferredframe2}
\Phi^{\text{{PF}}}_{r}=-\frac{1}{2}\alpha_{1}w_{r}U+\alpha_{2}w^{r}X_{,rr},
\end{equation}
\begin{equation}\label{preferredframe3}
\Phi^{\text{{PF}}}_{\theta}=-\frac{1}{2}\alpha_{1}w_{\theta}U+r\alpha_{2}w^{\theta}X_{,r},
\end{equation}
\begin{equation}\label{preferredframe4}
\Phi^{\text{{PF}}}_{\varphi}=-\frac{1}{2}\alpha_{1}w_{\varphi}U+\alpha_{2}w^{\varphi}r\sin^{2}\theta X_{,r}.
\end{equation}

Note that the assumptions about symmetry's configuration (spherical, in our case) must be the
 same for  the fields $U$, $X$ and $\Psi$,  as a consequence of eqs. \eqref{fieldU}-\eqref{fieldPsi}.
On the other hand, according to \eqref{preferredframe1}-\eqref{preferredframe4}, the preferred frame potentials,
$\left(\Phi^{\text{{PF}}},\Phi^{\text{{PF}}}_{r},\Phi^{\text{{PF}}}_{\theta},\Phi^{\text{{PF}}}_{\varphi}\right)$,
are not necessarily constrained to satisfy the same symmetry assumptions of configuration.
In fact, in our case all of them would depend on the angular variables $(\theta,\varphi)$ for the case in which $\alpha_{1}$ and $\alpha_{2}$
are non-vanishing and assumptions are not made about the preferred frame velocity $\vec{w}$.

In principle, by introducing a particular DF in \eqref{stressppn10}-\eqref{stressppn12}, we can determine the matter fields
and, consequently, solve the above equations to obtain the fields.
However it is convenient, at first, to take into account some fundamental concepts about the orbits in spherical models.

\section{Rotation curves outside a static spherical configuration }\label{sec3sub2}

Here we focus on the problem of circular motion restricted to the equatorial plane (i.e. equatorial circular orbits) of the configuration,
in order to study the behavior of rotation curves. In spherical coordinates, equatorial circular orbits satisfy
$\theta=\pi/2$, $\ddot{r}=\ddot{\theta}=\dot{r}=\dot{\theta}=0$ and the equations of
motion \eqref{eqmotion} can be cast as
\begin{multline}\label{circmotion}
\left.\left(\dot{\varphi}\right)^2\right|_{\theta=\frac{\pi}{2}} =-\frac{1}{r}\left.\left[U_{,r}+
\frac{\mathcal{W}_{,r}}{2c^2}-\frac{2\gamma UU_{,r}}{c^2}
\right.\right.\\\left.\left.
-\left(\Phi^{\text{{PF}}}_{r,\varphi}-\Phi^{\text{{PF}}}_{\varphi,r}\right)\frac{\sqrt{-rU_{,r}}}{rc^2}
-\frac{\gamma r\left(U_{,r}\right)^2}{c^2}\right]\right|_{\theta=\frac{\pi}{2}},
\end{multline}
\begin{equation}\label{condcircorbits1}
\frac{1}{c^{2}r^2}\left.\left[\frac{\mathcal{W}_{,\theta}}{2}
-\left(\Phi^{\text{{PF}}}_{\theta,\varphi}-\Phi^{\text{{PF}}}_{\varphi,\theta}\right)\dot{\varphi}\right]\right|_{\theta=\frac{\pi}{2}}=0,
\end{equation}
\begin{equation}\label{condcircorbits2}
\left.\ddot{\varphi}\right|_{\theta=\frac{\pi}{2}}=\frac{\left.\mathcal{W}_{,\varphi}\right|_{\theta=\frac{\pi}{2}}}{2c^2 r^{2}}.
\end{equation}
In the derivation of the above relations we use the fact that whenever $\left(r\dot{\varphi}\right)^2$ or $r\dot{\varphi}$ is
accompanied by $c^{-2}$ we can substitute it, in accordance with the PPN order,
by the Newtonian value \cite{ramos1}, i.e. $\left(r\dot{\varphi}\right)^2c^{-2}=-rU_{,r}c^{-2}$.

It is important to note that equations \eqref{condcircorbits1}-\eqref{condcircorbits2}
lead to conditions on the existence of circular orbits,
which can be expressed by the  following relations:
\begin{equation}\label{condition1}
\frac{\alpha_{2}}{c^2 r^{2}}w^{z}\left.w^{r}\right|_{\theta=\pi/2}\left(\frac{X_{,r}}{r}-X_{,rr}\right)=0,
\end{equation}
\begin{equation}\label{condition2}
\alpha_{2}\left.w^{\varphi}\right|_{\theta=\pi/2}X_{,rrr}=-\alpha_{1}U_{,r}\sqrt{-rU_{,r}}
\end{equation}
The first one of the above relations comes from (\ref{condcircorbits1}), after introducing the expressions for $\mathcal{W}$ and
$\Phi^{\text{{PF}}}$ in terms of $X$. The second one  comes from (\ref{condcircorbits2}), in a similar fashion.
An interesting consequence of condition (\ref{condition1}) is the restriction of the values of some post-Newtonian parameters.
Since the term ${X_{,r}}/{r}-X_{,rr}\neq0$ for all asymptotically flat models (i.e. solutions so that $\lim_{r\rightarrow\infty}X_{,r} =0$)
and assuming that conditions \eqref{condition1} and \eqref{condition2} must be satisfied for all choices of $\vec{w}$, we conclude that
\emph{only theories with $\alpha_{1}=\alpha_{2}=0$ can guarantee the existence of equatorial circular orbits in
static spherically symmetric configurations with no divergent fields}. In other words, only theories in which the angular momentum is
conserved globally (i.e. so that $\alpha_{1}=\alpha_{2}=0$ \cite{will1,will2,will3}) can admit equatorial circular orbits in spherical distributions.
From here on we focus on these cases.

Introducing the  restriction $\alpha_{1}=\alpha_{2}=0$ in the equation \eqref{circmotion} we obtain the following expression for the circular
velocity (i.e. the component $v^{\varphi}=r\dot{\varphi}$ for circular orbits with $\theta=\frac{\pi}{2}$, which we denote shortly as $v_c$):
\begin{equation}\label{rotationvelocity}
v_{c}=
\sqrt{-rU_{,r}\left[1+\frac{\alpha_{3} w^{2}}{2c^{2}}+\frac{\Psi_{,r}}{c^{2}U_{,r}}-\frac{\gamma
 rU_{,r}}{c^{2}}-\frac{2\left(\beta+\gamma\right)U}{c^{2}}\right]},
\end{equation}
which will be used later to sketch the rotation curves for different theories.

In order to investigate the behavior of circular velocity away from a finite distribution,
we first consider the case in which the matter is concentrated inside a sphere of radius $a$ and mass $M$,
and  solve the field equations \eqref{fieldU}-\eqref{fieldPsi} in the vacuum, i.e. when $\rho^{*}=P=\rho^{*}\Pi=0$ ($r> a$).
The expression for the fields $U$, $X$ and $\Psi$ can be cast as
$$
U=\frac{GM}{r},\qquad X=GMr+\frac{C_{1}}{r}+C_{2},
$$
$$
\Psi=-\frac{C_{3}}{r}+\frac{\xi GM C_{1}}{r^{4}},
$$
where $C_{1}$, $C_{2}$ and $C_{3}$ are constants to be determined from
boundary conditions. Introducing the above relations in \eqref{rotationvelocity}, we obtain
the corresponding expression for circular velocity:
\begin{equation}\label{noflat}
v_{c}=\sqrt{\frac{GM}{r}\left[1+\frac{1}{c^{2}}\left(A+\frac{B}{r}+\frac{C}{r^{3}}\right)\right]},
\end{equation}
where
\begin{eqnarray}
A=\frac{\alpha_{3}w^2}{2}-\frac{C_{3}}{ GM}, \qquad B=
-\left(\gamma+2\beta\right)GM, \nonumber \\  C=4\xi C_{1}.\label{noflat2}
\end{eqnarray}
According to (\ref{noflat}), circular velocity,
 for large values of $r$, exhibits the usual Keplerian fall off, as in Newtonian theory (i.e. proportional to $1/\sqrt{r}$).
Since most solutions, when far away from the center, could be well approximated to vacuum,
therefore \eqref{noflat} should be a good representative of the asymptotic behavior of circular velocities in the PPN approach.

No choice of the
PPN parameters can lead to the flattening of the rotation curves. Indeed, such statement is also true for values
of $r$ close to the radius $a$ of the mass distribution. This can be easily verified by finding the critic value of $r$ where
$v_c$ reaches its maximum. So, the values of $r$ for which $v_c$ is extreme (maximum or minimum) satisfy
the cubic equation
\begin{equation}\label{zeros}
\left(1+\frac{A}{c^2}\right)r^{3}+\frac{2B}{c^2}r^{2}+\frac{4C}{c^2}=0.
\end{equation}
The value $r=r_e$ for which $v_c$ reaches its maximum can be cast as
$$
r=\overset{0}{r_{e}}+\overset{l}{r_{e}}+...,
$$
where $l$ is the highest order of $c$ and is determined when substituting in eq. \eqref{zeros}.
From eq. \eqref{zeros} one may verify that
$$
\overset{0}{r_{e}}=0,\qquad\overset{\frac{2}{3}}{r_{e}}=\left({\frac{4C}{c^{2}}}\right)^{1/3},
$$
which means that $r_e =(4C/c^{2})^{1/3}$. Remembering that $C=4\xi C_{1}\sim4\xi a^{3}\epsilon c^{2}$, we have,
$$
r_{e}\sim({-16\xi\epsilon})^{1/3}a\ll a.
$$
This means that there are no solutions of \eqref{zeros} in the region $r>a$.
Therefore $dv_{c}/dr<0$ for all $r>a$ since $v_{c}$ and its derivatives are continuous
functions and the latter is true for large $r$.

Now we will abandon the void and consider situations in which some material content permeates the space,
in order to allow the possibility of obtaining a different behavior from the one captured by \eqref{noflat}.

\section{Rotation curves inside a static spherical configuration }\label{sec3sub3}

Let's go back to the metric of equations \eqref{metric00}-\eqref{metricij}.
Note that  $\alpha_{1}=\alpha_{2}=0$ if and only if $g_{\mu\nu}$ is spherically symmetric, i.e. $g_{0j}=0$,
which is necessary for the existence of circular orbits.
Therefore we can use the results of reference \cite{Gauy} to simplify the possible choices for a DF.
In such paper it was shown that when the space-time is static, spherically symmetric, asymptotically flat and the
 selfgravitating configuration is made of collisionless identical particles, the DF has the form
 $f=\xi(E)L^{2(k-1)}$, where $L$ is the angular momentum, $k$ is an integer and $\xi(E)$ is
 some function of the energy $E$. This implies that
 ${T^{\theta}}_{\theta}=k{T^{r}}_{r}$, which can be applied here,
since we are adopting the same assumptions about space-time and matter distribution.
If we also assume that the system constitutes a perfect fluid (a basic assumption of the PPN formalism),
i.e. ${T^{\theta}}_{\theta}={T^{r}}_{r}$,
we have to chose $k=1$ and, in consequence, $f=f(E)$.
This means that only ergodic DFs are allowed when dealing with static spherical configurations in the PPN formalism.

Here we consider  the so-called Polytropes \cite{ramos1,binneytremaineGD},
which are, from the perspective of the mathematical form of the DF, the simplest ergodic models of astrophysical
interest in galactic dynamics.

\subsection{Polytropes in PPN approach}\label{sec5}

 In Newtonian gravity, polytropes are spherical configurations defined by a DF of the form \cite{ramos1,binneytremaineGD}
\begin{equation}\label{politropesnewton}
	f(E)=\begin{cases}A_{n}\left(-E\right)^{n-3/2}\:\:\text{ for }\: E< 0\\
		0\:\:\text{ for }\:E\geq 0
	\end{cases},
\end{equation}
where $A_{n}$ is a constant and $n$ is a real number characterizing the different models,
usually known as polytropic index, which is constrained by the condition $n>1/2$.

Here we assume the same form \eqref{politropesnewton} for selfgravitating configurations
in metric theories, taking into account that energy can be cast as $E=E_{\text{N}}+E_{\text{PPN}}$,
where $E_{\text{N}}$ is the Newtonian contribution and $E_{\text{PPN}}$ represents
the PPN correction, so that $E_{\text{N}}\gg E_{\text{PPN}}$ (see appendix \ref{appendixenergy}).
These distributions
will be called here as \emph{PPN polytropes}.
The corresponding DF can be expanded (in a similar fashion as done in \cite{ramos1}, for the case of 1PN approximation)
 as the
sum of a Newtonian contribution, of order $0$ in $\epsilon$, and a PPN contribution,
of order $2$ in $\epsilon$:
$$
f=\overset{0}{f}+\overset{2}{f},
$$
where we have introduced,
\begin{equation}\label{f0}
\overset{0}{f}=A_{n}\left(-E_{\text{N}}\right)^{n-3/2},
\end{equation}
\begin{equation}\label{f2}
\overset{2}{f}=-A_{n}\left(n-\frac{3}{2}\right){E_{PPN}}\left(-E_{\text{N}}\right)^{n-5/2},
\end{equation}
for $E_{\text{N}}<0$ and $E_{\text{PPN}}<0$.

The above DF contributions determine $\rho^*$, $P$ and $\Pi$
through equations \eqref{stressppn10}-\eqref{stressppn12}. The integrals at the right hand side of
such equations must be limited to values of velocity
 such that $E<0$. This escape velocity, determined by condition $E=0$, is given by
\begin{equation}\label{ve}
v_{e}=\sqrt{2U}+\frac{\left[2\Psi+\Phi^{\text{PF}}-\left(4\gamma+5\right)U^{2}\right]}{2c^{2}\sqrt{2U}},
\end{equation}
but, in order to obtain expressions up to first order in $\epsilon$,
it is sufficient to take $v_{e}=\sqrt{2U}$ in the limits of integration.
So, by introducing \eqref{f0}, \eqref{f2} and \eqref{ve} in \eqref{stressppn10}-\eqref{stressppn12},
 we find the matter fields for the PPN polytropes, as functions of $U$,  $\psi$ and $\Phi^{\text{PF}}$:
\begin{equation}\label{rhopolytrope}
\rho^{*}=c_{n}U^{n},
\end{equation}
\begin{equation}\label{ppolytrope}
P=c_{n}\frac{U^{n+1}}{n+1},
\end{equation}
\begin{multline}\label{rhopipolytrope}
\rho^{*}\Pi=c_{n}U^{n+1}\left[3\gamma+\frac{3}{2}+\frac{27}{8\left(n+1\right)}+n\left(\frac{1}{2}-\beta\right)\right]
\\
+nc_{n}\left(\Psi+\frac{\Phi^{\text{PF}}}{2}\right)U^{n-1},
\end{multline}
where
$$
c_{n}=A_{n}\left(2\pi\right)^{\frac{3}{2}}\frac{\Gamma (n-1/2)}{\Gamma \left(1+n\right)}.
$$
Note that density $\rho^{*}$ and pressure $P$ follow exactly the same equations as their Newtonian counterparts, ensuring
 that they satisfy the polytropic equation of state \cite{binneytremaineGD}.

\subsection{Field Equations for PPN Polytropes}\label{sec4}

Introducing \eqref{rhopolytrope}-\eqref{rhopipolytrope}
in the field equations \eqref{fieldU}-\eqref{fieldPsi}, we obtain the following system of second order
differential equations:
\begin{equation}\label{campoU}
\frac{1}{\tilde{r}^2}\frac{\mathrm{d}}{\mathrm{d}\tilde{r}}\left( \tilde{r}^2\frac{\mathrm{d}\widetilde{U}}{\mathrm{d}\tilde{r}}\right) = - \widetilde{U}^{n},
\end{equation}
\begin{equation}\label{campoX}
\frac{1}{\tilde{r}^2}\frac{\mathrm{d}}{\mathrm{d}\tilde{r}}\left( \tilde{r}^2\frac{\mathrm{d}\widetilde{X}}{\mathrm{d}\tilde{r}}\right)  = 2 \widetilde{U},
\end{equation}
\begin{multline}\label{campoPsi}
\frac{1}{\tilde{r}^2}\frac{\mathrm{d}}{\mathrm{d}\tilde{r}}\left( \tilde{r}^2\frac{\mathrm{d}\widetilde{\Psi}}{\mathrm{d}\tilde{r}}\right)=b_{n}\frac{{U_{0}}^{2}}{\Psi_{0}}\widetilde{U}^{n+1}-e_{n}\frac{{U_{0}}}{\Psi_{0}}\widetilde{U}^{n}-d_{n}\widetilde{\Psi} \widetilde{U}^{n-1}
\\
+4\frac{{U_{0}}^{2}}{\Psi_{0}}\xi\left[\frac{\widetilde{U}^{n}}{\tilde{r}}\frac{\mathrm{d}\widetilde{X}}{\mathrm{d}\tilde{r}}+\frac{\mathrm{d}\widetilde{U}}{\mathrm{d}\tilde{r}}\left(\frac{\mathrm{d}\widetilde{U}}{\mathrm{d}\tilde{r}}+\frac{3}{\tilde{r}^2}\frac{\mathrm{d}\widetilde{X}}{\mathrm{d}\tilde{r}}-\frac{2\widetilde{U}}{\tilde{r}}\right)\right].
\end{multline}
where $U_{0}$ and $\Psi_{0}$ are the values of $U$ and $\Psi$ at the center of the configuration, respectively,
and we have defined,
\begin{equation}\label{camposTiu}
\widetilde{U}:=U/U_{0},\;\;\widetilde{\Psi}:=\Psi/\Psi_{0},\;\;\widetilde{X}:=C_{n}{U_{0}}^{n-2}X,
\end{equation}
\begin{equation}\label{rTiu}
\tilde{r}:=r\sqrt{C_{n}{U_{0}}^{n-1}},
\end{equation}
along with the constants
\begin{equation}
C_{n}=4\pi G c_{n}, \qquad d_{n}=\left(1+\zeta_{3}\right)n=\frac{2e_{n}}{\alpha_{3}w^2}. \label{constantes}
\end{equation}
\begin{eqnarray}
b_{n}&=&\left(2\beta-1-\zeta_{2}-6\xi\right)-\frac{\left(3\gamma+3\zeta_{4}-2\xi\right)}{\left(n+1\right)}
\nonumber\\
&&-\left(1+\zeta_{3}\right)\left[3\gamma+\frac{3}{2}+\frac{27}{8\left(n+1\right)}+\frac{n}{2}-\beta n\right],
\label{constantes2}
\end{eqnarray}

Equation \eqref{campoU} has simple analytic solutions for the cases $n=0,1,5$ \cite{ramos1,binneytremaineGD},
being $n=5$ the well known  case of Plummer model \cite{ramos1,binneytremaineGD,plummer1,plummer2}.
For other values of $n$, Eq. \eqref{campoU} requires a numerical solution.
On the other hand, equations \eqref{campoX} and \eqref{campoPsi} are much more complicated,
even in the cases $n=0,1,5$, where $U$ has analytical expression. However, as we will show in
the following subsection, one can find an asymptotic solution for large $r$. In subsection \ref{numerical}
we will obtain numerical solutions for the system \eqref{campoU}-\eqref{campoPsi}, by using the initial conditions
$$
\widetilde{U}\left(0\right)=\widetilde{X}\left(0\right)=\widetilde{\Psi}\left(0\right)=1, \;\widetilde{U}'\left(0\right)=\widetilde{X}'\left(0\right)=\widetilde{\Psi}'\left(0\right)=0,
$$
reflecting the assumption  that $\widetilde{U}$, $\widetilde{X}$ and $\widetilde{\Psi}$ have maximum value in the center of
the configuration.

Since parameter $\epsilon$ can be regarded a representative of the gravitational strength, we also assume that
$U_{0}=\epsilon c^{2}$, $\Psi_{0}=\left(U_{0}\right)^{2}$ and $w=\epsilon c$, for numerical and asymptotic solutions.
For numerical solutions we consider only $0\leq\epsilon\leq 0.15$. Larger values of $\epsilon$ may exceed the validity limit
of the PPN approximation.

On the other hand,  we only accept solutions that result in an asymptotically flat space-time. Therefore any solution leading
to something different from
$$
\lim\limits_{r\rightarrow\infty}\mathrm{g}_{00}=-1,\;\lim\limits_{r\rightarrow\infty}\mathrm{g}_{ij}=1,
$$
will be discarded, since it does not satisfy the basic assumptions of the PPN approximation.

%-----------------------------------------------------------------------------------

\subsection{The Approximated Solution}\label{sec4sub2}

In order to understand the importance of the contribution of each of the parameters appearing in field equations,
specially in relation with the behavior of solutions far from the center of configuration,
 we consider here
an asymptotic solution of  \eqref{campoX}-\eqref{campoPsi}, based on the Plummer's polytrope (i.e. the case
$n=5$ of eq. \eqref{campoU} \cite{binneytremaineGD,plummer1,plummer2}),
\begin{equation}\label{PlummerU}
\widetilde{U}=\frac{1}{\sqrt{1+\frac{\tilde{r}^{2}}{3}}}.
\end{equation}
For large radius, $\tilde{r}\gg \sqrt{3}$, we have $\widetilde{U}\approx\sqrt{3}/\tilde{r}$, which means that, according to
\eqref{campoX}, we can write
$$
\frac{\mathrm{d}\widetilde{X}}{\mathrm{d}\tilde{r}}\approx \sqrt{3},\;\text{  for  }\;r\gg\sqrt{3},
$$
which, in turn, implies that eq. \eqref{campoPsi} reduces to
$$
\frac{\mathrm{d}}{\mathrm{d}\tilde{r}}\left( \tilde{r}^2\frac{\mathrm{d}\widetilde{\Psi}}{\mathrm{d}\tilde{r}}\right)+9d_{5}\frac{\widetilde{\Psi}}{\tilde{r}^{2}}
=\frac{27\left(b_{5}-4\xi\right)}{\tilde{r}^{4}}-\frac{9\sqrt{3}\alpha_{3}d_{5}}{2\tilde{r}^{3}}
+\frac{24\xi}{\tilde{r}^{2}},
$$
for $\tilde{r}\gg \sqrt{3}$. The solution of the above equation can be written as
\begin{align}\label{PlummerPsi}
\widetilde{\Psi}\left(\tilde{r}\right)=c_1 &\cos \left(\frac{3 \sqrt{d_{5}}}{\tilde{r}}\right)-c_2 \sin \left(\frac{3 \sqrt{d_{5}}}{\tilde{r}}\right)-\frac{ \sqrt{3} \alpha_{3}}{2 \tilde{r}}\nonumber
\\
&+\frac{3 \left(b_{5}-4\xi\right)}{ {d_{5}} \tilde{r}^2}-\frac{2 \left(b_{5}-4\xi\right)}{3 {d_{5}}^2 }+\frac{8 \xi }{3 {d_{5}}}.
\end{align}
Since we are interested in solutions consistent with the requirement of asymptotically flat space-time, we demand that
$\lim\limits_{\tilde{r}\rightarrow\infty}\widetilde{\Psi}\left(\tilde{r}\right)=0$, leading to the following relation:
$$
c_1=\frac{2 b_{5}-8 \left(d_{5}+1\right) \xi}{3 {d_{5}}^2}.
$$
On the other hand, $c_{2}$ has to remain undetermined since no other boundary condition can be applied here.
Regardless, one can extract additional information from the corresponding expression for the  velocity of circular orbits:
\begin{multline*}
v_{c}= c\sqrt{\epsilon\frac{\sqrt{3}}{\tilde{r}}}\left\{1+\epsilon\frac{\sqrt{3}}{\tilde{r}}
\left[\frac{2\left(b_{5}-4\xi\right)}{d_{5}}-\left(\gamma+2\beta\right)\right]
\right.\\\left.
-\epsilon\sqrt{3 d_{5}}\left[c_{1} \sin \left(\frac{3 \sqrt{d_{5}}}{\tilde{r}}\right)
+c_2 \cos \left(\frac{3 \sqrt{d_{5}}}{\tilde{r}}\right)\right]\right\}^{1/2}.
\end{multline*}
Note that the second term inside the brackets should be insignificantly small, because it involves terms multiplied by $\epsilon \sqrt{3}/\tilde{r}$.
 Therefore it is only relevant to the behavior of the curves the third term ($c_{1} \sin \left({3 \sqrt{d_{5}}}/{\tilde{r}}\right)+
 c_2 \cos \left({3 \sqrt{d_{5}}}/{\tilde{r}}\right)$). For large values of $d_{5}$ the curves may present significant modifications
 from the Newtonian ones. One may even find, depending on the constants $c_{1}$ and $c_{2}$, curves resembling flat rotation curves.
 Remembering that $d_{5}=5(1+\zeta_{3})$
we conclude that the only significant contributions will arise from positive values of $\zeta_{3}$.

\subsection{The Numerical Solution}\label{numerical}

\begin{figure}[ht]
	
	\includegraphics[scale=0.3]{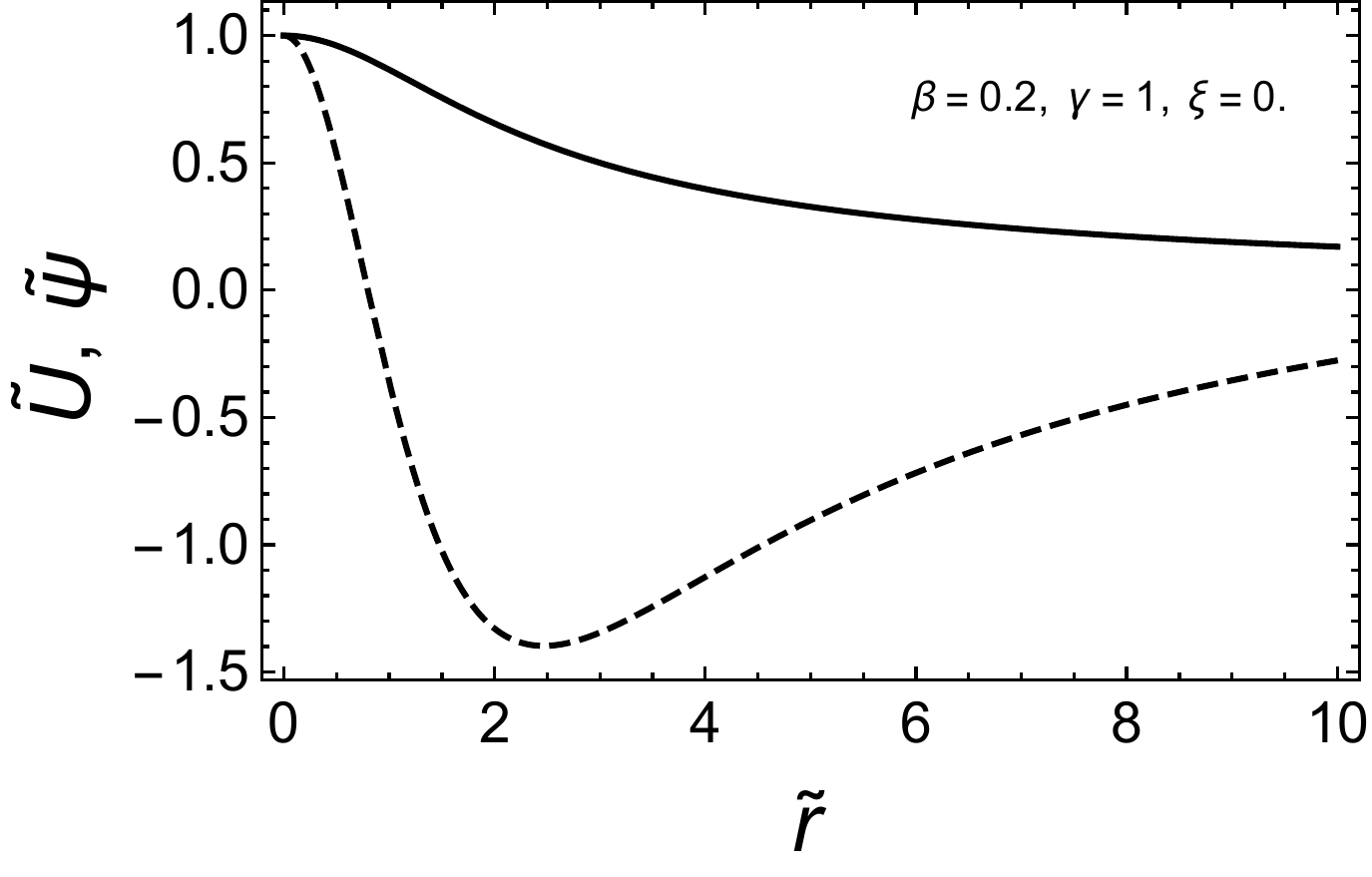}
	\includegraphics[scale=0.3]{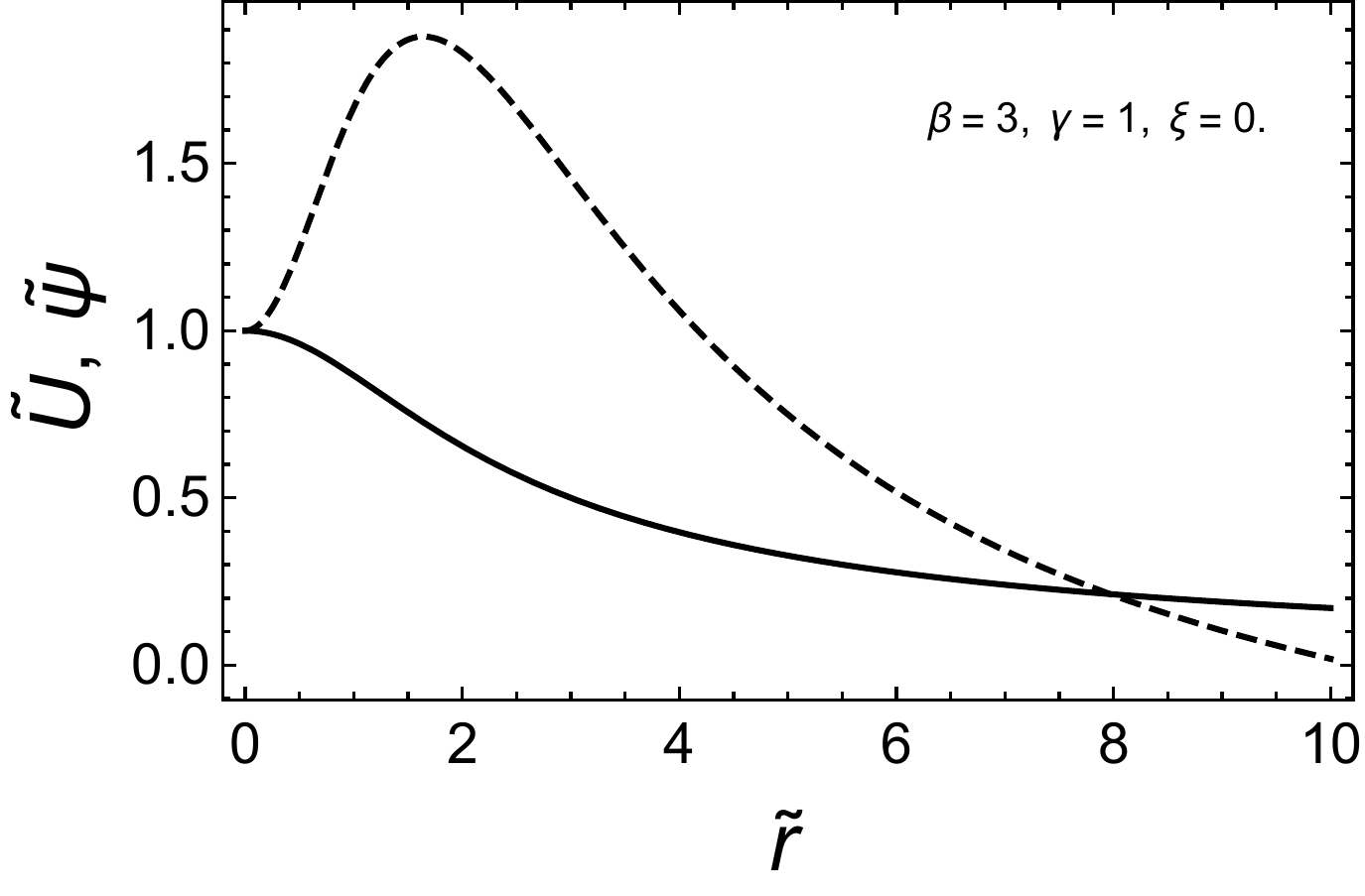}
	
	\includegraphics[scale=0.3]{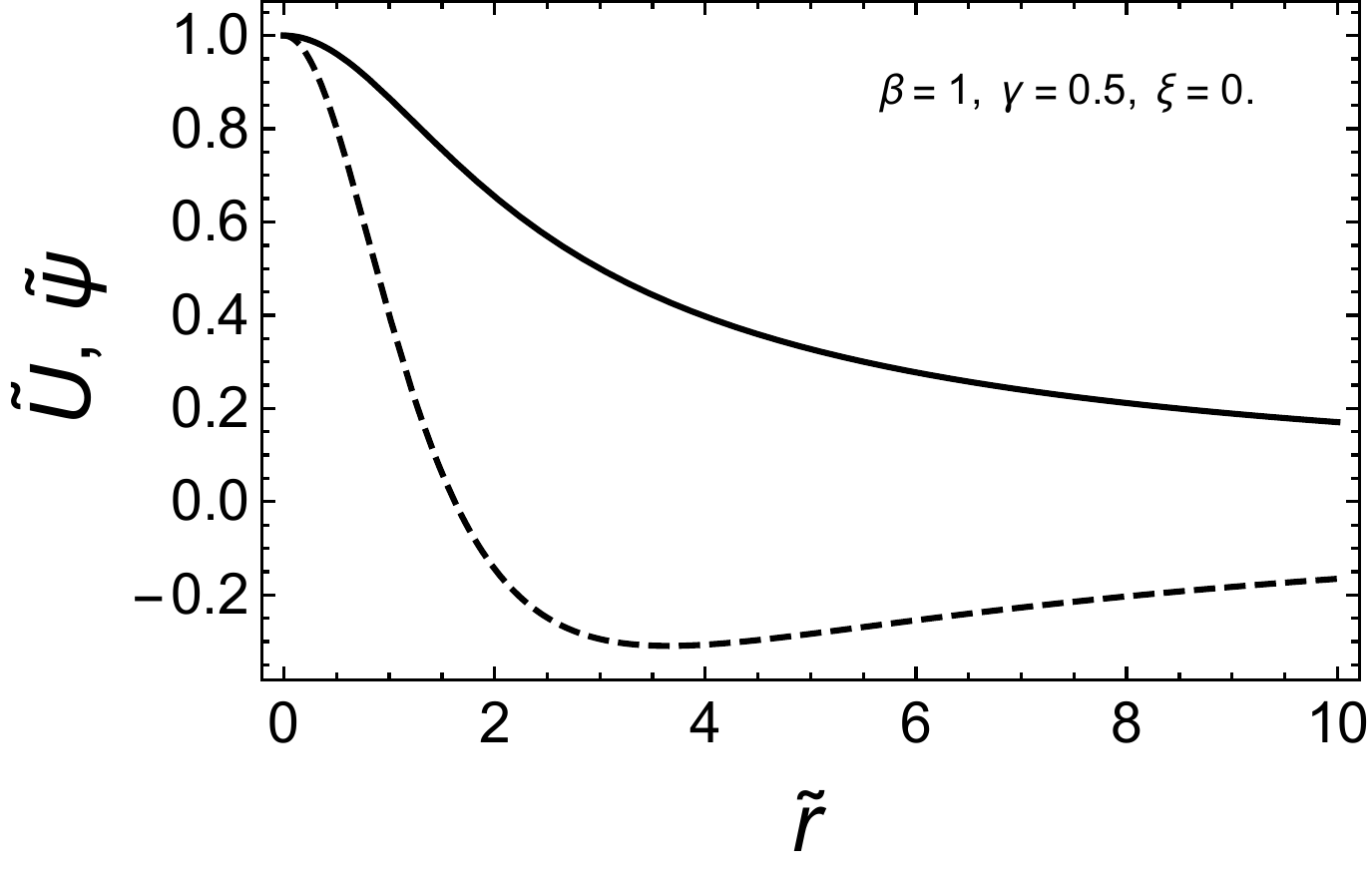}
	\includegraphics[scale=0.3]{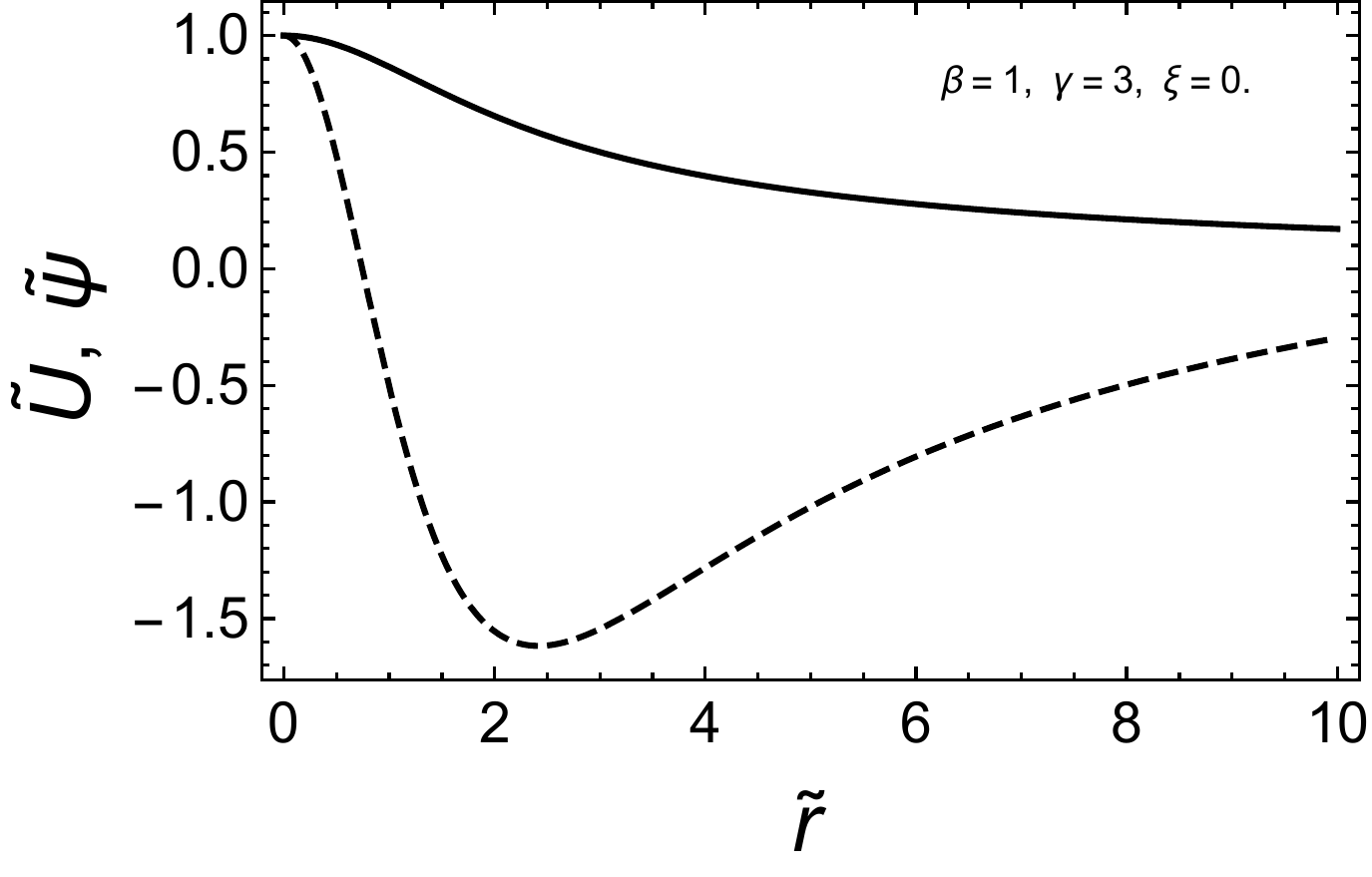}
	
	\includegraphics[scale=0.3]{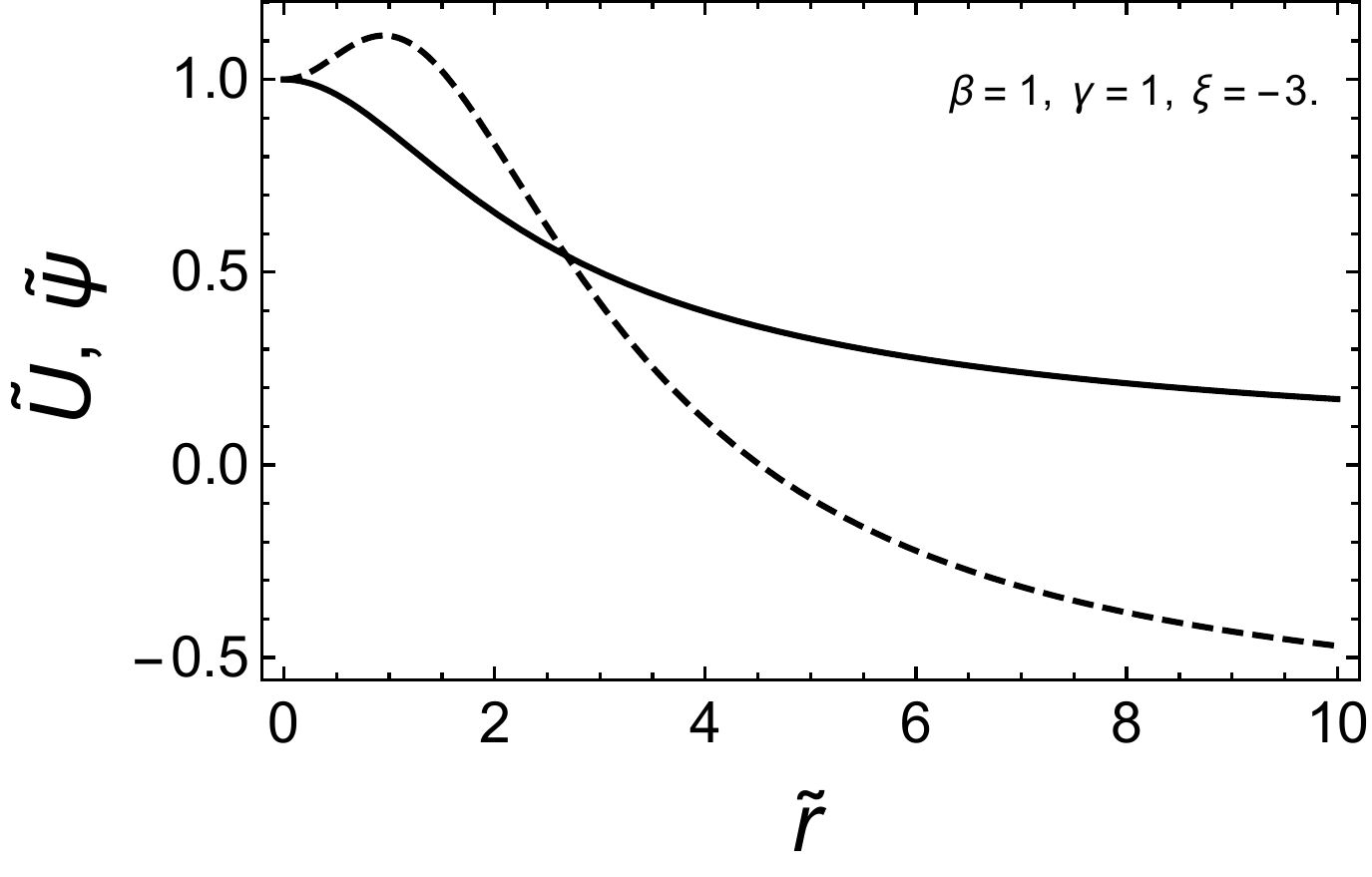}
	\includegraphics[scale=0.3]{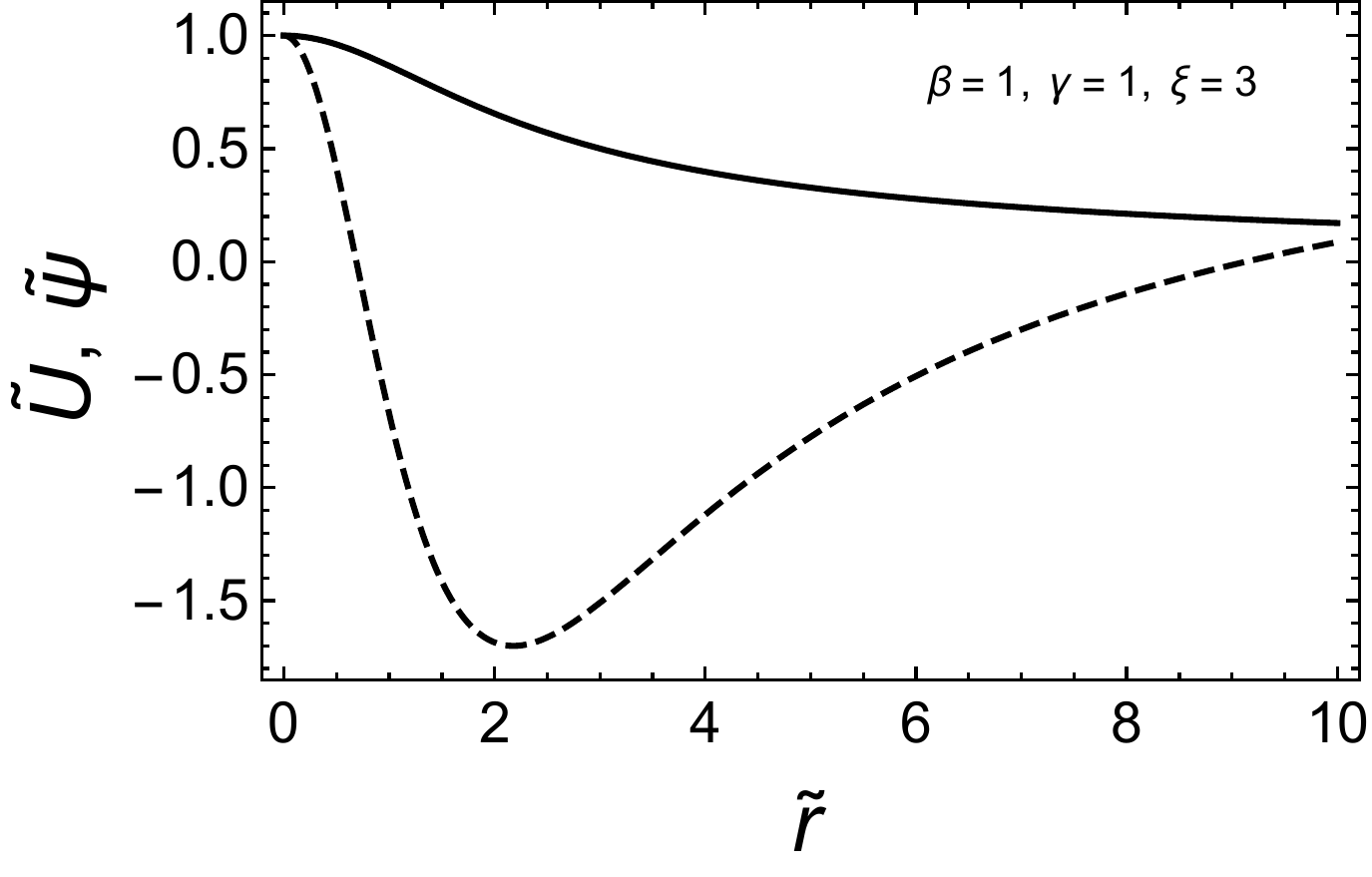}

	\caption{We present the gravitational fields $\widetilde{U}$ and $\widetilde{\Psi}$ for a politropic model with $n=5$ for fully conservative theories $\left(\zeta_{1}=\zeta_{2}=\zeta_{3}=\zeta_{4}=\alpha_{3}=0\right)$. The top figures are for $\beta$ different from $1$, the central ones for $\gamma$ different from $1$ and the lowest are for a $\xi$ different from $0$. The continuous line represents $\widetilde{U}$ and the traced one $\widetilde{\Psi}$.
	}\label{camposcons}
\end{figure}

\begin{figure}[ht]
	\includegraphics[scale=0.3]{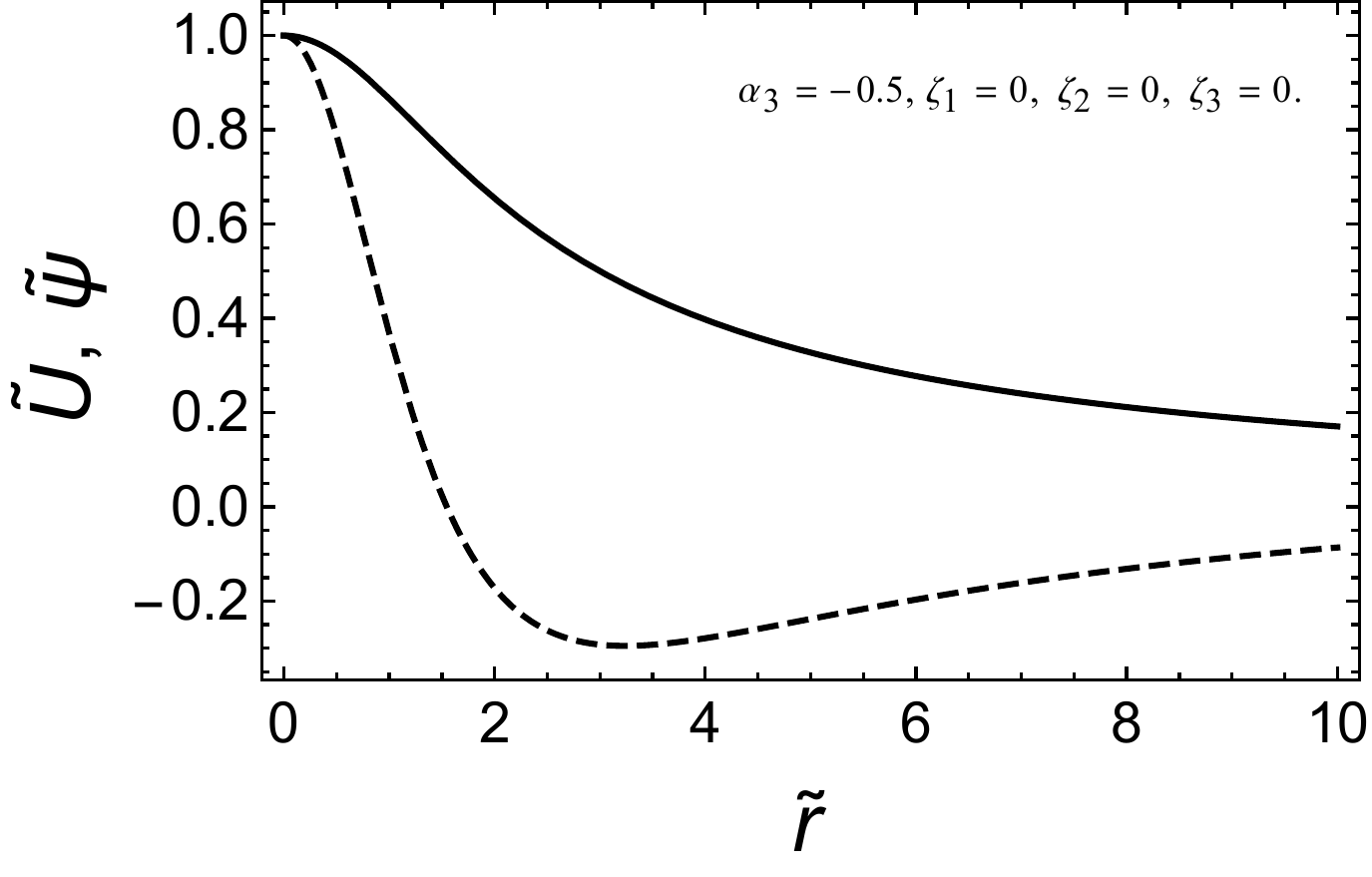}
	\includegraphics[scale=0.3]{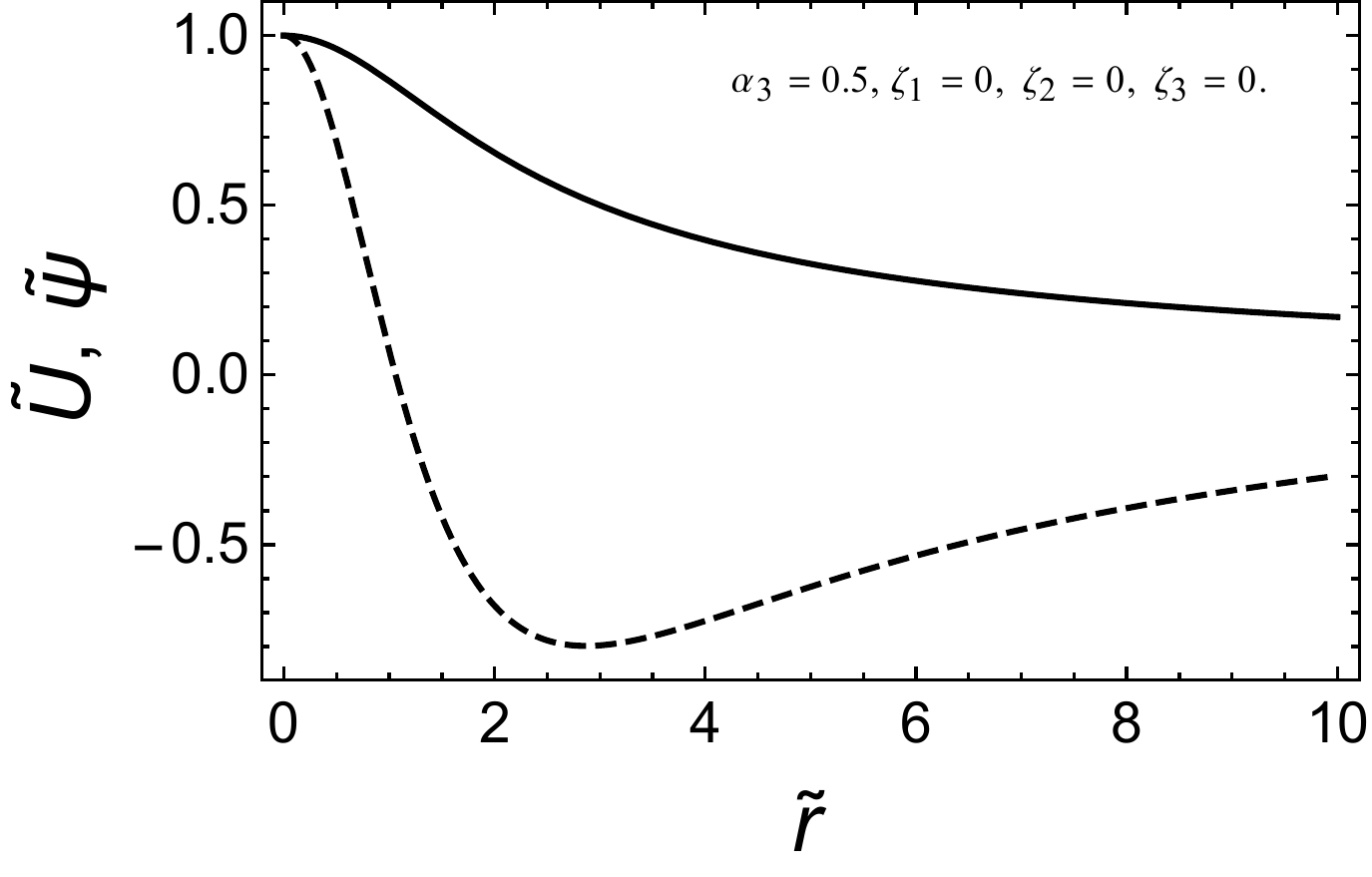}
	
	\includegraphics[scale=0.3]{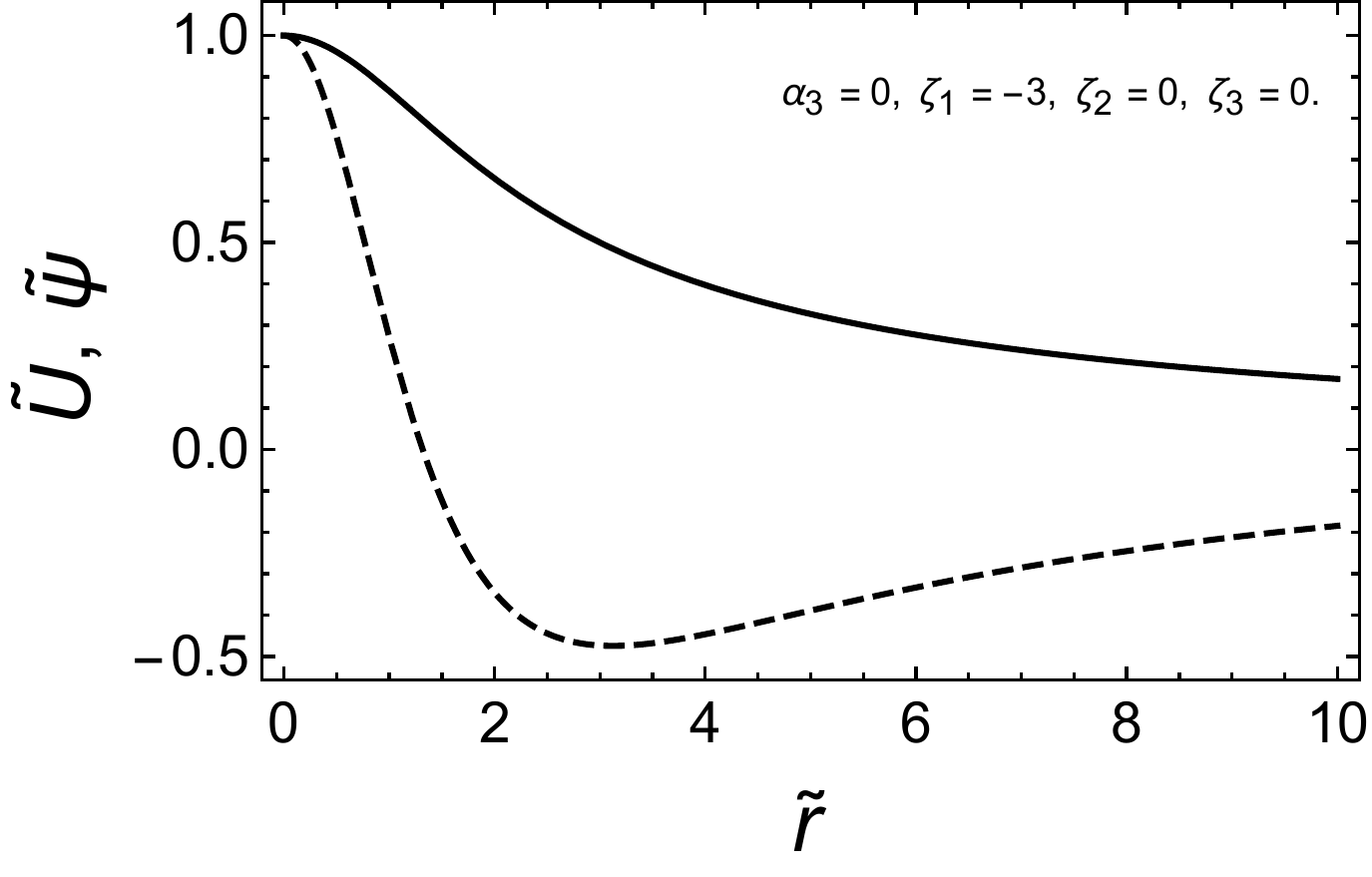}
	\includegraphics[scale=0.3]{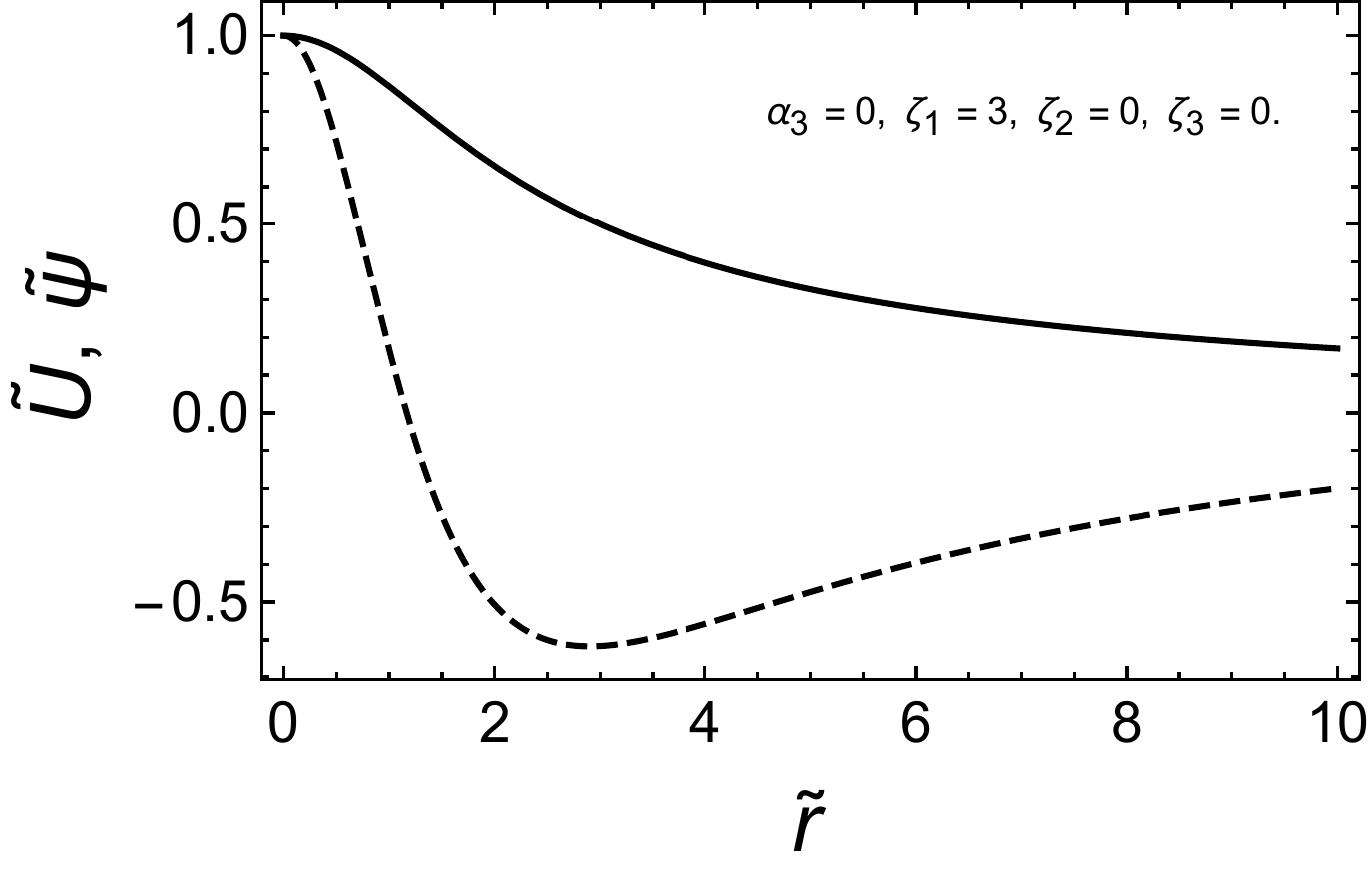}
	
	\includegraphics[scale=0.3]{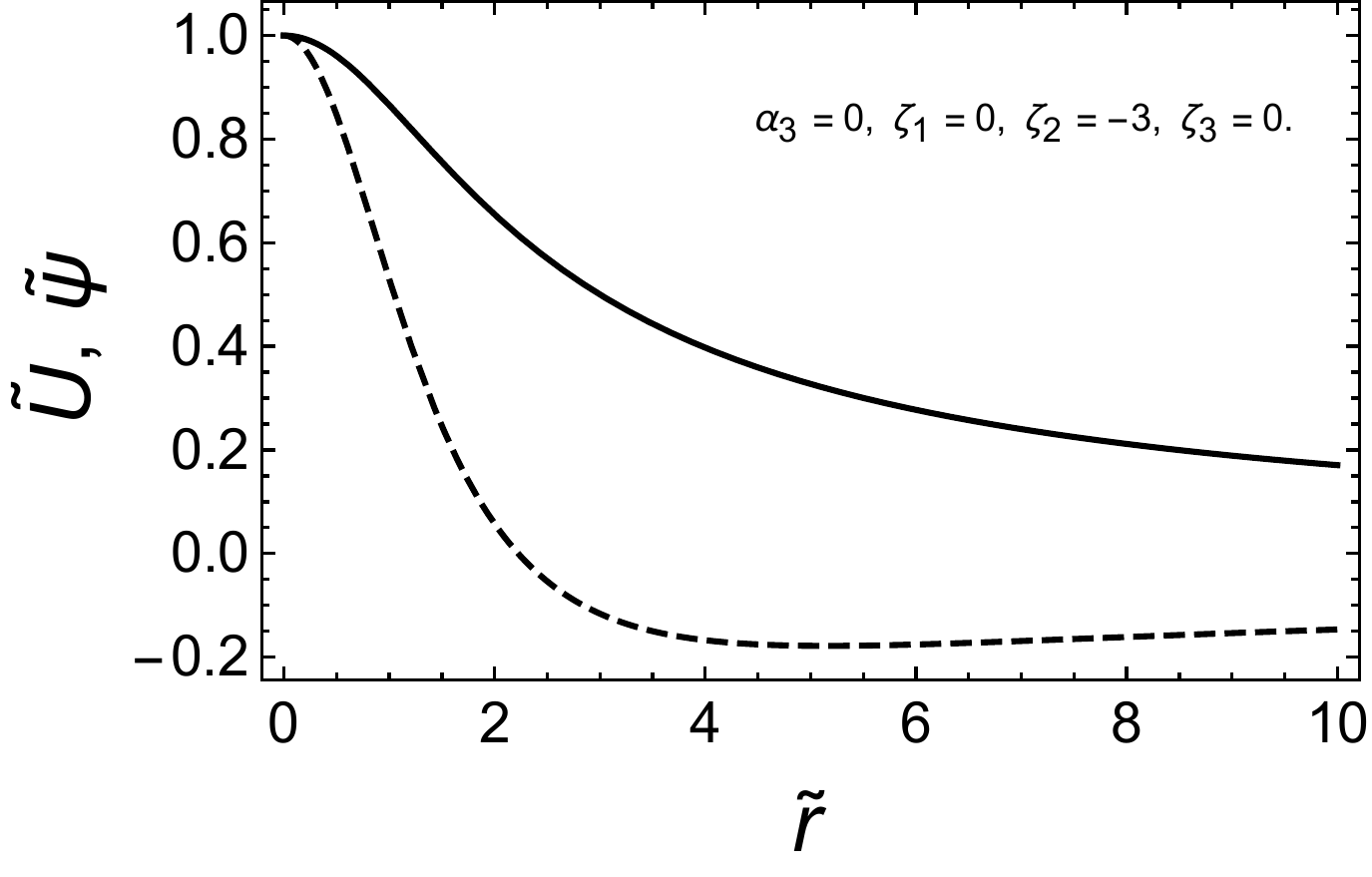}
	\includegraphics[scale=0.3]{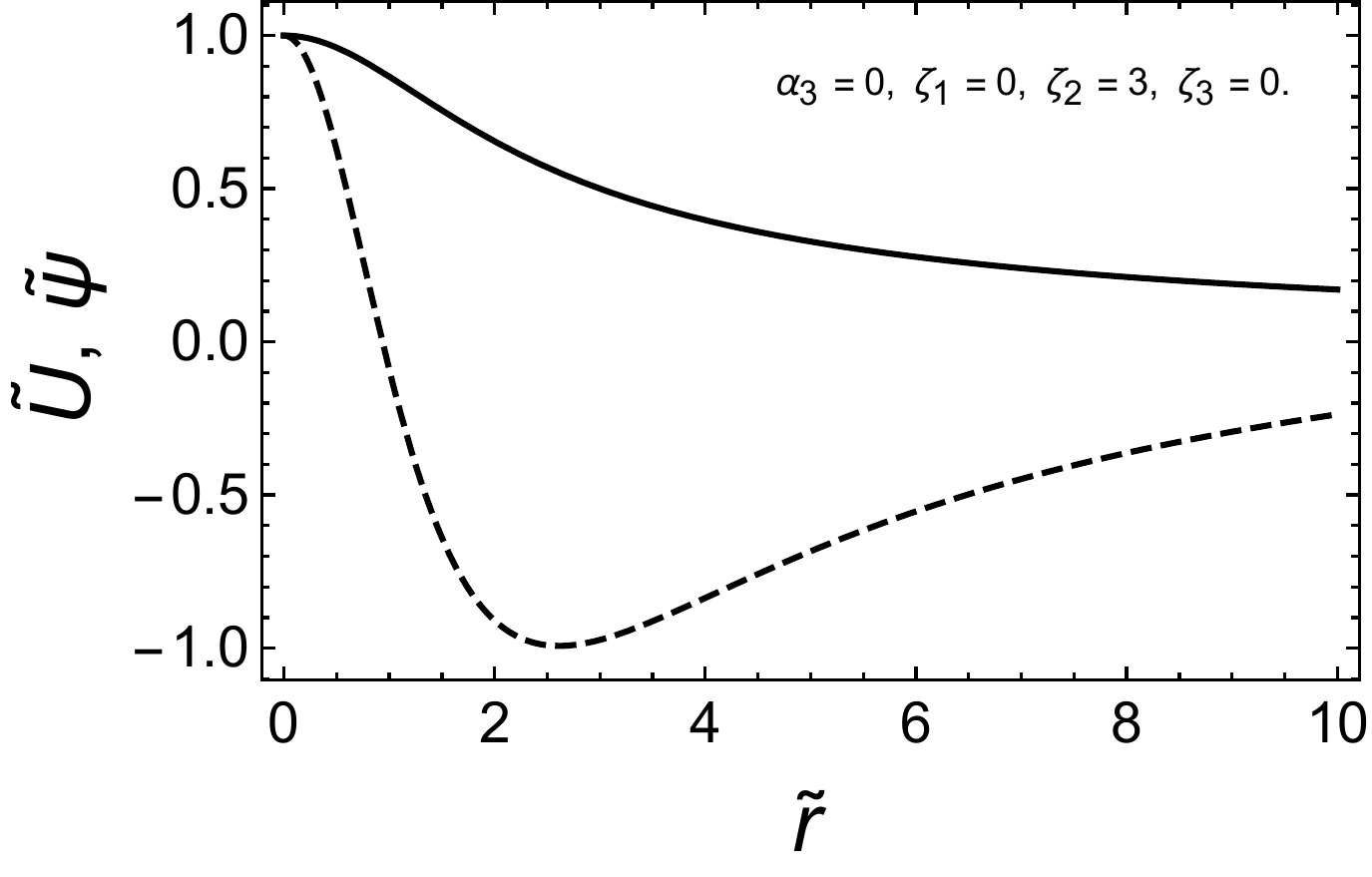}
		
	\includegraphics[scale=0.3]{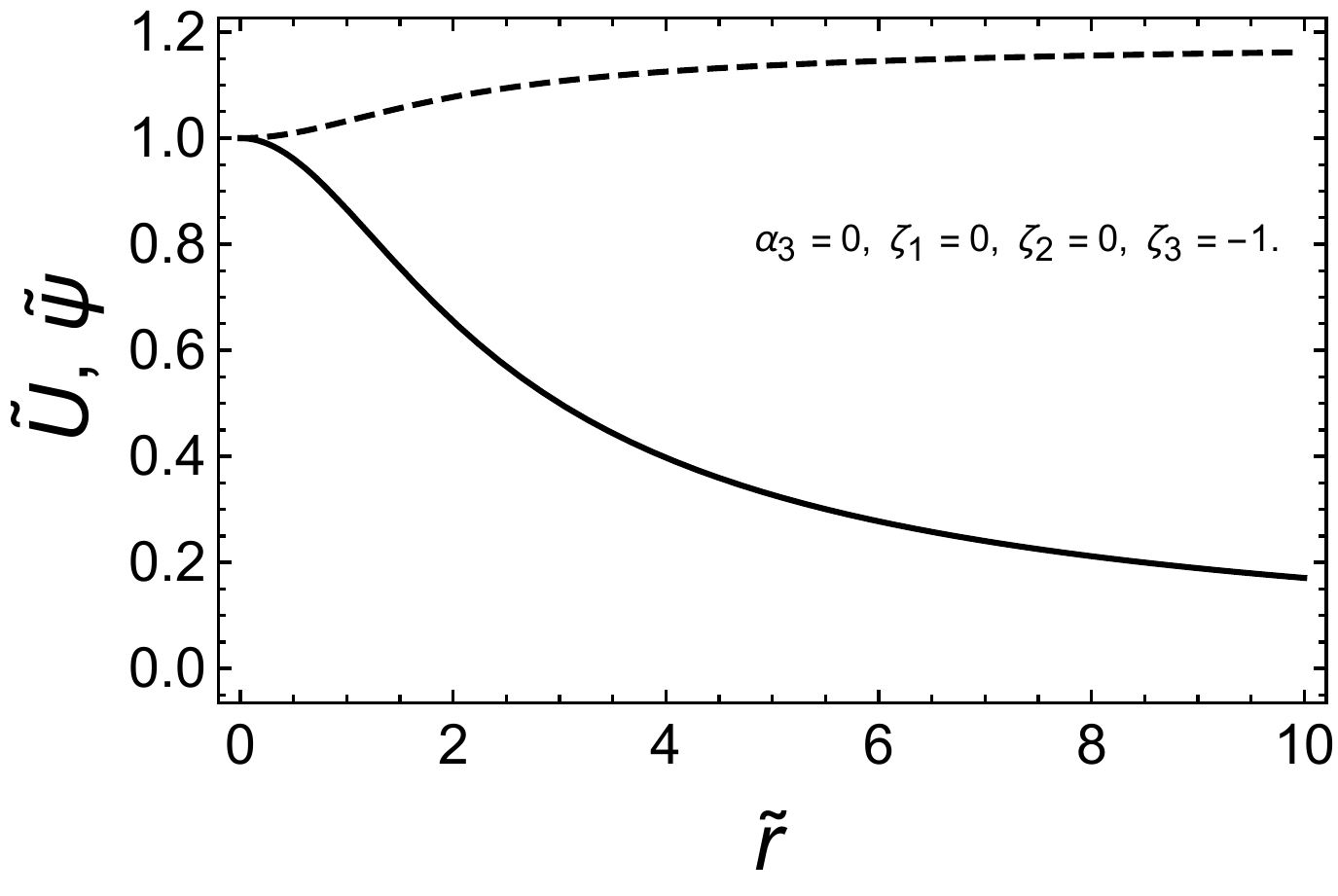}
	\includegraphics[scale=0.3]{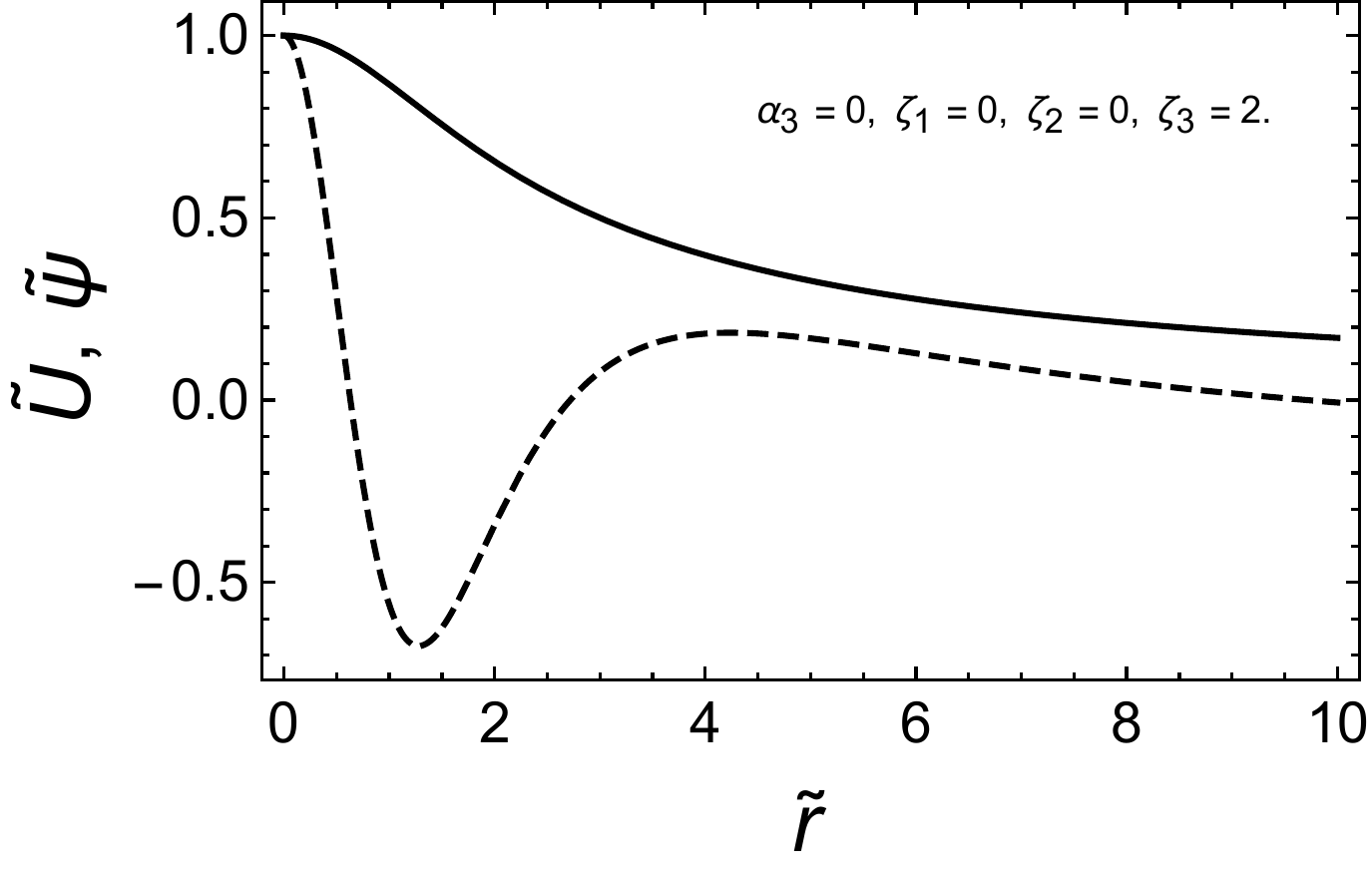}
	
	\caption{We present the gravitational fields $\widetilde{U}$ and $\widetilde{\Psi}$ for a politropic model with $n=5$ for non-conservative theories. For simplicity, here we consider $\gamma=\beta=1 \text{ and }\xi=0$. From top to bottom, we alternate $\alpha_{3}$, $\zeta_{1}$, $\zeta_{2}$ and $\zeta_{3}$ to be different form $0$, respectively. The continuous line represents $\widetilde{U}$ and the traced one $\widetilde{\Psi}$.
	}\label{camposnoncons}
\end{figure}

To solve the field equations \eqref{campoU}-\eqref{campoPsi}
for the entirety of space  we use a fourth order Runge-Kutta method.
In figures \ref{camposcons} and \ref{camposnoncons}, we show numerical solutions for 
$\widetilde{U}$ and $\widetilde{\Psi}$ for PPN Polytropes with politropic index $n=5$, for various values of the PPN parameters.

\begin{figure}[ht]
	
	\includegraphics[scale=0.3]{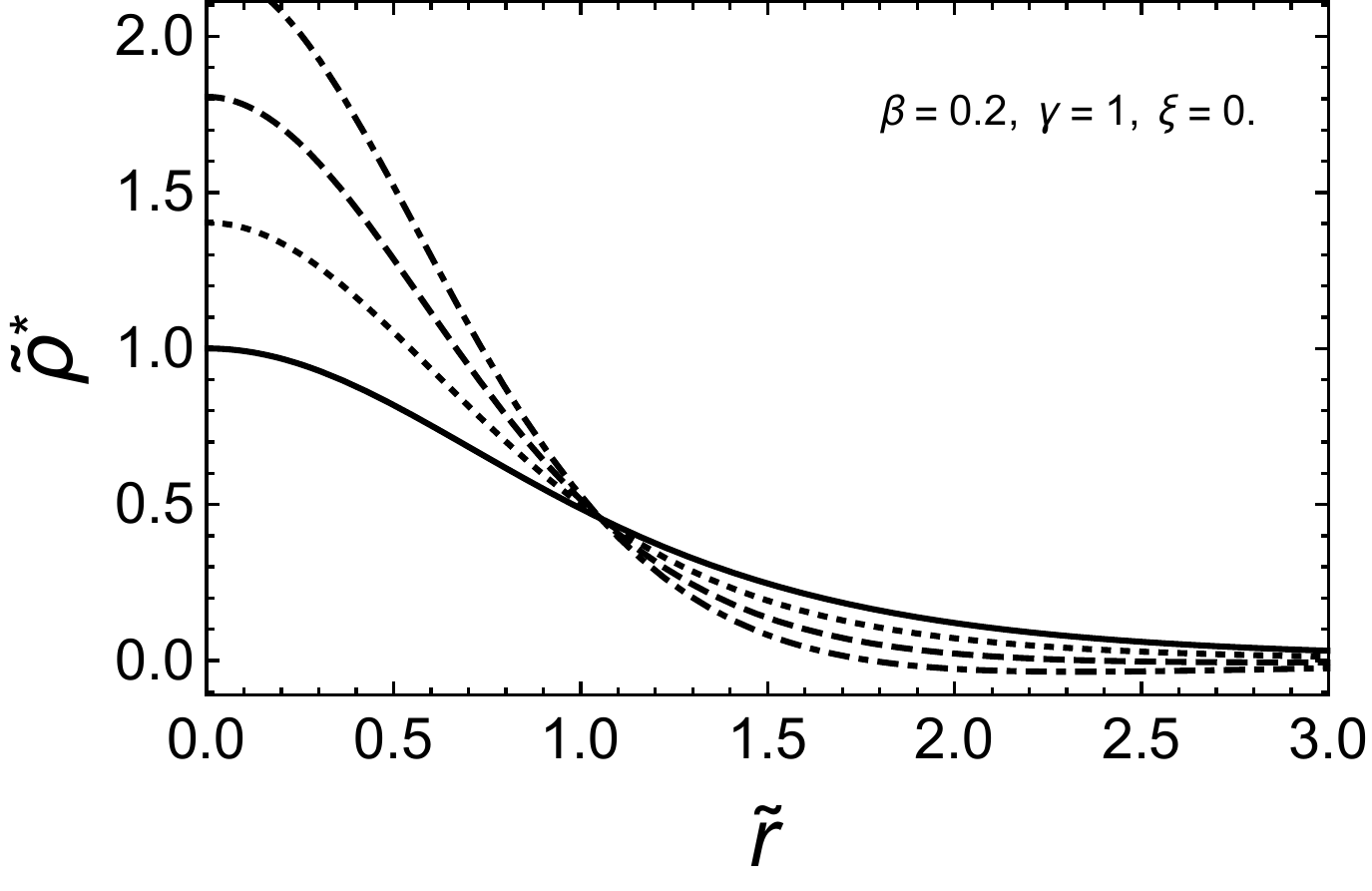}
	\includegraphics[scale=0.3]{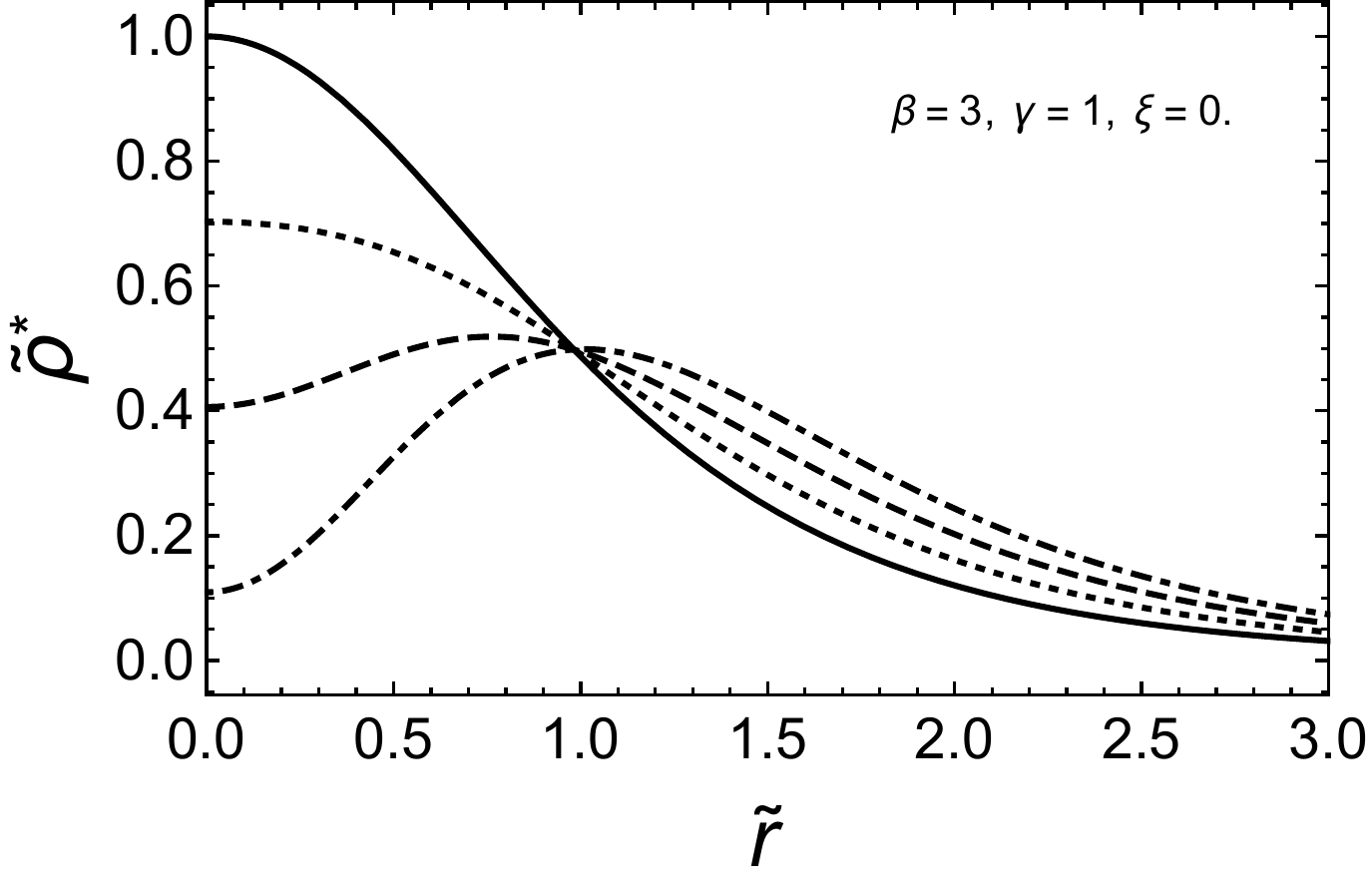}
	
	\includegraphics[scale=0.3]{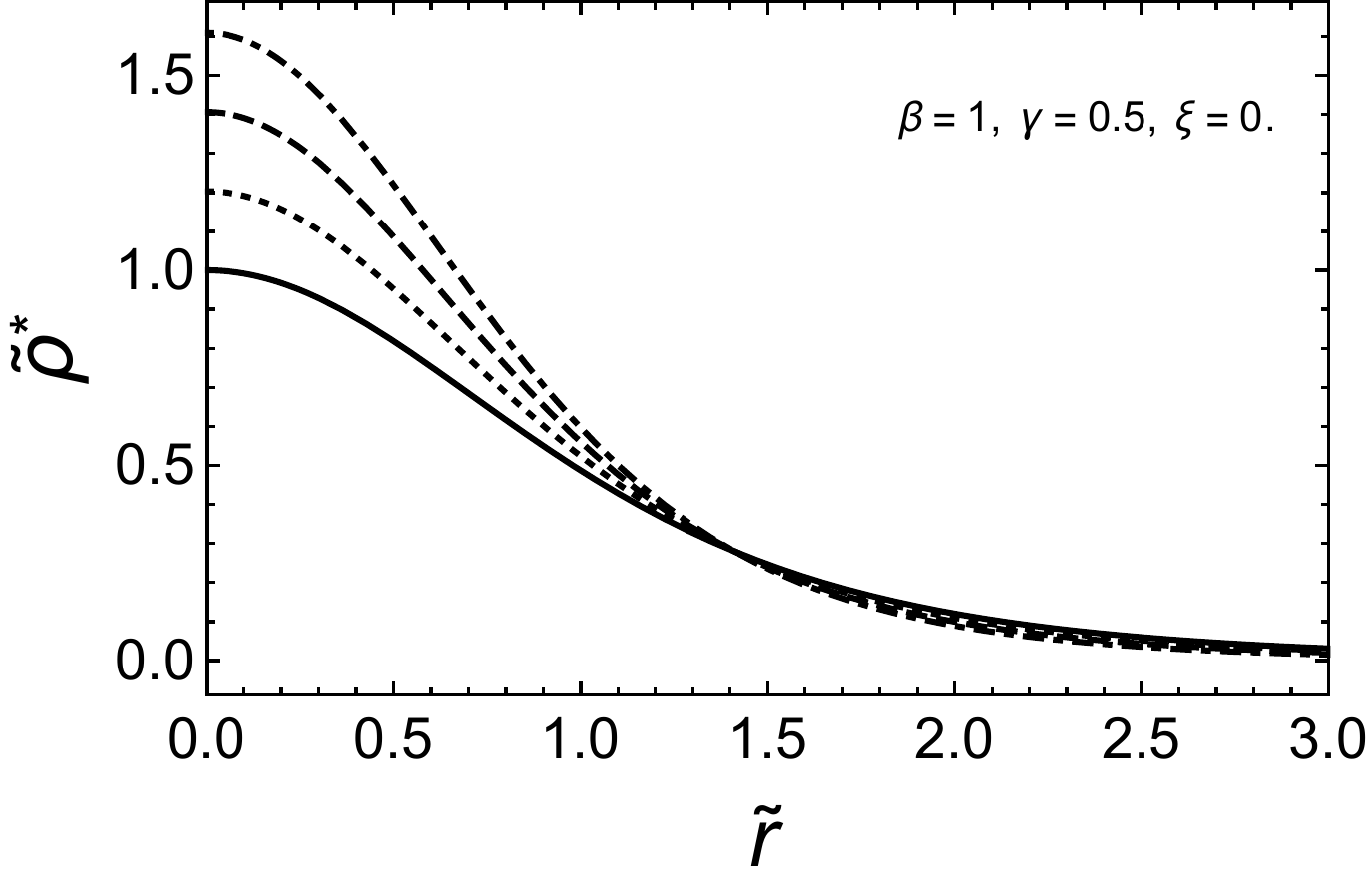}
	\includegraphics[scale=0.3]{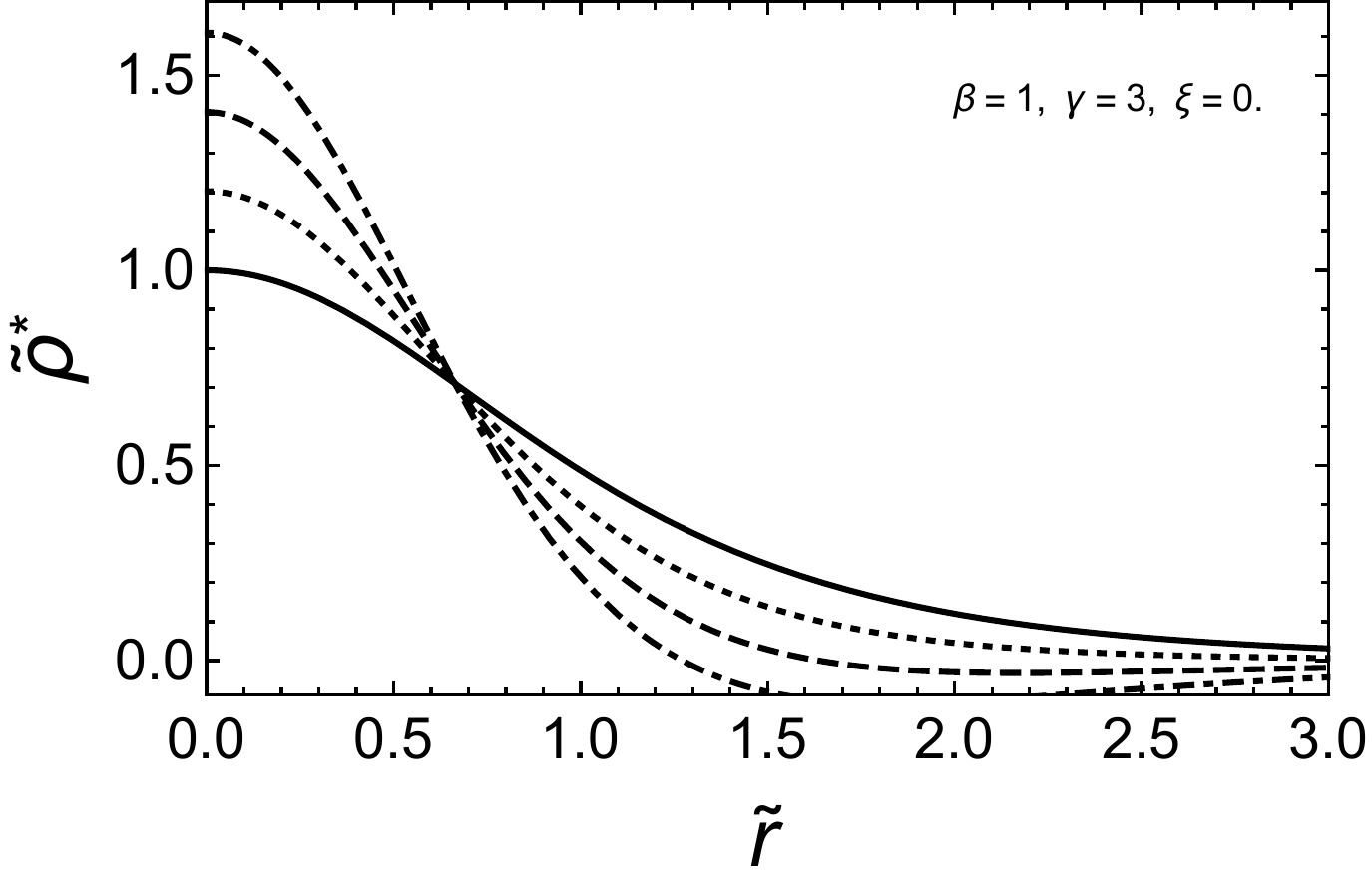}
		
	\includegraphics[scale=0.3]{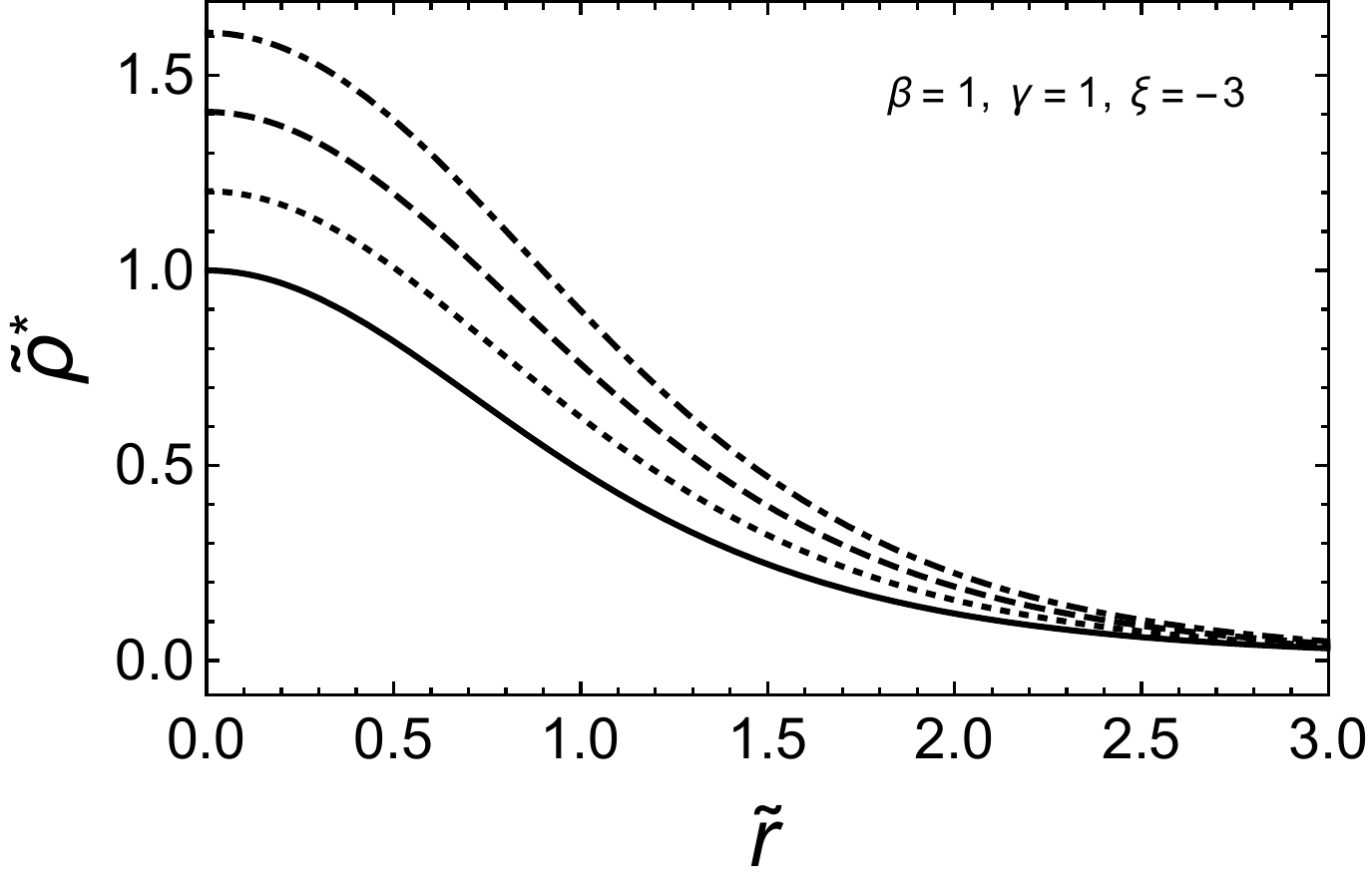}
	\includegraphics[scale=0.3]{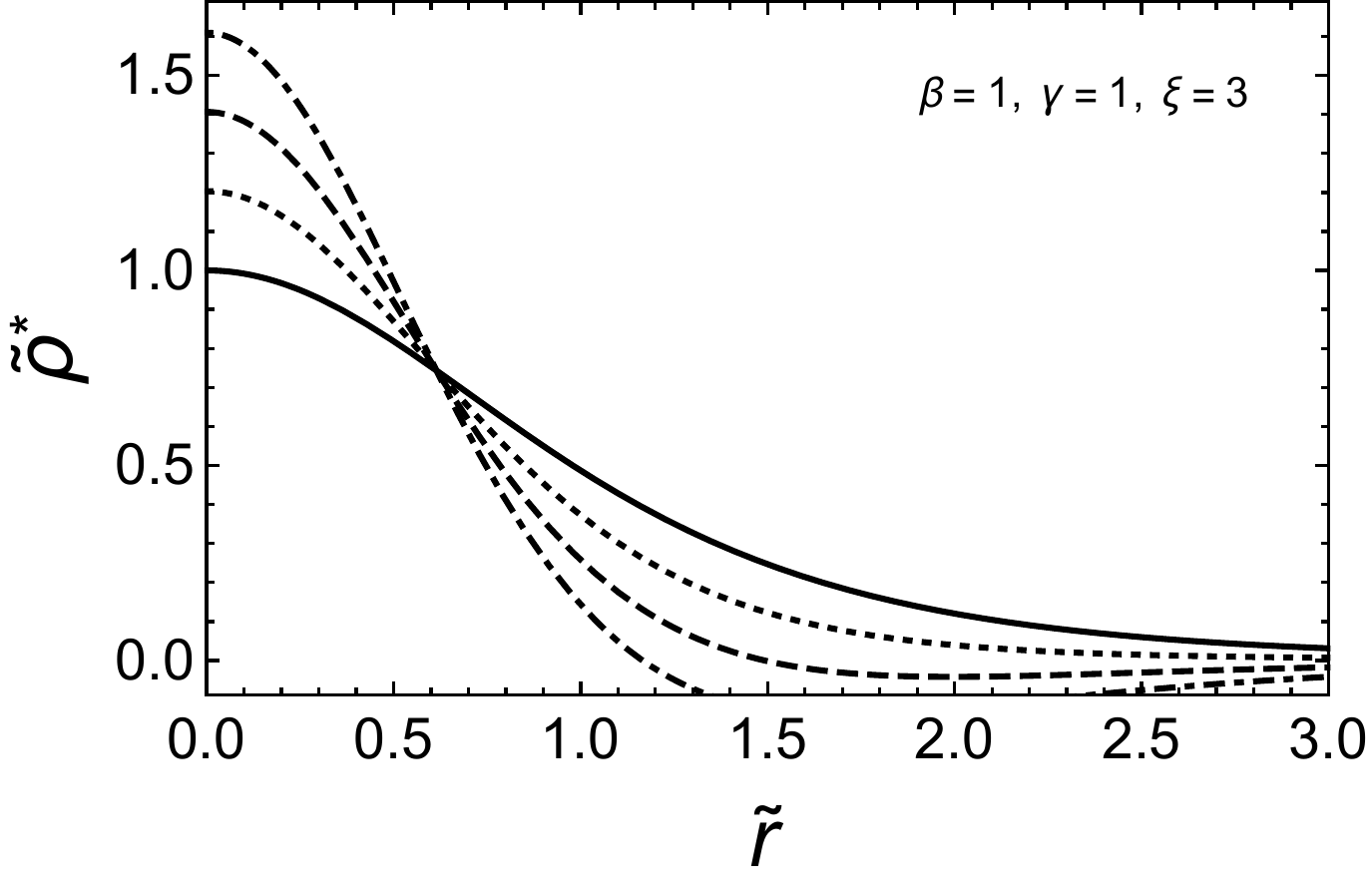}

	\caption{We present the effective density $\widetilde{\rho}^{*}$ for a polytopic model with index $n=5$ for fully conservative theories $\left(\zeta_{1}=\zeta_{2}=\zeta_{3}=\zeta_{4}=\alpha_{3}=0\right)$. The top figures are for $\beta$ different from $1$, the middle ones for $\gamma$ different from $1$ and the lower ones for $\xi$ different from $0$. The continuous line represents the Newtonian curve and the dotted, dashed and dash-dotted are PPN curves with $\epsilon=0.05$, $\epsilon=0.1$ and $\epsilon=0.15$, respectively.
	}\label{densidadecons}
\end{figure}

\begin{figure}[ht]
	
	\includegraphics[scale=0.3]{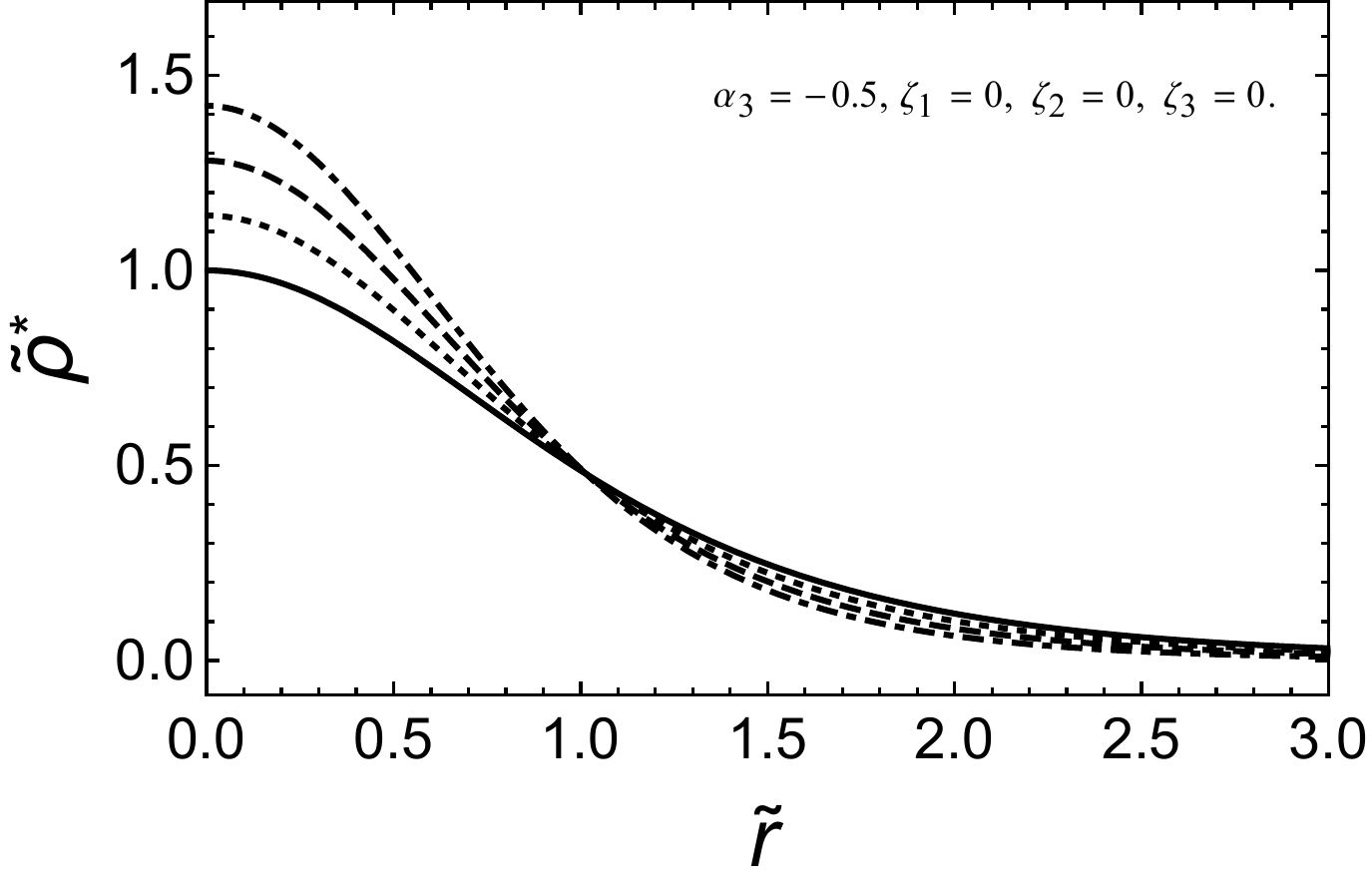}
	\includegraphics[scale=0.3]{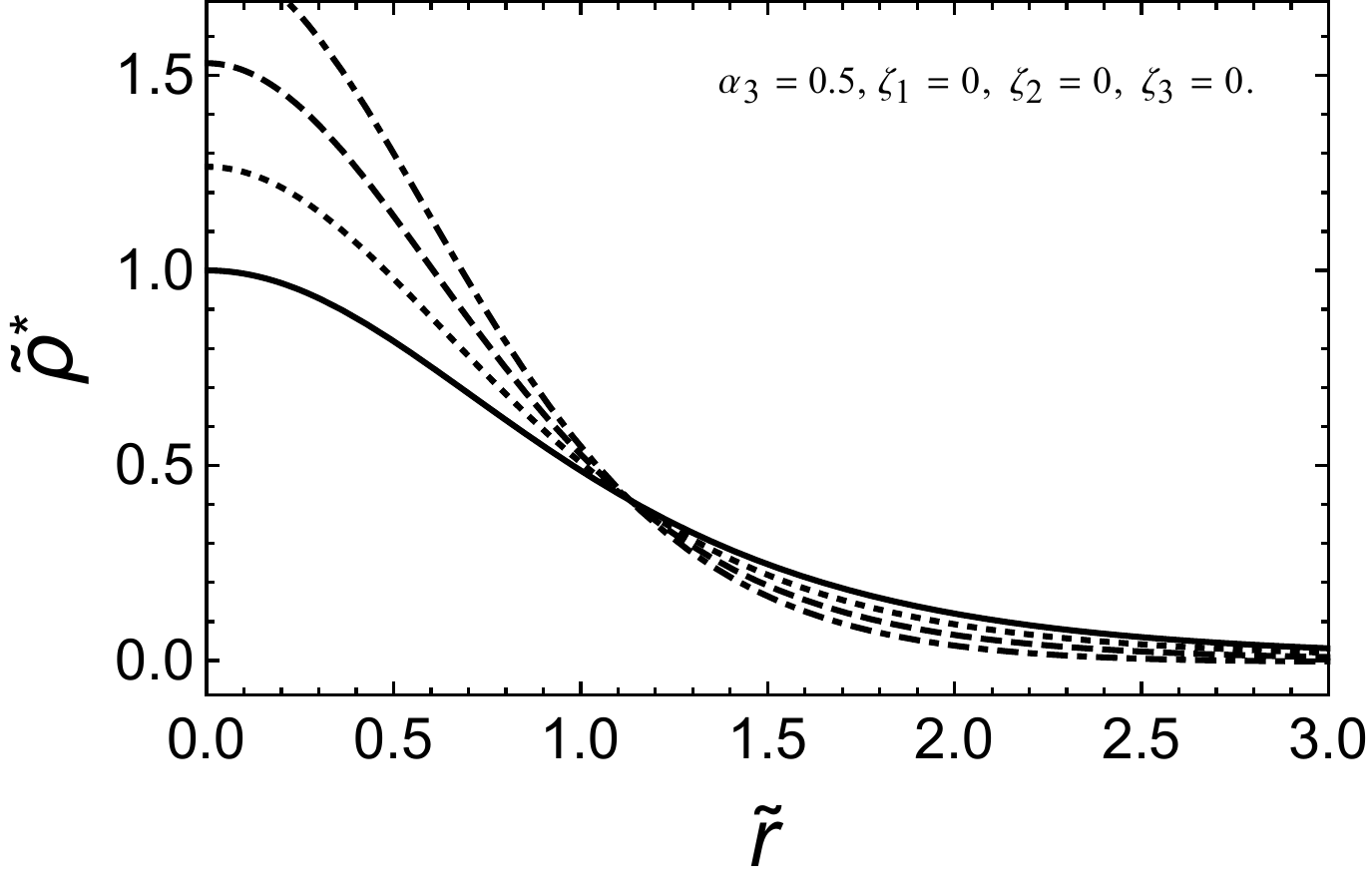}

	\includegraphics[scale=0.3]{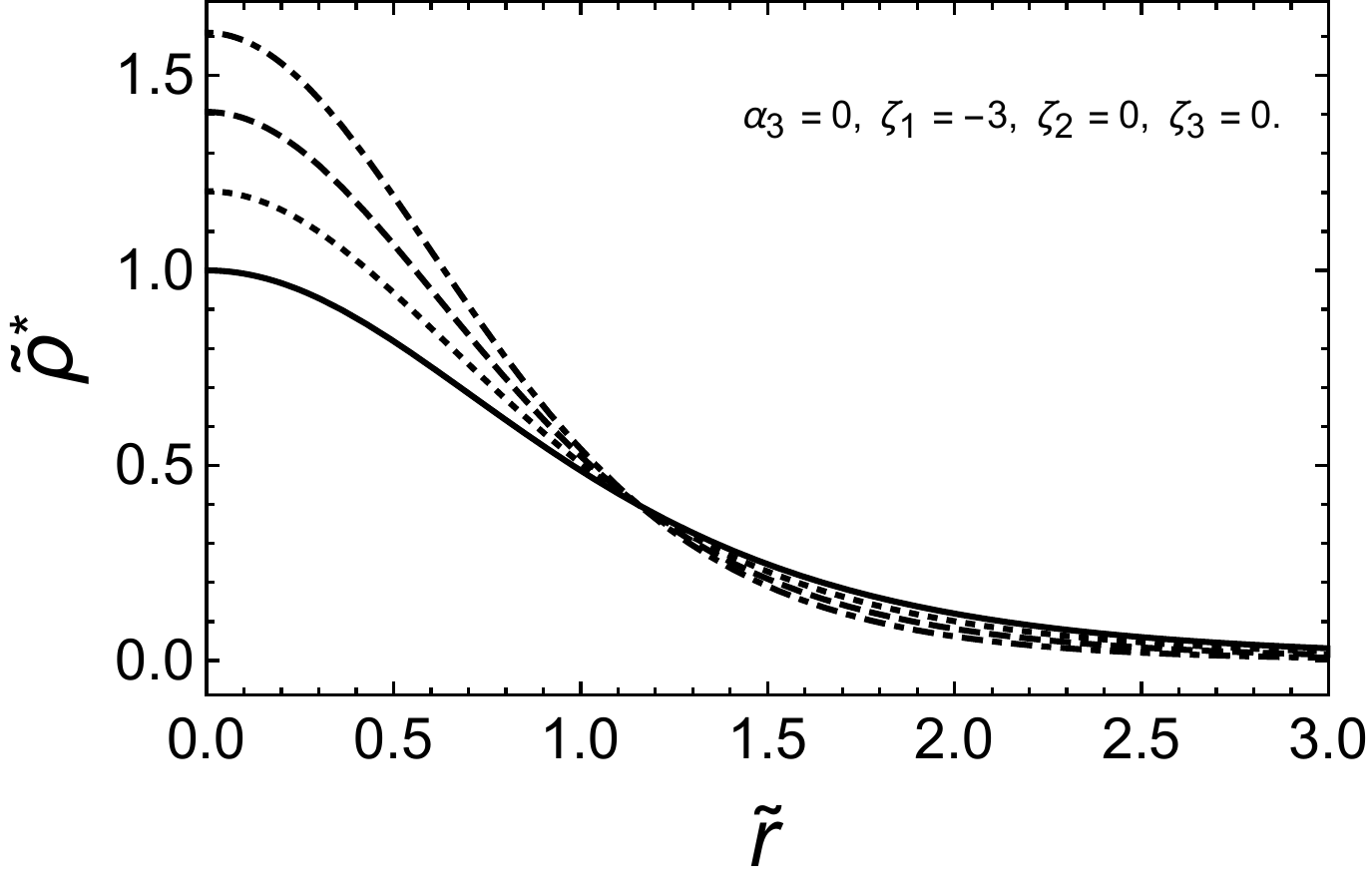}
	\includegraphics[scale=0.3]{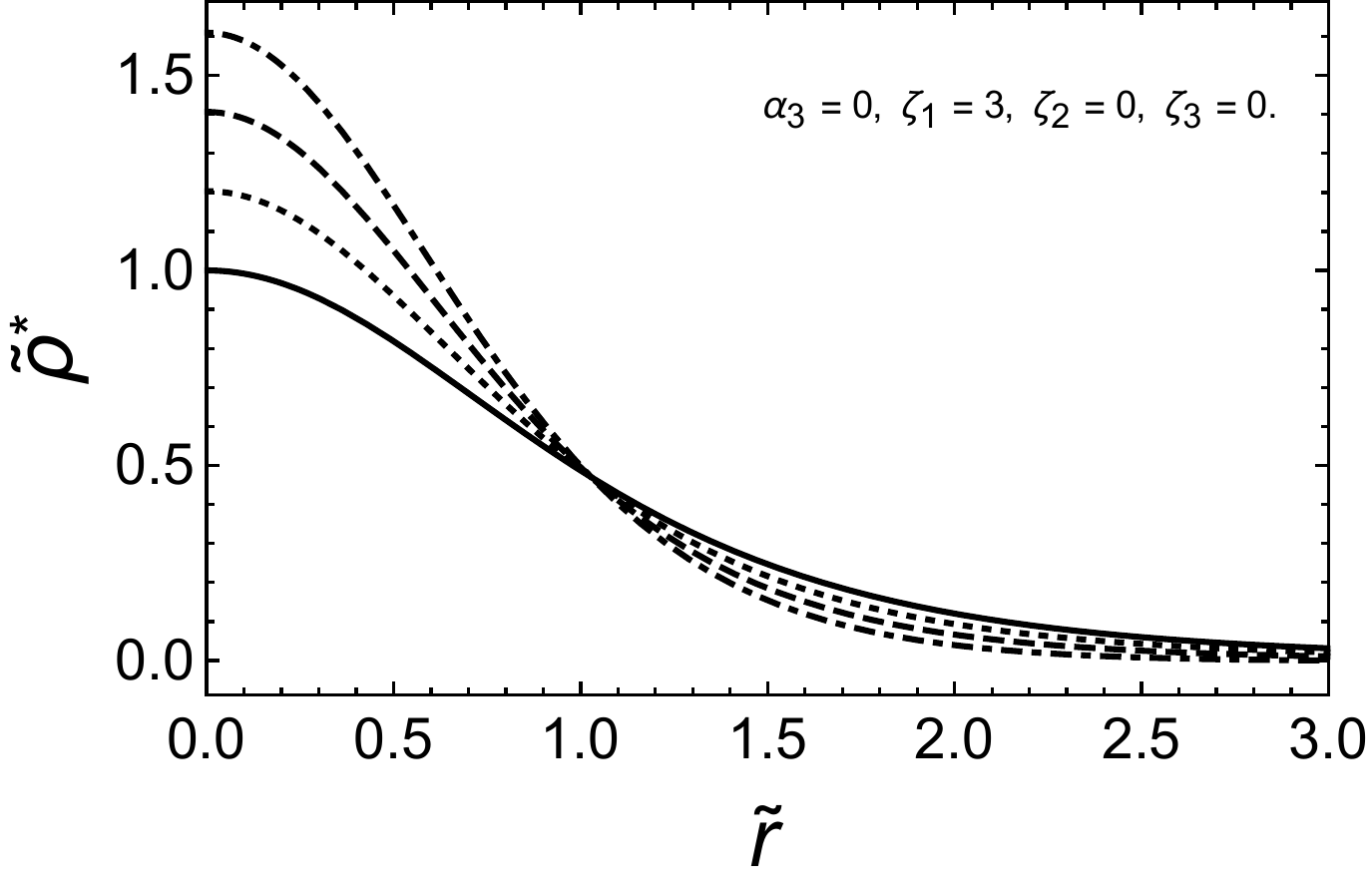}

	\includegraphics[scale=0.3]{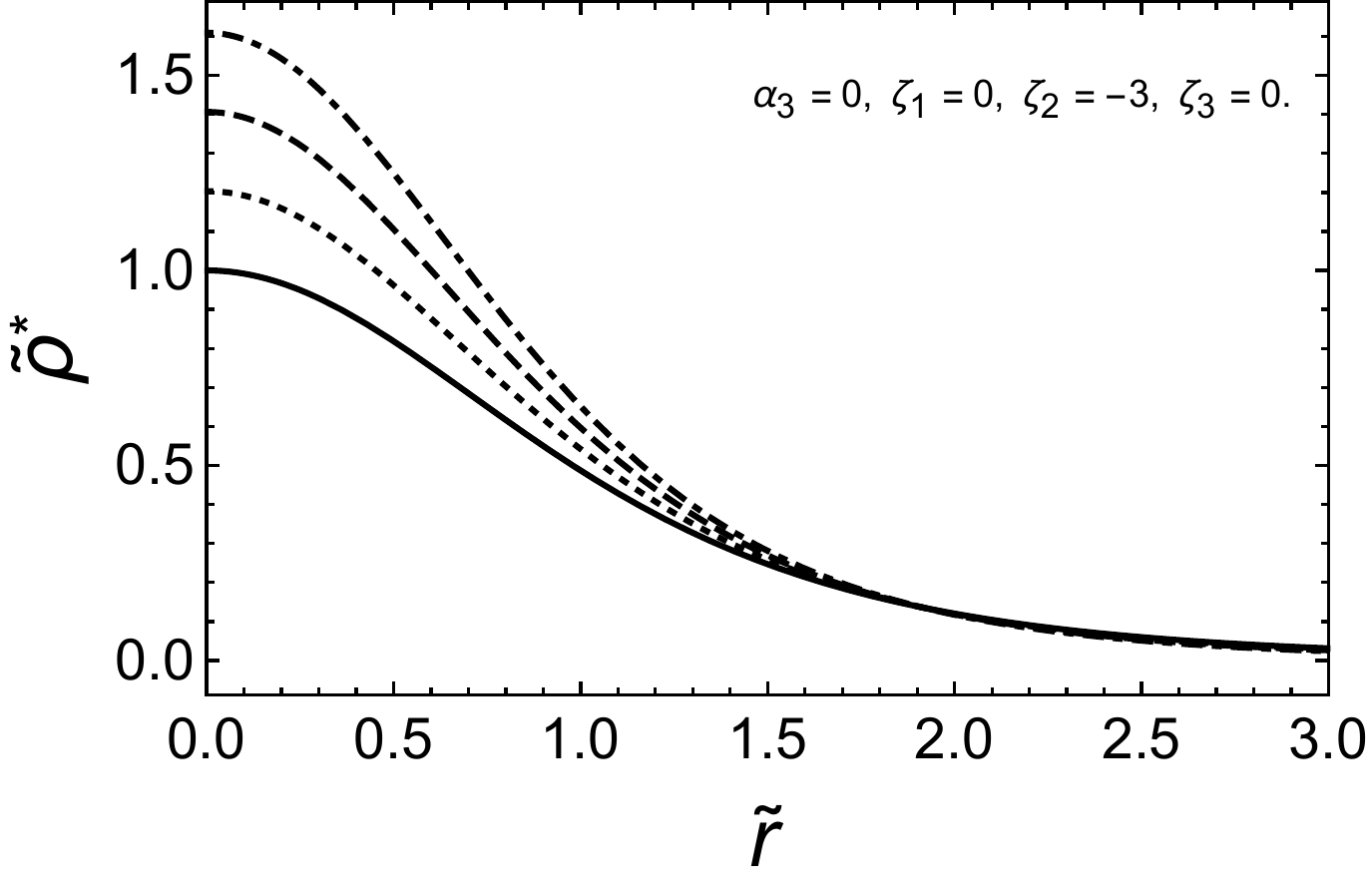}
	\includegraphics[scale=0.3]{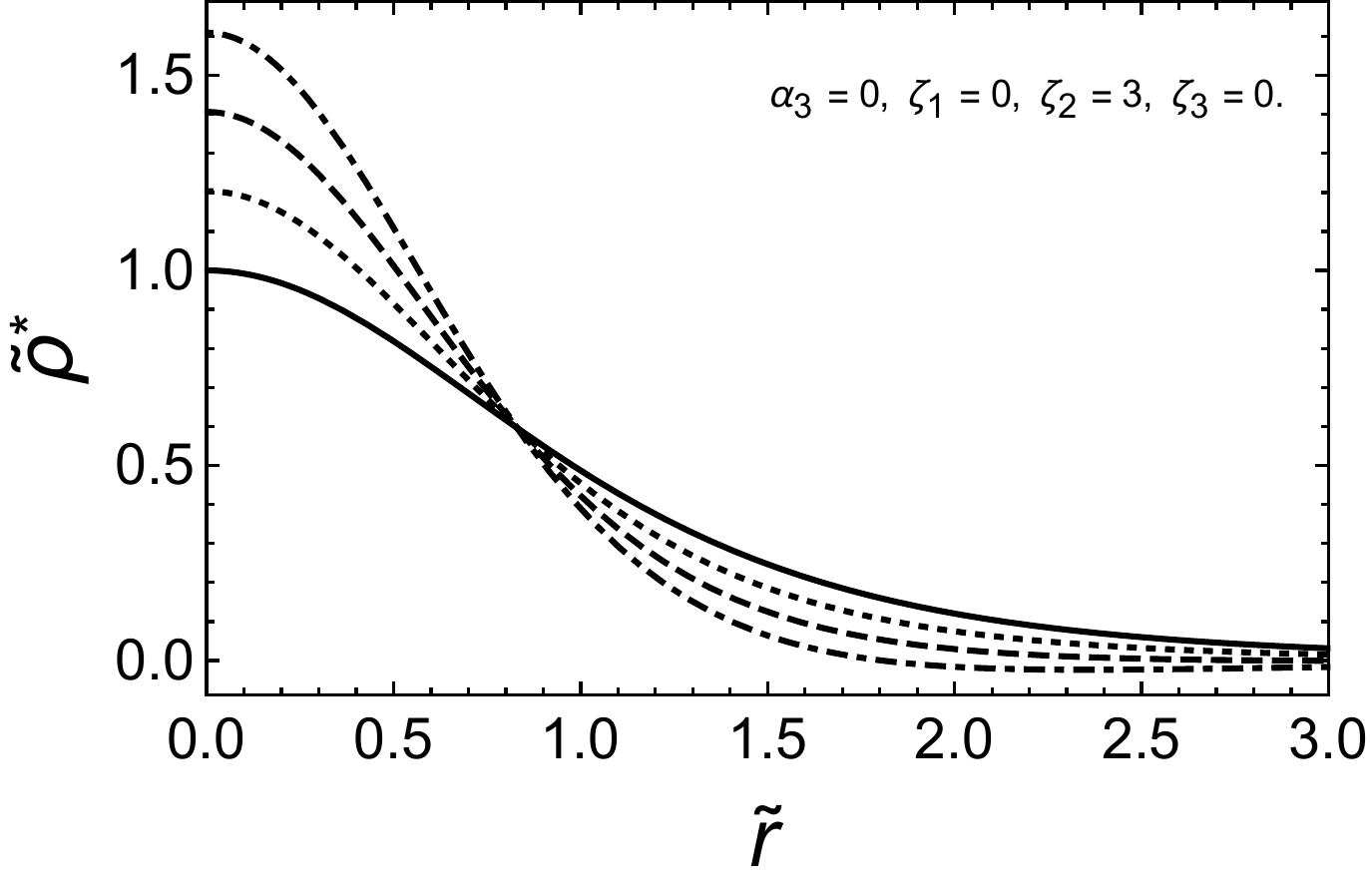}

	\includegraphics[scale=0.3]{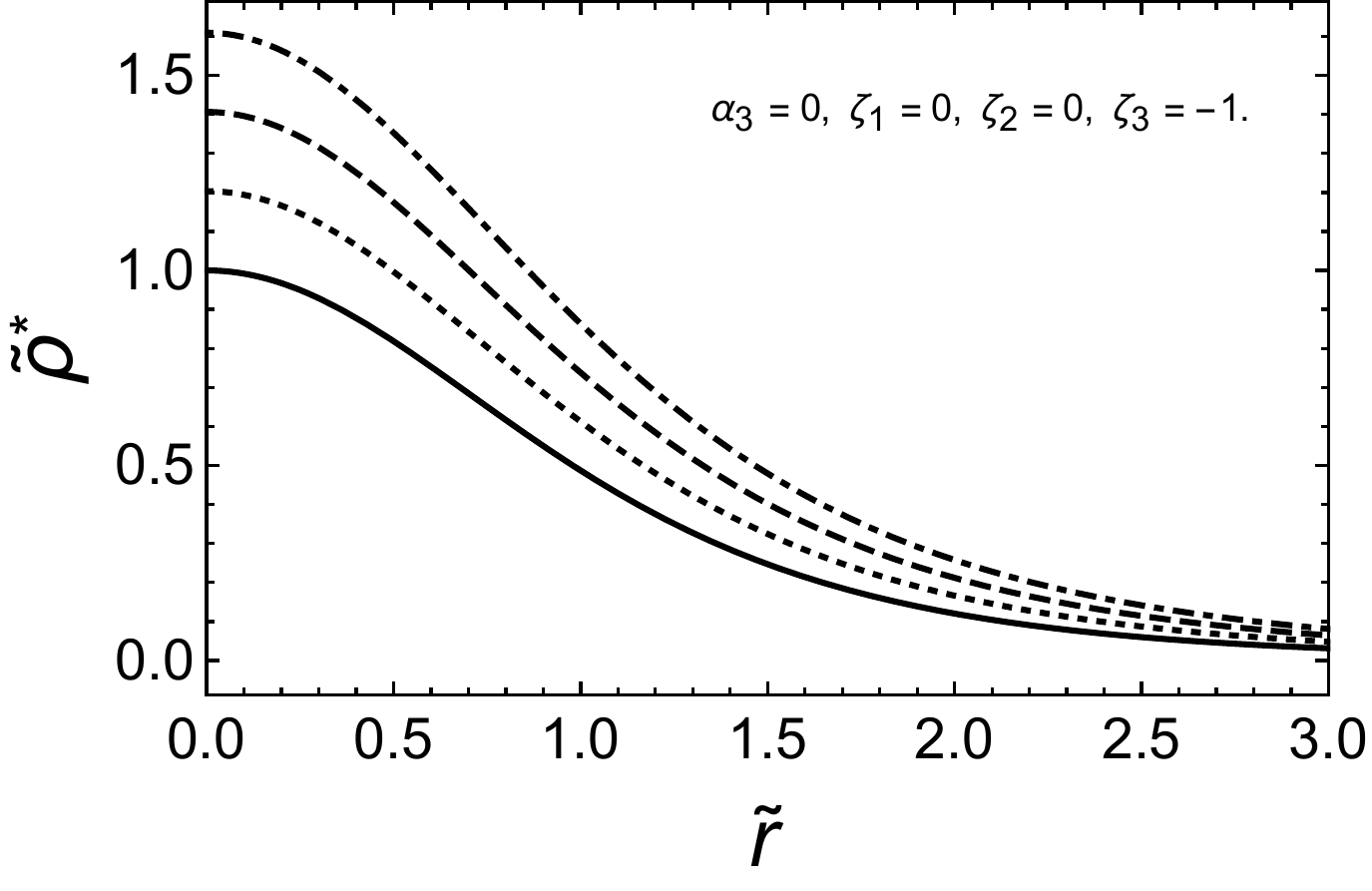}
	\includegraphics[scale=0.3]{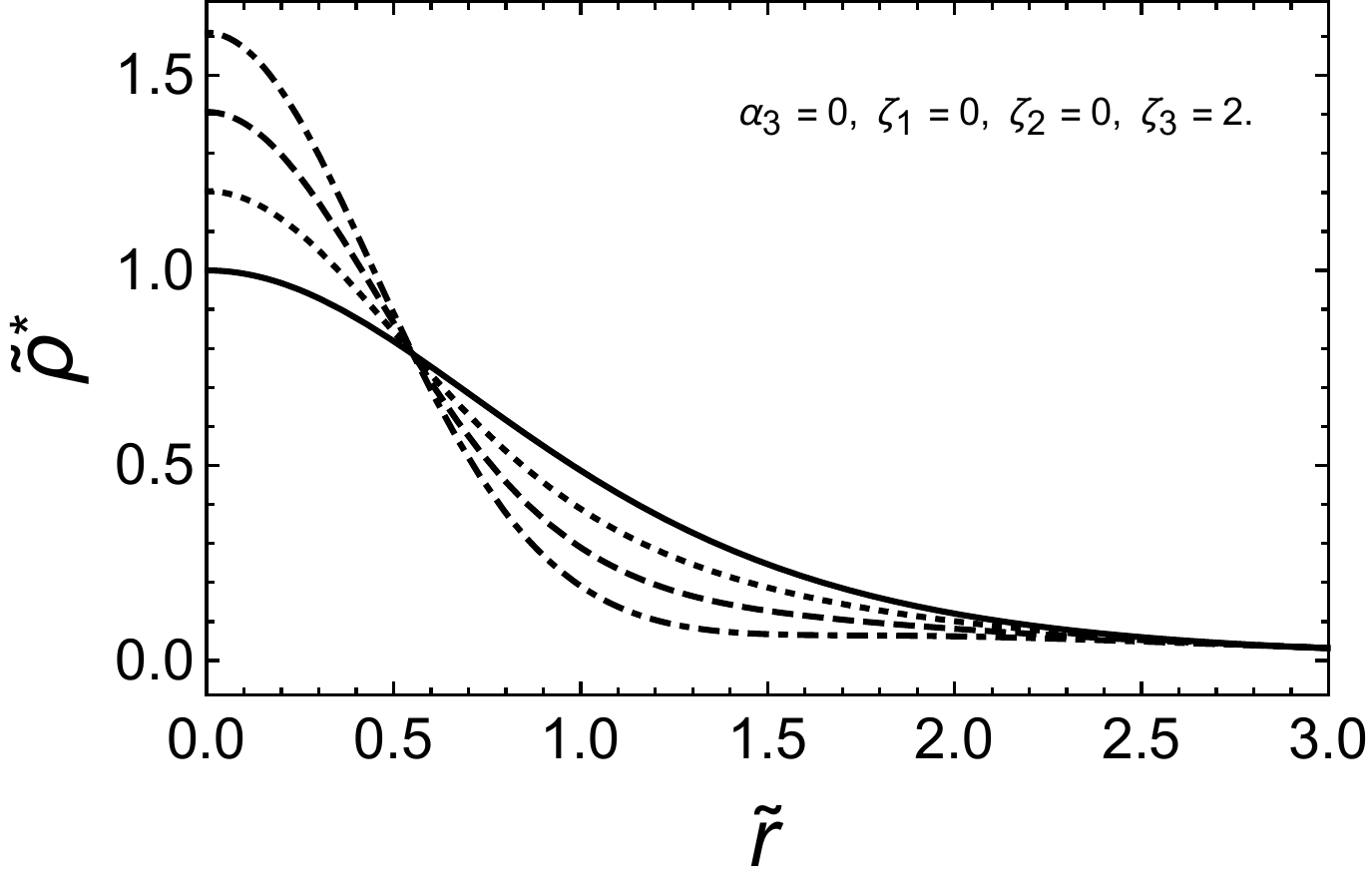}

	\caption{We present the effective density $\widetilde{\rho}^{*}$ for a polytopic model with index $n=5$ for non-conservative theories, here we regard the parameters $\gamma=\beta=1 \text{ and }\xi=0$. From top to bottom, we take alternatively $\alpha_{3}$, $\zeta_{1}$, $\zeta_{2}$ and $\zeta_{3}$ different from $0$, respectively. The continuous line represents the Newtonian curve and the dotted, dashed and dash-dotted are PPN curves with $\epsilon=0.05$, $\epsilon=0.1$ and $\epsilon=0.15$, respectively.
	}\label{densidadenoncons}
\end{figure}

The behavior of the matter fields is shown in figures \ref{densidadecons} and \ref{densidadenoncons}, through the effective density 
$\widetilde{\rho}$ for each model, defined as
\begin{align*}
\widetilde{\rho}^{*}&=\frac{-4\pi G{T}}{\left(U_{0}\right)^{n}C_{n}}=\frac{-4\pi G{T^{\mu}}_{\mu}}{\left(U_{0}\right)^{n}C_{n}}
\\
&=\frac{4\pi G\rho^{*}}{\left(U_{0}\right)^{n}C_{n}}\left\{1+\frac{1}{c^{2}}\left[\Pi-3\gamma U\right]\right\}-\frac{12\pi G P}{\left(U_{0}\right)^{n}C_{n}}
\\
&=\widetilde{U}^n\left(1+\frac{\epsilon n\alpha _3}{2}\right)+\epsilon n \widetilde{\Psi}\widetilde{U}^{n-1}
\\
&\qquad+\epsilon\widetilde{U}^{n+1}\left[\left(\frac{1}{2}-\beta \right) n+\frac{3}{8 (n+1)}+\frac{3}{2}\right].
\end{align*}

The overall behavior turn out to be similar for other politropic indexes and other combinations of the PPN parameters.

%------------------------------------------------------------------------------------------------
\subsection{The PPN corrections to the rotation curves}

\begin{figure}[ht]
\includegraphics[scale=0.3]{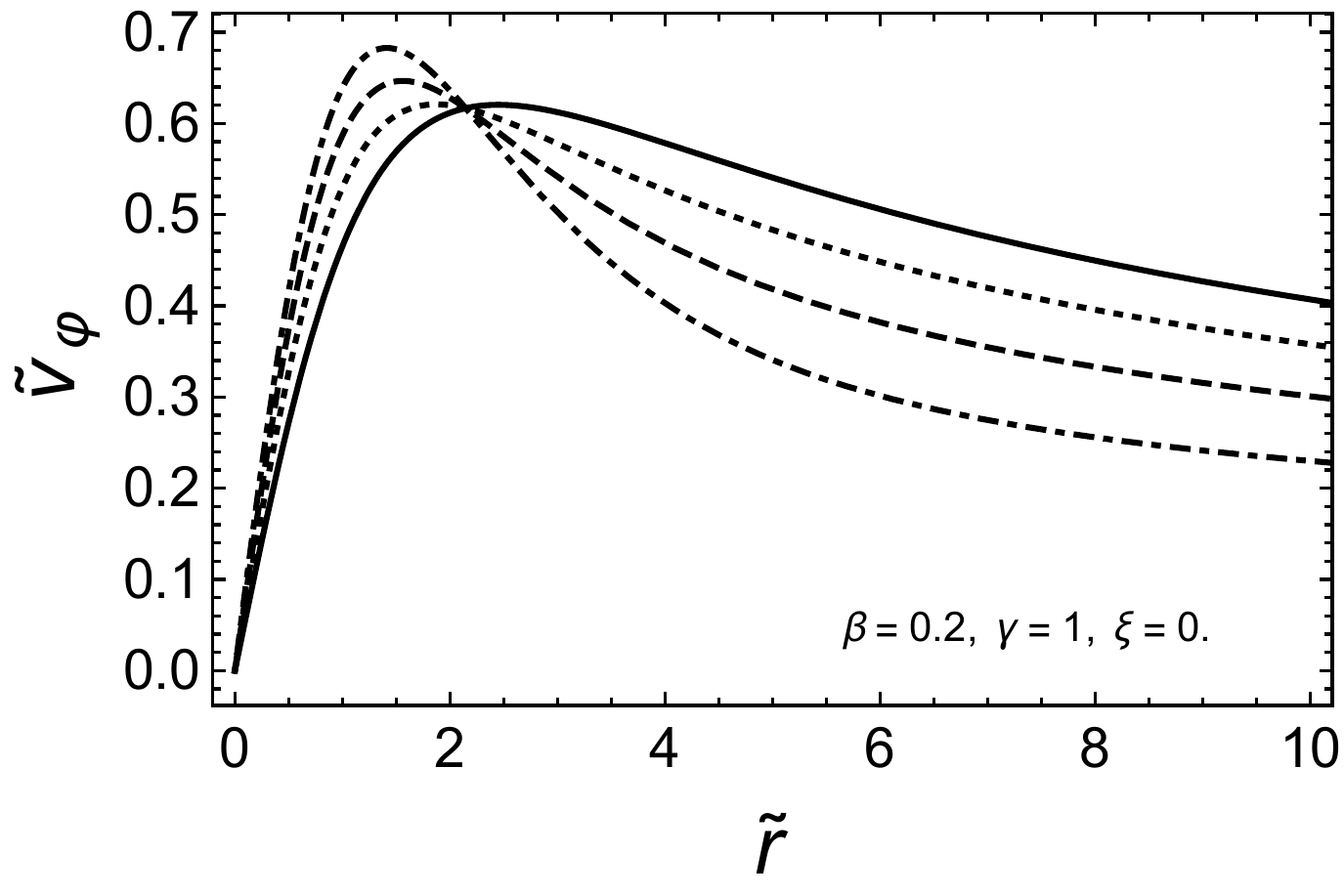}
\includegraphics[scale=0.3]{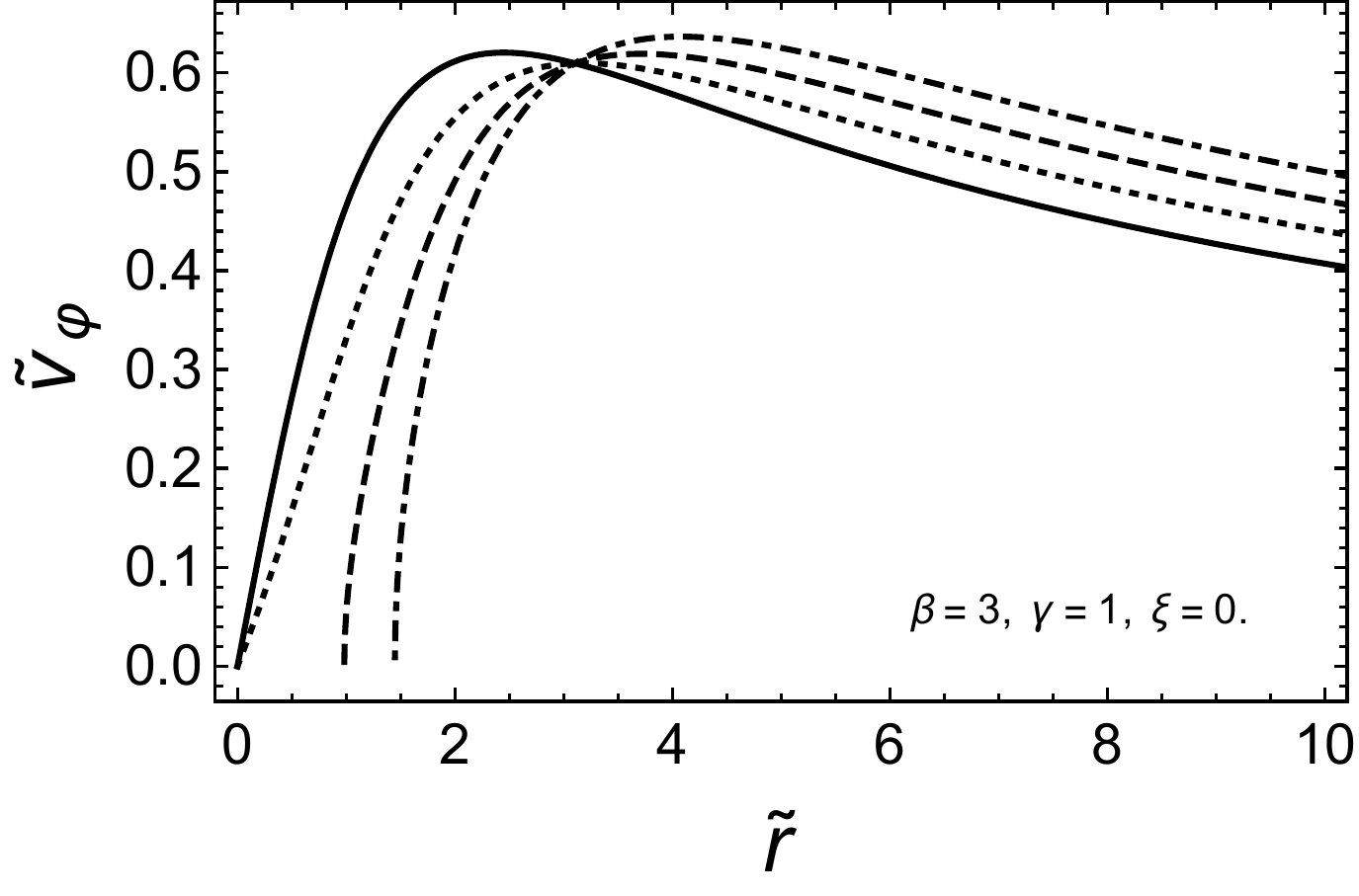}

\includegraphics[scale=0.3]{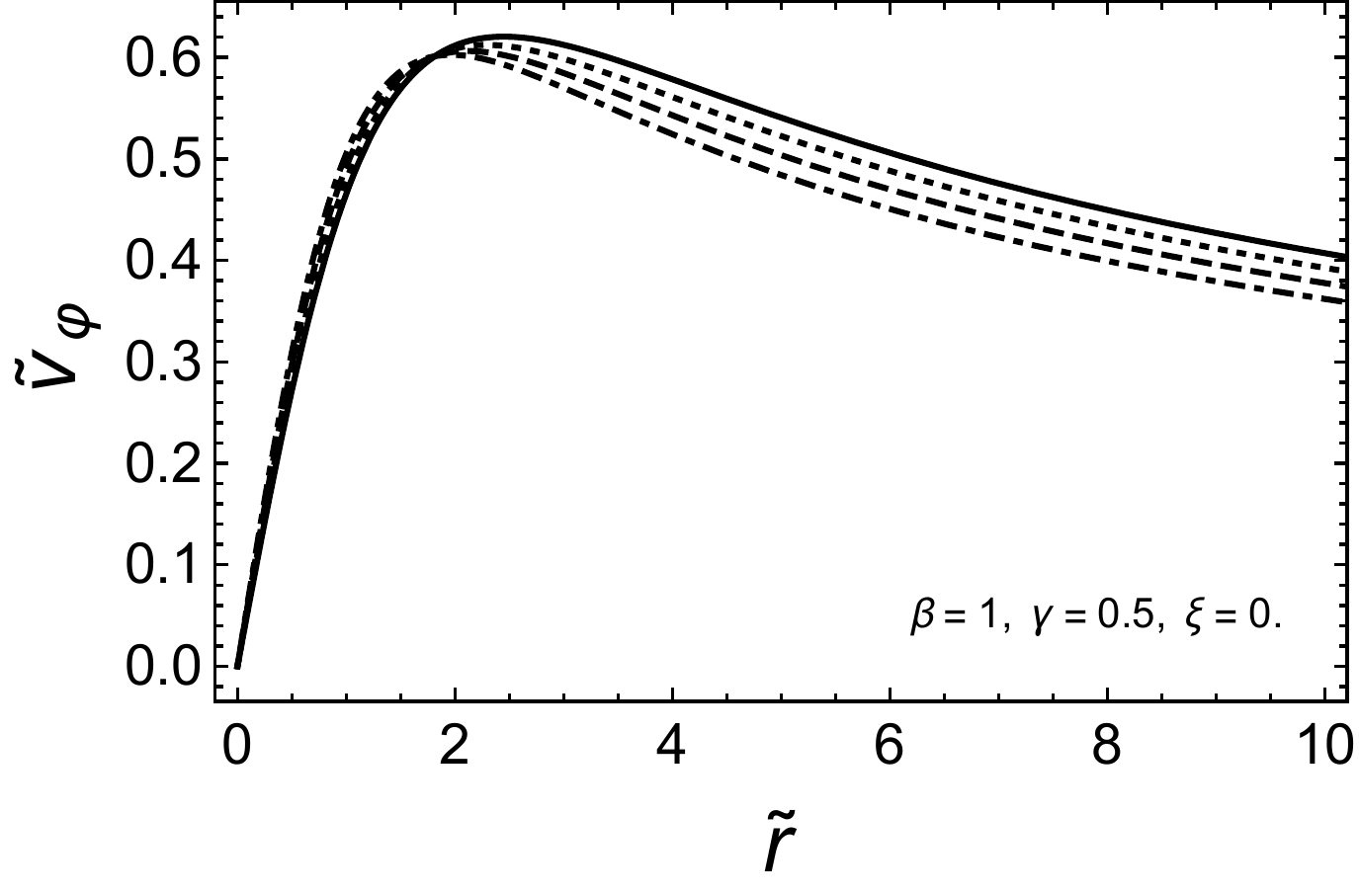}
\includegraphics[scale=0.3]{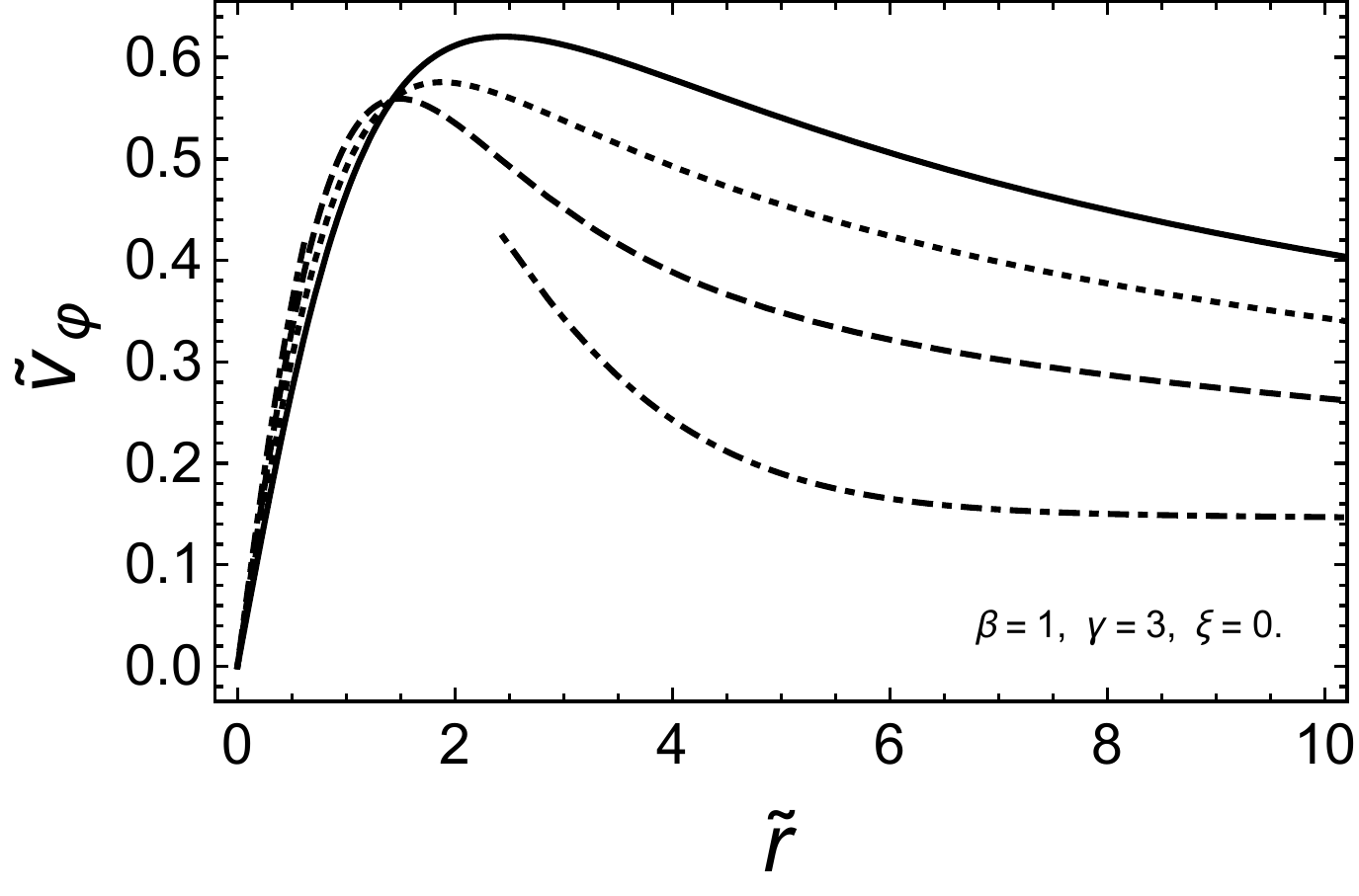}

\includegraphics[scale=0.3]{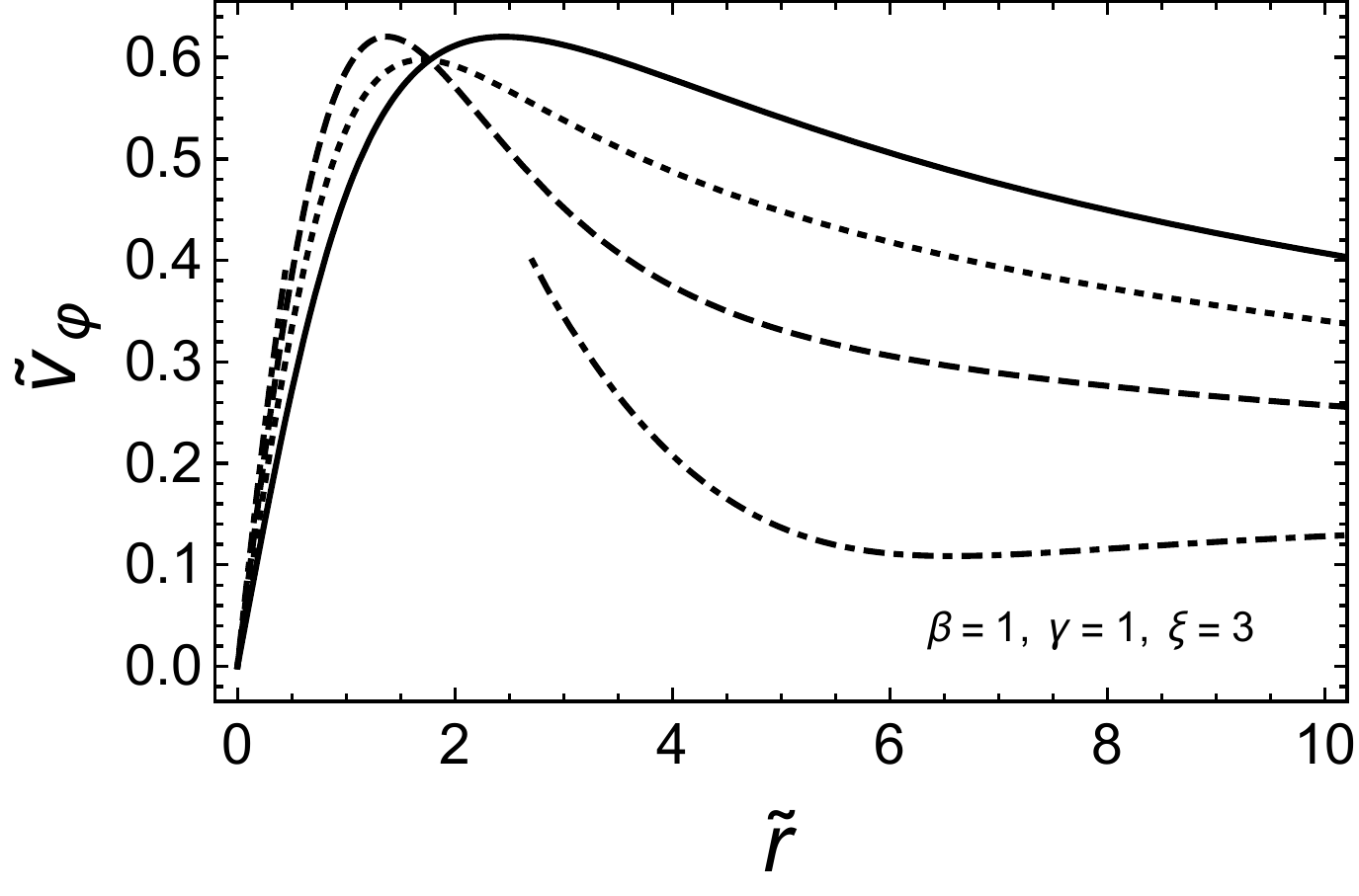}
\includegraphics[scale=0.3]{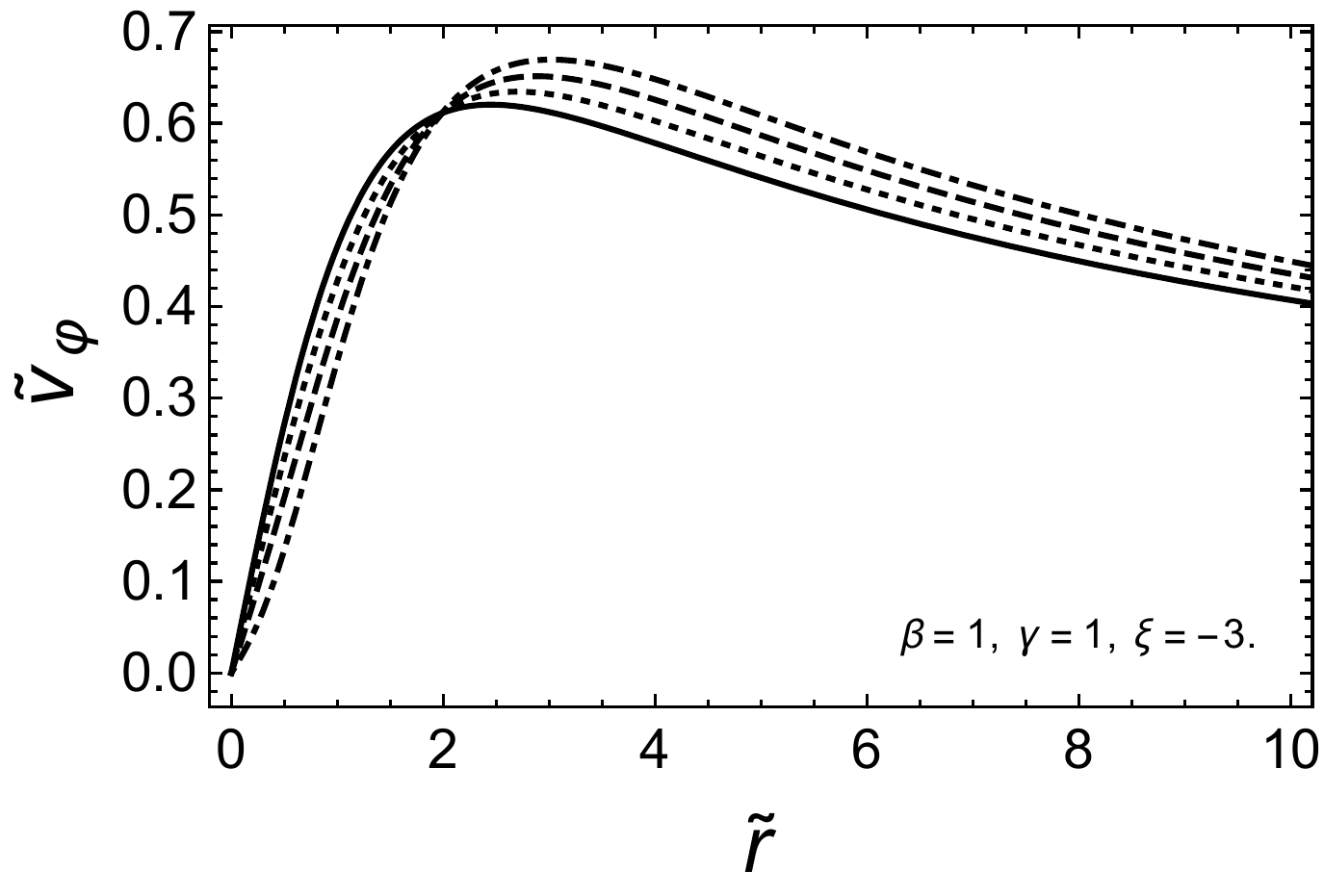}

	\caption{We present the rotation curves $\left(\tilde{v}^{\varphi}={v}^{\varphi}/\sqrt{U_{0}}\right)$ for a polytropic model with index $n=5$ for fully conservative theories $\left(\zeta_{1}=\zeta_{2}=\zeta_{3}=\zeta_{4}=\alpha_{3}=0\right)$. The top figures are for $\beta$ different from $1$, the middle ones for $\gamma$ different from $1$ and the lower ones for $\xi$ different from $0$. The continuous line represents the Newtonian curve and the dotted, dashed and dash-dotted are PPN curves with $\epsilon=0.05$, $\epsilon=0.1$ and $\epsilon=0.15$, respectively.
	}\label{rotationcurve1}
\end{figure}

\begin{figure}[ht]
	\includegraphics[scale=0.3]{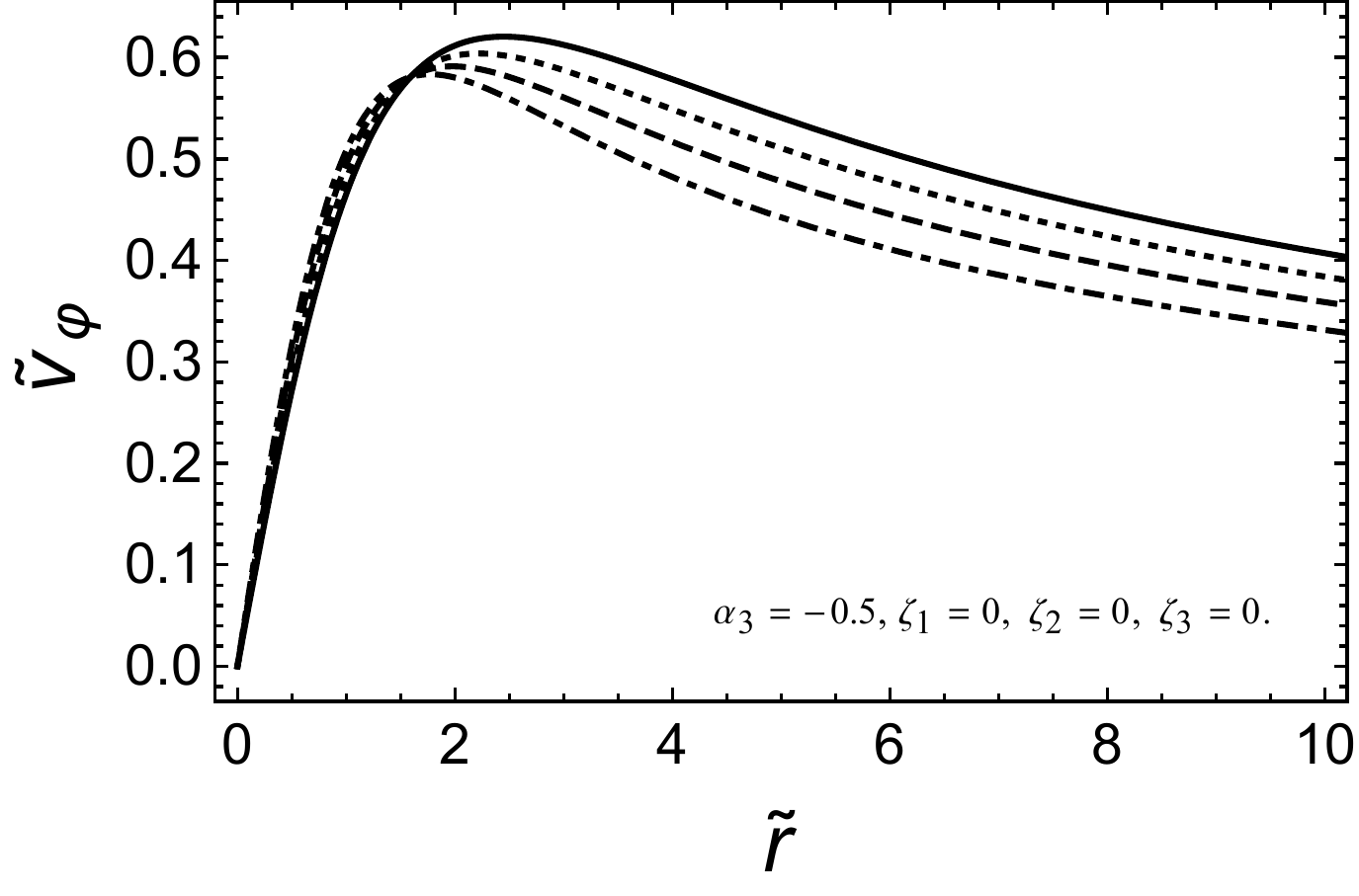}
	\includegraphics[scale=0.3]{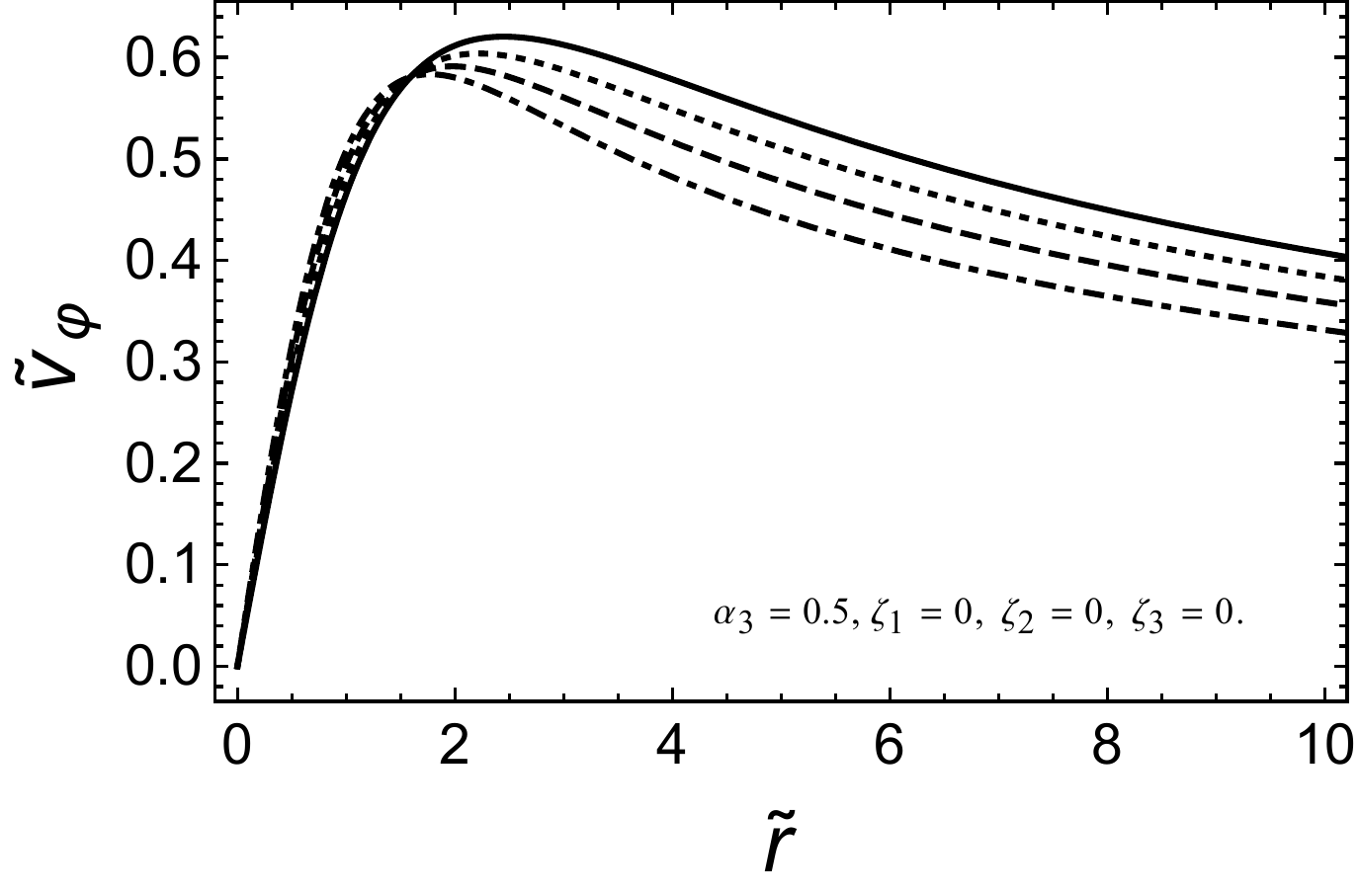}
	
	\includegraphics[scale=0.3]{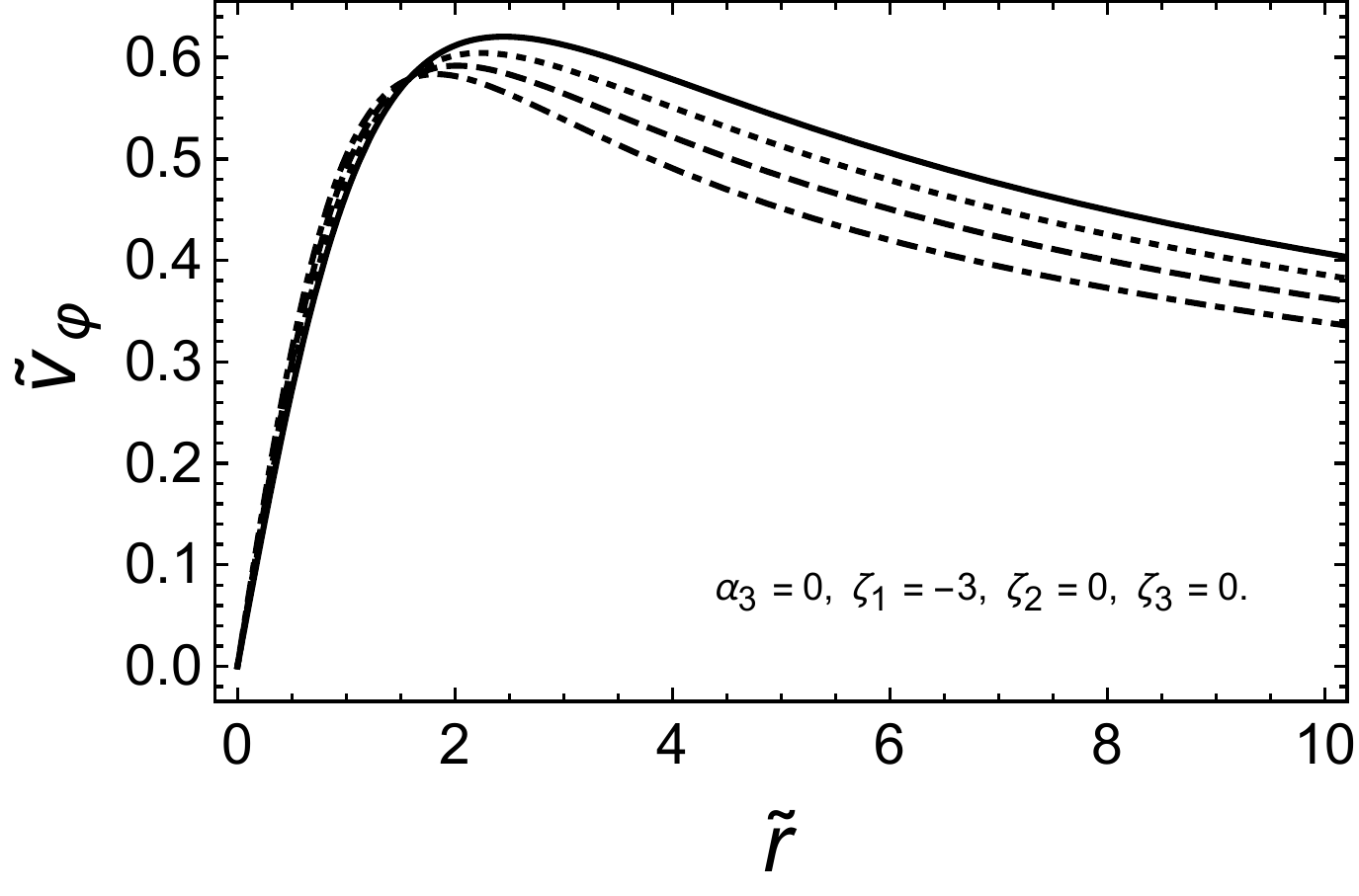}
	\includegraphics[scale=0.3]{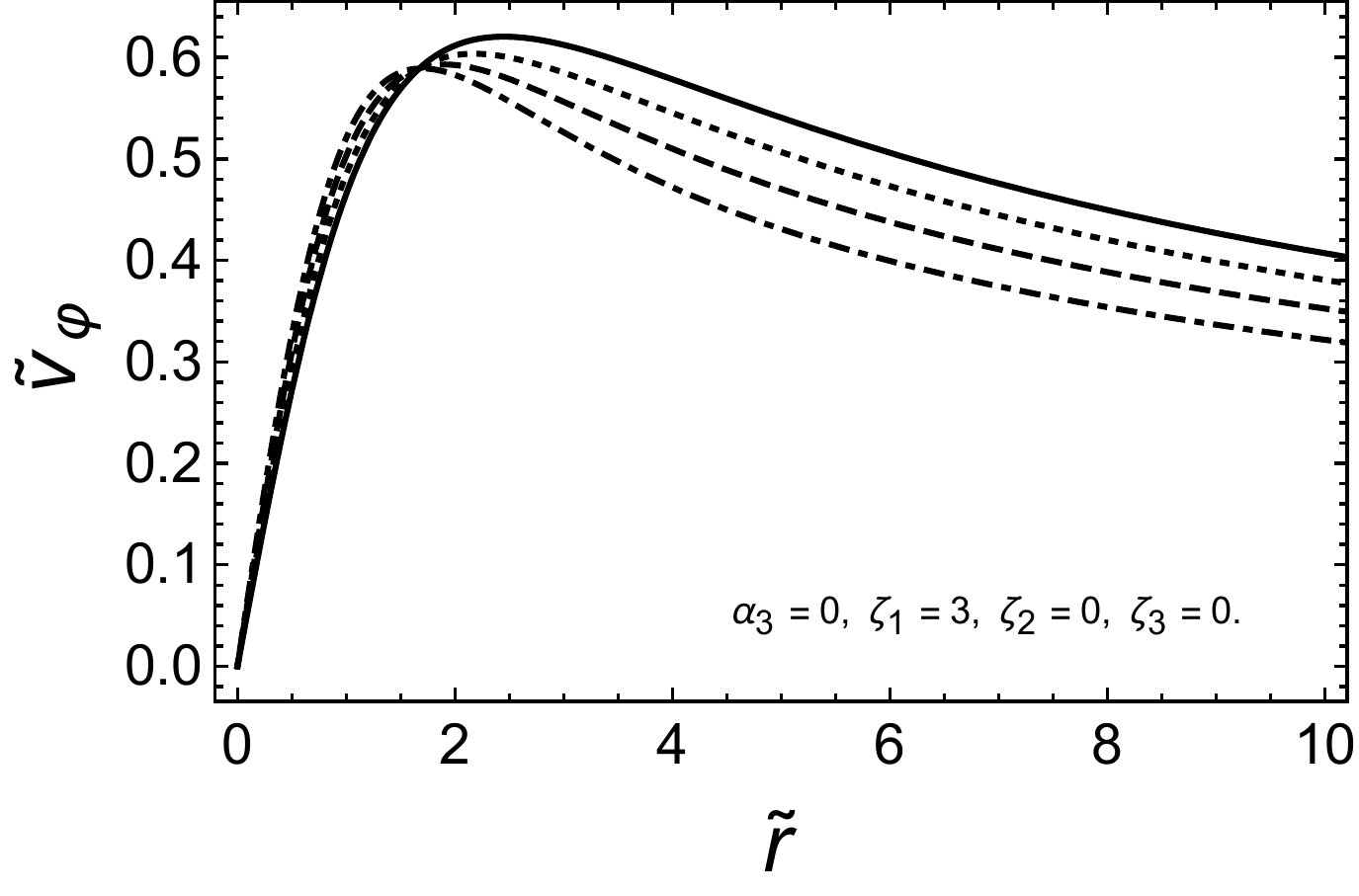}
	
	\includegraphics[scale=0.3]{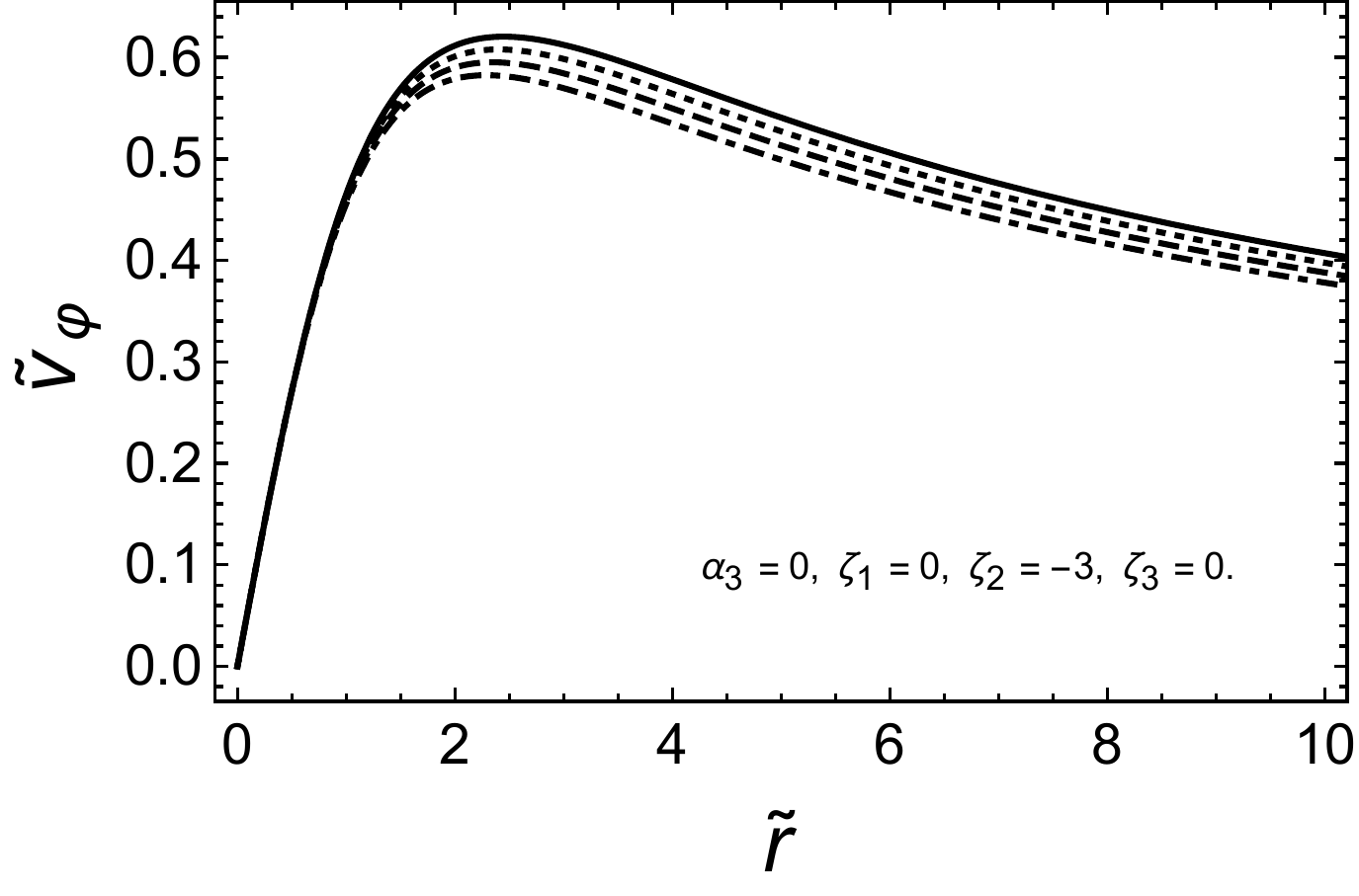}
	\includegraphics[scale=0.3]{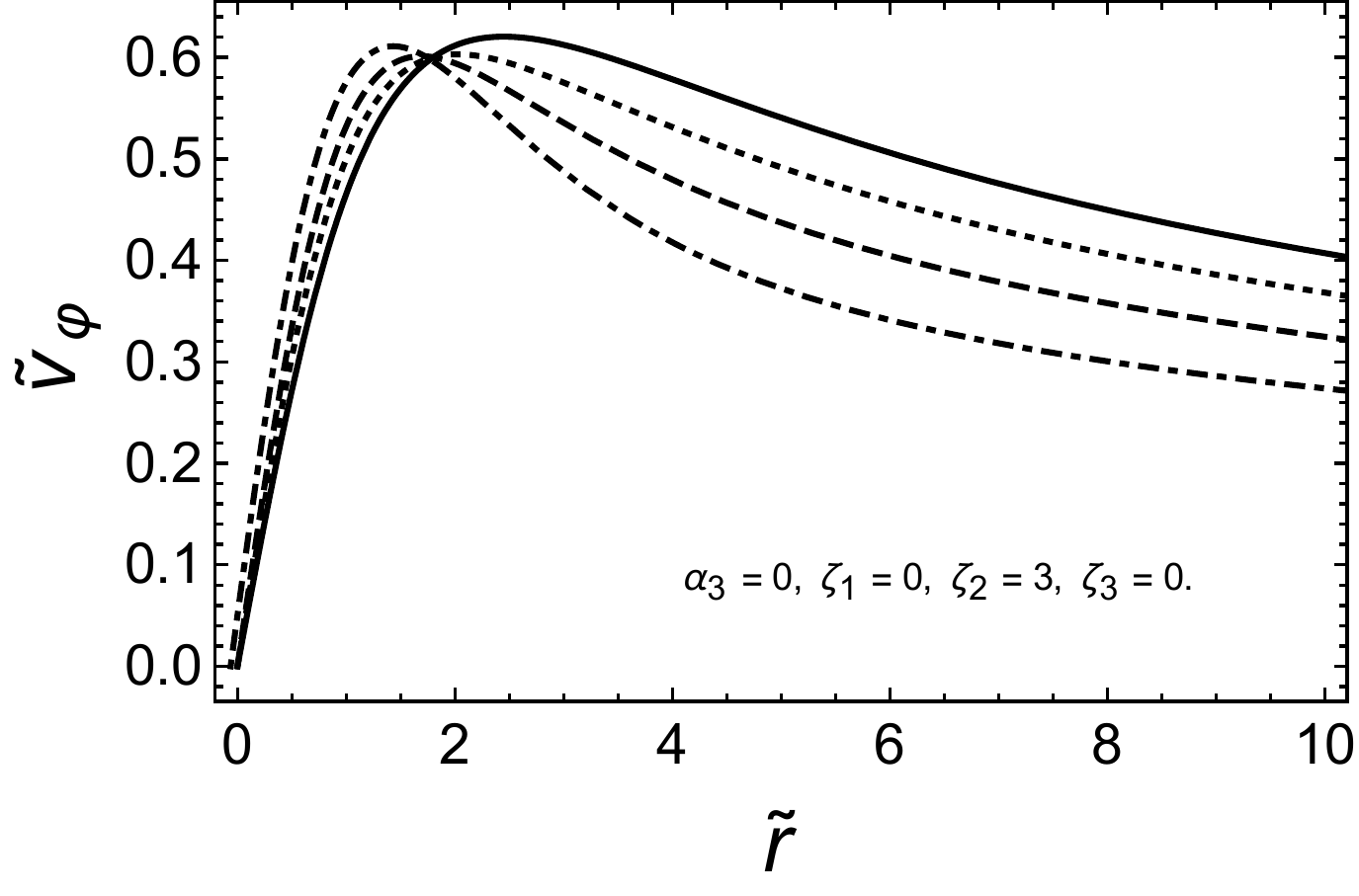}
	
	\includegraphics[scale=0.3]{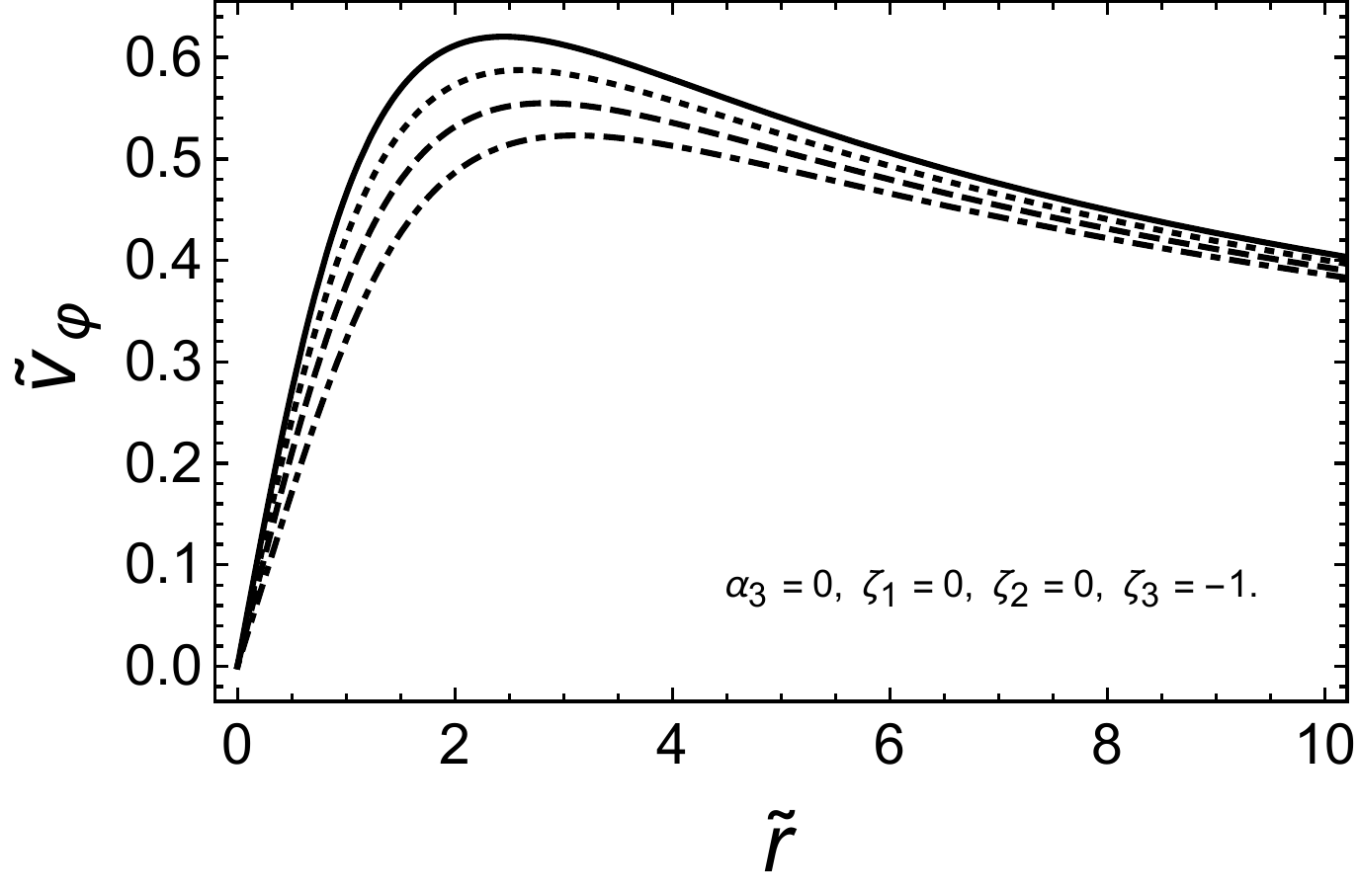}
	\includegraphics[scale=0.3]{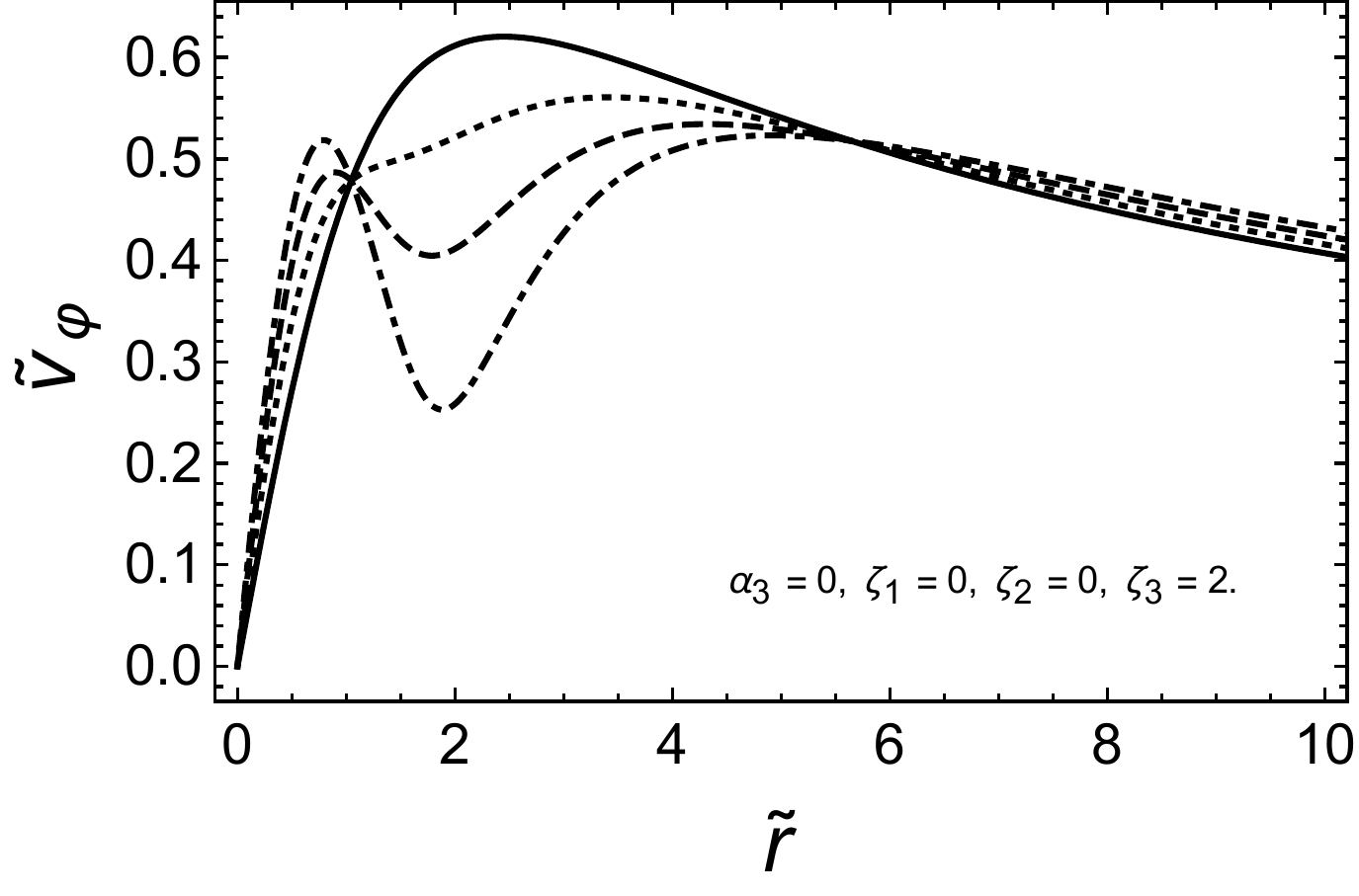}
	
	\caption{We present the rotation curves $\left(\tilde{v}^{\varphi}={v}^{\varphi}/\sqrt{U_{0}}\right)$ for a polytropic model with index $n=5$ for non-conservative theories, here we regard the parameters $\gamma=\beta=1 \text{ and }\xi=0$. From top to bottom, we take alternatively $\alpha_{3}$, $\zeta_{1}$, $\zeta_{2}$ and $\zeta_{3}$ different from $0$, respectively. The continuous line represents the Newtonian curve and the dotted, dashed and dash-dotted are PPN curves with $\epsilon=0.05$, $\epsilon=0.1$ and $\epsilon=0.15$, respectively.
	}\label{rotationcurve2}
\end{figure}

With the solutions at hand, we sketch the rotation curves for a polytropic model with index $n=5$ 
for several values of the PPN parameters (see figures \ref{rotationcurve1} and \ref{rotationcurve2}). 
The rotation curves for $n\neq5$ are either similar to the ones of $n=5$ or have unphysical properties.

Although we only present some choices for the PPN parameters (it is not easy to develop a 
systematic approach to test all possible combination of values for the PPN parameters)
it is reasonable to conclude, from the curves we presented, Figs. \ref{rotationcurve1} and \ref{rotationcurve2}, 
that only the ones with $\zeta_{3}>0$ have some resemblance to flattened curves (as expected from the analysis of the asymptotic
solution of previous section). 
The outcome corresponding to other combinations, including other politropes with $n\neq 5$, are alike Figs. \ref{rotationcurve1} and \ref{rotationcurve2}.
In particular, the curves with $\zeta_{3}=2$ present some flat behavior, 
but have lower values of velocity than the Newtonian one (except at some critical value around $\tilde{r}=6$). 
So, for this model to be a plausible explanation of the rotation curves of galaxies, it must have less mass than the Newtonian one. This can be seen
 by comparing the density of mass $\rho$ with the trace of the stress-energy tensor $-T=-{T^{\mu}}_{\mu}$. This is shown in Fig. \ref{densidadenoncons} and
 we note that $-T$ falls much faster, 
with the radius, than the density of mass, therefore explaining why the rotation curves for the PPN have 
a lower first local maximum value of velocity and much sooner than the Newtonian one.
%mudei
This means that a theory with $\zeta_{3}>0$ does present some interesting corrections to the rotation curves. 
But, as it should be, the usual Keplerian falloff is still found for $\tilde{r}>10$.

On the other hand, we note that choices involving $\zeta_{3}\leq0$ (even for high values of the parameter $\epsilon$
and other exotic choices),
have the same behavior than the Newtonian models far from the center of the configuration.

\section{Conclusion}
%in progress%
In this paper, the rotation curves of static spherically symmetric models were studied in the framework of the PPN approximation. 
We first started addressing vacuum solutions and later we focus on  politropic models (see \cite{ramos1} for a similar approach 
in the post-Newtonian context).

When the matter fields, endowed with spherical symmetry, are assumed static  one can readily notice that the PPN metric is not necessarily spherically symmetric (a direct consequence of the preferred frame potentials). Therefore, circular orbits will only exist if and only if the parameters $\alpha_{1}$ and $\alpha_{2}$ are null. This means, as their heuristic meaning imply, that we excluded the effects of non-conservation of global angular momenta of the PPN scheme. For all calculations that followed Sec. \ref{sec3sub2} we assumed $\alpha_{1}=\alpha_{2}=0$.

As a first setup for testing the circular orbits in the PPN approach, we solved field eqs. \eqref{fieldU}-\eqref{fieldPsi} in vacuum outside some static spherically symmetrical distribution of matter of finite radius and presented, to PPN order, the velocity of circular orbits (eq. \eqref{noflat}). We argued that independently of the choice of the PPN parameters such solution can not lead to flat curves. Our argument is based on the simple property that, to PPN order, such solutions have necessarily $\mathrm{d}v/\mathrm{d}r<0$ for all $r>a$. Therefore a Keplerian falloff, similar to the ones found in Newtonian gravity, will still be observed in this description of gravity.

%ainda em progresso%
To construct astrophysical models, when matter fields are present, we applied a statistical description, following the assumptions of Subsec. \ref{sec3}.
In such cases, the system is fully described by a DF that satisfy the CBE, which in the context of the PPN approximation takes the form of eq. \eqref{vlasovppn}.
In statical spherical models the previous assumptions and the results of Subsec. \ref{sec3sub2} implies (see Subsec. \ref{sec3sub3}) that the DF must be a function of only the energy like integral of motion.
The Politropes are a family of models whose DF is a particular case that satisfy the previously stated requirement.
In the context of the PNP approach, the Politropes can be constructed by a simple extension of their Newtonian DF, this leads to the complete specification of the matter fields.
Therefore one can determine all the gravitational fields from \eqref{campoU}-\eqref{campoPsi}.

To solve eqs. \eqref{campoU}-\eqref{campoPsi} we employed: I) approximated methods (Sec. \ref{sec4sub2}); II) numerical techniques (Sec. \ref{numerical}).
In \ref{sec4sub2}, we solved approximately the field equations (for $n=5$) when the radius is much larger than a characteristic radius of the model ($\tilde{r}\gg \sqrt{3}$).
The approximation had the setback of not providing proper boundary conditions, but the velocity of circular orbits did present some terms with behavior different from the Newtonian solution.
From the approximated solutions we observed that the larger modifications will generally come from large values of the parameter $\zeta_{3}$. But to be certain, since we lack proper boundary conditions, one needs the solution for the entirety of space, which can only be achieved by a numerical route.

Using numerical techniques (fourth order Runge-Kutta method) we were able to find the gravitational and matter fields for all choices of parameters, Figs. \ref{camposcons}, \ref{camposnoncons}, \ref{densidadecons} and \ref{densidadenoncons}. In Figs. \ref{rotationcurve1} and \ref{rotationcurve2} we present the rotation curves for several values of the Post-Newtonian parameters, and concluded that many of the choices lead to trivial corrections to the rotation curves, sometimes leading to smaller values of the circular velocities or an even higher descent rate, with the radius, then the Newtonian models.

There was a single choice of the parameters that lead to different behavior. By choosing $\zeta_{3}>0$ we found some flat rotation curves and with the hassle of lower values of circular velocity. But by analyzing the effective density ($-T$) we realize this should be expected, since the amount of matter in this model is significantly less than in its Newtonian counterpart.
Therefore a theory with $\zeta_{3}>0$, as far as our analysis goes, does present interesting modifications to the Keplerian behavior.
All other choices of the parameters comes short in explaining the rotation curves of galaxies.  Since only one of the parameters presents real significance one may conclude that gravity theories compatible with the PPN scheme are not very successful as alternatives to the dark matter hypothesis in galactic dynamics.

Here, we do not claim that the PPN approach contains all the potentials needed to describe theories with SM.
In \cite{avilez,PhysRevD.99.084009,McManus_2017} was already shown that the parameterized version of the Vainshteinian screening requires more potentials (which describes the effects of the mechanism).
In this paper, we only considered the potentials present in the standard PPN approach, we excluded the Vainshteinian ones (or from other SMs) that may emerge from a more complete formalism.
But it is to be expected that for $r\gg r_{v}$ the approximation should be equivalent to the PPN approaximation. Nonetheless, we recognize that all theories (with or without a SM) have, at least, some of the standard PPN potentials.
Even though the work by \citeauthor{avilez} \cite{avilez}, and the further extensions by \cite{PhysRevD.99.084009,McManus_2017}, does outline the main ideas for constructing the PPNV (Parametrized Post-Newtonian Vainshteinian) formalism, we will leave the complete analysis of this problem for when the Post-Newtonian expansion of theories with SMs are better understood.

\acknowledgments
H.M.G. is grateful for the financial support provided by
CNPq (Grant No. 132389/2017-7 and No. 141924/2019-5).

\appendix

\section{Derivation of the Field Equations}
Expressions \eqref{invariant1} and \eqref{invariant2} are easily determined from their integral forms (see \cite{will2}). But this is not the case for expressions \eqref{appendix1} and \eqref{appendix2}, they require more attention.

In this section we present a derivation of the field equations \eqref{appendix1} and \eqref{appendix2} from the integral forms found in reference \cite{will2}.

\subsection*{Derivation of the Field Equation for $\phi_{6}$}
Starting from the integral form of $\phi_{6}$ \cite{will2}:
\begin{align*}
\phi_{6}&=G\int\rho^{*'}{u'}_{j}{u'}_{k}\frac{\left(\boldsymbol{x}-\boldsymbol{x}'\right)^{j}\left(\boldsymbol{x}-\boldsymbol{x}'\right)^{k}}{{\left|\boldsymbol{x}-\boldsymbol{x}'\right|}^{3}}{\mathrm{d}}^{3}x'
\\
&=-G\int\rho^{*'}\left[\boldsymbol{u}'\cdot\left(\boldsymbol{x}-\boldsymbol{x}'\right) \boldsymbol{u}'\cdot\nabla{\left|\boldsymbol{x}-\boldsymbol{x}'\right|}^{-1}\right]{\mathrm{d}}^{3}x'.
\end{align*}
We want to reduce this to its differential form, to remove the integral we take the laplacian of the integral form, leading to:
\begin{multline*}
\nabla^{2}\phi_{6}=-G\int\rho^{*'}\big[3u^{2}4\pi\delta\left(\boldsymbol{x}-\boldsymbol{x}'\right)+2\left(\boldsymbol{u}'\cdot\nabla\right)^{2}{\left|\boldsymbol{x}-\boldsymbol{x}'\right|}^{-1}
\big.\\\big.
-4\pi\left(\boldsymbol{u}'\cdot\nabla\right)\boldsymbol{u}'\cdot\left(\boldsymbol{x}-\boldsymbol{x}'\right)\delta\left(\boldsymbol{x}-\boldsymbol{x}'\right)\big]{\mathrm{d}}^{3}x',
\end{multline*}
which follows from
\begin{multline*}
\nabla^{2} \left[\boldsymbol{u}'\cdot\left(\boldsymbol{x}-\boldsymbol{x}'\right) \boldsymbol{u}'\cdot\nabla{\left|\boldsymbol{x}-\boldsymbol{x}'\right|}^{-1}\right]=
\\
3u^{2}4\pi\delta\left(\boldsymbol{x}-\boldsymbol{x}'\right)+2\left(\boldsymbol{u}'\cdot\nabla\right)^{2}{\left|\boldsymbol{x}-\boldsymbol{x}'\right|}^{-1}
\\
-4\pi\left(\boldsymbol{u}'\cdot\nabla\right)\boldsymbol{u}'\cdot\left(\boldsymbol{x}-\boldsymbol{x}'\right)\delta\left(\boldsymbol{x}-\boldsymbol{x}'\right).
\end{multline*}
From this it's straightforward to show that
\begin{align*}
\nabla^{2}\phi_{6}&=3\nabla^{2}\phi_{1}-2G\int\rho^{*'}\left(\boldsymbol{u}'\cdot\nabla\right)^{2}\left(\frac{1}{\left|\boldsymbol{x}-\boldsymbol{x}'\right|}\right){\mathrm{d}}^{3}x'
\\
&=3\nabla^{2}\phi_{1}-2G\partial_{ij}\int\frac{\rho^{*'}{u^{i}}'{u^{j}}'}{\left|\boldsymbol{x}-\boldsymbol{x}'\right|}{\mathrm{d}}^{3}x'.
\end{align*}

The Laplacian of the previous expression results in \eqref{appendix1}.

\subsection*{Derivation of the Field Equation for $\phi_{w}$}
The integral form of $\phi_{w}$ can be presented in different forms \cite{will2}. Here, we start from the identity, see \cite{will2},
$$
\phi_{w}=-U^2-\phi_{2}-\nabla U\cdot\nabla X+G\nabla\cdot\int\frac{\rho^{*'}}{\left|\boldsymbol{x}-\boldsymbol{x}'\right|}\nabla'X'{\mathrm{d}}^{3}x'.
$$
Again, we wish to reduce this to a partial differential equation. To achieve this we start from the laplacian of the previous expression, resulting in:
\begin{multline}\label{appendix4}
\nabla^{2}\phi_{w}=-\nabla^{2}U^2-\nabla^{2}\phi_{2}-\nabla^{2}\left(\nabla U\cdot\nabla X\right)
\\
+G\nabla^{2}\left(\nabla\cdot\int\frac{\rho^{*'}}{\left|\boldsymbol{x}-\boldsymbol{x}'\right|}\nabla'X'{\mathrm{d}}^{3}x'\right).
\end{multline}
Now the identities,
$$
\nabla^{2}\left(\nabla U\cdot\nabla X\right)=2U_{,ij}X_{,ij}+2\nabla U\cdot\nabla U-4\pi G\nabla X\cdot\nabla\rho^{*},
$$
$$
\nabla^{2} U^2=2\nabla^{2}\phi_{2}+2\nabla U\cdot\nabla U,
$$
$$
G\nabla^{2}\left(\nabla\cdot\int\frac{\rho^{*'}}{\left|\boldsymbol{x}-\boldsymbol{x}'\right|}\nabla'X'{\mathrm{d}}^{3}x'\right)=2\nabla^{2}\phi_{2}-4\pi G\nabla X\cdot\nabla\rho^{*},
$$
can be substituted in \eqref{appendix4}, which implies in:
$$
\nabla^{2}\phi_{w}=-2\nabla^{2}U^2+3\nabla^{2}\phi_{2}-2U_{,ij}X_{,ij},
$$
this is exactly \eqref{appendix2}.

\section{A derivation of eq. (\ref{vlasovppn})}\label{appendix3}
Take the map $\left(x^{\mu},V^{i}\right)\rightarrow\left(x^{\mu},v^{i}\left(x^{\mu},V^{i}\right)\right)$, where
\begin{equation}\label{trivelocity}
v^{i}:=\frac{\mathrm{d}x^{i}}{\mathrm{d}t}=\frac{cV^{i}}{V^{0}\left(x^{\mu},V^{i}\right)},
\end{equation}
and rewrite \eqref{vlasov} in the new variables $\left(x^{\mu},v^{i}\right)$, then the partial derivatives of $f$ are expressed as
$$
\left(\frac{\partial f}{\partial x^{\mu}}\right)_{V^{i}}=\left(\frac{\partial f}{\partial x^{\mu}}\right)_{v^{i}}+\left(\frac{\partial f}{\partial v^{i}}\right)_{x^{\mu}}\left(\frac{\partial v^{i}}{\partial x^{\mu}}\right)_{V^{i}},
$$
$$
\left(\frac{\partial f}{\partial V^{j}}\right)_{x^{\mu}}=\left(\frac{\partial f}{\partial v^{i}}\right)_{x^{\mu}}\left(\frac{\partial v^{i}}{\partial V^{j}}\right)_{x^{\mu}}.
$$
A straightforward calculation, using eq. \eqref{trivelocity}, lead us to express
\begin{equation}\label{derivative1}
\left(\frac{\partial v^{i}}{\partial x^{\mu}}\right)_{V^{i}}=\frac{-cV^{i}}{\left({V^{0}}\right)^{2}}\left(\frac{\partial V^{0}}{\partial x^{\mu}}\right)_{V^{i}},
\end{equation}
\begin{equation}\label{derivative2}
\left(\frac{\partial v^{i}}{\partial V^{j}}\right)_{x^{\mu}}=\frac{c}{V^{0}}\left[{\delta^{i}}_{j}-\frac{V^{i}}{V^{0}}\left(\frac{\partial V^{0}}{\partial V^{j} }\right)_{x^{\mu}}\right].
\end{equation}

Now $f$, as defined, represent the probability density of a single massive particle, therefore the possible values $V^{\mu}$ can take is restricted to the positive light cone, where the shell condition holds:
\begin{equation}{\label{positivecone}}
g_{\mu\nu}V^{\mu}V^{\nu}=-c^{2},
\end{equation}
also $V^{0}>0$. This restriction completely defines $V^{0}$ in terms of $\left(x^{\mu},V^{i}\right)$ or $\left(x^{\mu},v^{i}\right)$, which to PPN order can be expressed by
\begin{multline*}
V^{0}=c\left\{1+\frac{U}{c^2}+\frac{\mathcal{W}}{2c^4}+\frac{\mathcal{Q}_{j}v^{j}}{c^4}+\frac{3v^4}{8c^4}+\frac{3U^2}{2c^4}
\right.\\\left.
+\left[1+\frac{\left(2\gamma+3\right) U}{c^2}\right]\frac{v^2}{2c^2}+O\left(c^{-6}\right)\right\},
\end{multline*}
also one can determine, from the latter and \eqref{trivelocity}, $V^{i}$ as an expression of $(x^{\mu},v^{i})$.

Some straightforward calculations determines the right-hand side of \eqref{derivative1} and \eqref{derivative2} only in terms of the new variables $(x^{\mu},v^{i})$ and to PPN order, as follows:
$$
\left(\frac{\partial v^{i}}{\partial x^{\mu}}\right)_{V^{k}}=-v^{i}\frac{U_{,\mu}}{c^2},
$$
$$
\left(\frac{\partial v^{i}}{\partial V^{j}}\right)_{x^{\mu}}=\left(1-\frac{U}{c^2}-\frac{v^2}{2c^2}\right)\delta^{i}_{j}-\frac{v^{i}v^{j}}{c^2}.
$$

With all these results in mind we can determine each term of the Vlasov equation in terms of $(x^{\mu},v^{i})$ to PPN order:
\begin{multline*}
\frac{V^{0}}{c}\left(\frac{\partial f}{\partial t}+\frac{\partial f}{\partial v^{i}}\frac{\partial v^{i}}{\partial t}\right)=\left[1+\frac{U}{c^2}+\frac{v^2}{2c^2}\right]\frac{\partial f}{\partial t}\\-v^{i}\frac{\partial f}{\partial v^{i}}\frac{U_{,t}}{c^2},
\end{multline*}
\begin{multline*}
V^{i}\left(\frac{\partial f}{\partial x^{i}}+\frac{\partial f}{\partial v^{j}}\frac{\partial v^{j}}{\partial x^{i}}\right)=v^{i}\left[1+\frac{U}{c^2}+\frac{v^2}{2c^2}\right]\frac{\partial f}{\partial x^{i}}
\\
-v^{i}v^{j}\frac{\partial f}{\partial v^{j}}\frac{U_{,i}}{c^2},
\end{multline*}
\begin{multline*}
\Gamma^{j}_{00}V^{0}V^{0}\frac{\partial f}{\partial v^{i}}\frac{\partial v^{i}}{\partial V^{j}}=\frac{\partial f}{\partial v^{i}}\left\{-\frac{1}{2}\frac{\mathcal{W}_{,i}}{c^2}+\frac{\mathcal{Q}_{i,t}}{c^2}+\frac{2\gamma}{c^2}UU_{,i}
\right.\\\left.
-U_{,i}\left(1+\frac{U}{c^2}+\frac{v^2}{2c^2}\right)+U_{,j}\frac{v^{i}v^{j}}{c^2}\right\},
\end{multline*}
$$
\Gamma^{j}_{l0}V^{l}V^{0}\frac{\partial f}{\partial v^{i}}\frac{\partial v^{i}}{\partial V^{j}}=\frac{1}{2c^2}\left[\mathcal{Q}_{j,l}+2\gamma U_{,t}\delta_{lj}-\mathcal{Q}_{l,j}\right]v^{l}\frac{\partial f}{\partial v^{j}},
$$
$$
\Gamma^{j}_{lp}V^{l}V^{p}\frac{\partial f}{\partial v^{i}}\frac{\partial v^{i}}{\partial V^{j}}=\gamma\left[U_{,l}\delta_{jp}+U_{,p}\delta_{jl}-U_{,j}\delta_{lp}\right]\frac{v^{l}v^{p}}{c^2}\frac{\partial f}{\partial v^{j}}.
$$

Substituting each of these expressions in eq. \eqref{vlasov} we find the Vlasov equation to PPN order:
\begin{multline*}
\left[1+\frac{U}{c^2}+\frac{v^2}{2c^2}\right]\left(\frac{\partial f}{\partial t}+v^{i}
\frac{\partial f}{\partial x^{i}}\right)+U_{,i}\frac{\partial f}{\partial v^{i}}
\\
-\frac{\partial f}{\partial v^{i}}\frac{v^{i}}{c^2}U_{,t}\left(1+2\gamma\right)-
\frac{\partial f}{\partial v^{i}}\frac{v^{i}v^{j}}{c^2}U_{,j}\left(2+2\gamma\right)
\\
-\frac{\partial f}{\partial v^{i}}\left\{-\frac{1}{2}\frac{\mathcal{W}_{,i}}{c^2}+
\frac{\mathcal{Q}_{i,t}}{c^2}+\frac{2\gamma-1}{c^2}UU_{,i}\right\}
\\
-\left[\mathcal{Q}_{j,l}-\mathcal{Q}_{l,j}\right]\frac{v^{l}}{c^2}
\frac{\partial f}{\partial v^{j}}+\left(\gamma+\frac{1}{2}\right) U_{,j}\frac{v^{2}}{c^2}\frac{\partial f}{\partial v^{j}}=0,
\end{multline*}
which is exactly eq. \eqref{vlasovppn}.
\section{Derivation of the conditions (\ref{condition1}) and (\ref{condition2})}\label{appendixc}
We start from eq. \eqref{condcircorbits1} and calculate each term,
\begin{align*}
\left.\mathcal{W}_{,\theta}\right|_{\theta=\frac{\pi}{2}}&=2\alpha_{2}\left(X_{,rr}-\frac{X_{,r}}{r}\right)\left.\left(w^{r}{w^{r}}_{,\theta}\right)\right|_{\theta=\frac{\pi}{2}}\\
&=-2\alpha_{2}\left(X_{,rr}-\frac{X_{,r}}{r}\right)w^{z}\left.w^{r}\right|_{\theta=\frac{\pi}{2}},
\end{align*}
$$
\left.\Phi^{\text{{PF}}}_{\theta,\varphi}\right|_{\theta=\frac{\pi}{2}}=\left(\alpha_{2}\frac{X_{,r}}{r}-\frac{\alpha_{1}}{2}U\right)\left.w_{\theta,\varphi}\right|_{\theta=\frac{\pi}{2}}=0,
$$
$$
\left.\Phi^{\text{{PF}}}_{\varphi,\theta}\right|_{\theta=\frac{\pi}{2}}=\left(\alpha_{2}\frac{X_{,r}}{r}-\frac{\alpha_{1}}{2}U\right)\left.w_{\varphi,\theta}\right|_{\theta=\frac{\pi}{2}}=0.
$$

Therefore substituting the previous results in eq. \eqref{condcircorbits1} we find condition \eqref{condition1}.

The next condition \eqref{condition2} we find by comparing eq. \eqref{circmotion} with \eqref{condcircorbits2}. First we take the derivative with time $\left(t\right)$ of eq. \eqref{circmotion}:
\begin{multline}\label{derivativeofcirc}
\left.\ddot{\varphi}\right|_{\theta=\frac{\pi}{2}} =\frac{1}{2}\left.\left[\left(\Phi^{\text{{PF}}}_{r,\varphi\varphi}-\Phi^{\text{{PF}}}_{\varphi,r\varphi}\right)\frac{\sqrt{-rU_{,r}}}{r^2c^2}
-\frac{\mathcal{W}_{,r\varphi}}{2rc^2}
\right]\right|_{\theta=\frac{\pi}{2}}.
\end{multline}
Now calculate each term of eq. \eqref{derivativeofcirc} and eq. \eqref{condcircorbits2}, leading to the following:
\begin{align*}
\left.\Phi^{\text{{PF}}}_{r,\varphi\varphi}\right|_{\theta=\frac{\pi}{2}}&=\left(\alpha_{2}\frac{X_{,r}}{r}-\frac{\alpha_{1}}{2}U\right)\left.w_{r,\varphi\varphi}\right|_{\theta=\frac{\pi}{2}}
\\
&=-\left(\alpha_{2}\frac{X_{,r}}{r}-\frac{\alpha_{1}}{2}U\right)\left.w_{r}\right|_{\theta=\frac{\pi}{2}},
\end{align*}
\begin{align*}
\left.\Phi^{\text{{PF}}}_{\varphi,r\varphi}\right|_{\theta=\frac{\pi}{2}}&=\left(\alpha_{2}X_{,r}-\frac{\alpha_{1}}{2}rU\right)_{,r}\left.\left(\frac{w_{\varphi}}{r}\right)_{,\varphi}\right|_{\theta=\frac{\pi}{2}}
\\
&=-\left(\alpha_{2}X_{,rr}-\frac{\alpha_{1}}{2}U-\frac{\alpha_{1}}{2}rU_{,r}\right)\left.w_{r}\right|_{\theta=\frac{\pi}{2}},
\end{align*}
\begin{align*}
\left.\mathcal{W}_{,\varphi}\right|_{\theta=\frac{\pi}{2}}&=2\alpha_{2}\left(X_{,rr}-\frac{X_{,r}}{r}\right)\left.\left(w^{r}{w^{r}}_{,\varphi}\right)\right|_{\theta=\frac{\pi}{2}}
\\
&=2\alpha_{2}\left(X_{,rr}-\frac{X_{,r}}{r}\right)\left.\left(w_{r}\frac{w_{\varphi}}{r}\right)\right|_{\theta=\frac{\pi}{2}},
\end{align*}
$$
\left.\mathcal{W}_{,r\varphi}\right|_{\theta=\frac{\pi}{2}}=2\alpha_{2}\left(X_{,rrr}-\frac{X_{,rr}}{r}+\frac{X_{,r}}{r^2}\right)\left.\left(w_{r}\frac{w_{\varphi}}{r}\right)\right|_{\theta=\frac{\pi}{2}}.
$$
Since \eqref{derivativeofcirc}=\eqref{condcircorbits2} we must have
$$
\left.\left[\left(\Phi^{\text{{PF}}}_{r,\varphi\varphi}-\Phi^{\text{{PF}}}_{\varphi,r\varphi}\right)\sqrt{-rU_{,r}}
-\frac{r\mathcal{W}_{,r\varphi}}{2}
\right]\right|_{\theta=\frac{\pi}{2}}=\left.\mathcal{W}_{,\varphi}\right|_{\theta=\frac{\pi}{2}},
$$
now substitute the previous results to find:
\begin{multline*}
\left(\alpha_{2}X_{,rr}-\alpha_{2}\frac{X_{,r}}{r}-\frac{\alpha_{1}}{2}rU_{,r}\right)\sqrt{-rU_{,r}}\left.w_{r}\right|_{\theta=\frac{\pi}{2}}
\\
-\alpha_{2}rX_{,rrr}\left.\left(w_{r}\frac{w_{\varphi}}{r}\right)\right|_{\theta=\frac{\pi}{2}}
\\
=\alpha_{2}\left(X_{,rr}-\frac{X_{,r}}{r}\right)\left.\left(w_{r}\frac{w_{\varphi}}{r}\right)\right|_{\theta=\frac{\pi}{2}},
\end{multline*}
here the choice of $\vec{w}$ shouldn't influence the result, therefore by employing condition \eqref{condition1} in the previous expression we find
$$
\alpha_{2}\left.w^{\varphi}\right|_{\theta=\pi/2}X_{,rrr}=-\alpha_{1}U_{,r}\sqrt{-rU_{,r}},
$$
which is exactly \eqref{condition2}.

\section{Energy like integral of motion}\label{appendixenergy}

In Sect. \ref{sec3sub3} we show that, for the situations analyzed here, it is sufficient to consider DFs depending only
on energy.
Therefore we need to determine an expression for the conserved energy in the PPN formalism. To do so we follow the same procedure used in \cite{ramos1}, which consist of starting from the Lagrangian,
$$
\mathcal{L}=\frac{1}{2}\mathrm{g}_{\mu\nu}V^{\mu}V^{\nu},
$$
which, for the PPN metric \eqref{metric00}-\eqref{metricij}, takes the form
\begin{multline}
\mathcal{L}=\frac{m^2c^2}{2}\left[\left(-1+\frac{2U}{c^2}+\frac{\mathcal{W}}{c^4}\right)\dot{t}^{2}
\right.\\\left.
+\left(1+\frac{2\gamma U}{c^2}\right)\delta_{ij}\frac{V^{i}V^{j}}{c^2}\right].
\end{multline}
where $\dot{t}=\mathrm{d}x^{0}/\mathrm{d}\tau$.
Since the metric does not depend on $x^{0}$, then ${\partial \mathcal{L}}/{\partial \dot{t}}$ must be a conserved quantity.

By  defining the energy to be, \cite{ramos1},
$$
E:=-\frac{1}{m^{2}}\frac{\partial \mathcal{L}}{\partial \dot{t}}-c^{2},
$$
we find, to the order of the PPN, that $E$ can be cast as
$$
E=E_{\text{N}}+E_{\text{PPN}},
$$
where $E_{\text{N}}$ represents the Newtonian expression for the energy,
$$
E_{\text{N}}=-U+\frac{v^2}{2},
$$
and $E_{\text{PPN}}$ is the PPN correction
$$
E_{\text{PPN}}=\left(\gamma+\frac{1}{2}\right)\frac{Uv^2}{c^2}-\frac{\Psi}{c^2}-
\left(\frac{1}{2}-\beta\right)\frac{U^{2}}{c^{2}}-\frac{\Phi^{\text{PF}}}{2c^2}+\frac{3v^4}{8c^2}.
$$
As it is usual, one can assume that $E_{\text{N}}\gg E_{\text{PPN}}$.

\footnotesize
%\bibliography{stabgr_refs}

%\bibliographystyle{physrev}
%\bibliographystyle{h-physrev4}
%\bibstyle{h-physrev4}
%\bibliographystyle{apsrev}

\end{document}